\documentclass[useAMS,usenatbib]{mn2e}
\usepackage{graphicx}
\usepackage{subfigure}
\newcommand       \be           {\begin{equation}}
\newcommand       \ee           {\end{equation}}
\newcommand       \bea          {\begin{eqnarray}}
\newcommand       \eea          {\end{eqnarray}}

\def\simlt{\mathrel{\hbox{\rlap{\hbox{\lower4pt\hbox{$\sim$}}}\hbox{$<$}}}}
\def\simgt{\mathrel{\hbox{\rlap{\hbox{\lower4pt\hbox{$\sim$}}}\hbox{$>$}}}}

\def\simlt{\mathrel{\hbox{\rlap{\hbox{\lower4pt\hbox{$\sim$}}}\hbox{$<$}}}}
\def\simgt{\mathrel{\hbox{\rlap{\hbox{\lower4pt\hbox{$\sim$}}}\hbox{$>$}}}}
\def\lesssim{\mathrel{\hbox{\rlap{\hbox{\lower4pt\hbox{$\sim$}}}\hbox{$<$}}}}

\def\gtrsim{\mathrel{\hbox{\rlap{\hbox{\lower4pt\hbox{$\sim$}}}\hbox{$>$}}}}

\def\simlt{\mathrel{\hbox{\rlap{\hbox{\lower4pt\hbox{$\sim$}}}\hbox{$<$}}}}
\def\simgt{\mathrel{\hbox{\rlap{\hbox{\lower4pt\hbox{$\sim$}}}\hbox{$>$}}}}

\title[]{The Proto-Magnetar Model for Gamma-Ray Bursts} \author[B.~D.~Metzger, D.~Giannios, T.~A.~Thompson, N.~Bucciantini, $\&$ E.~Quataert]{B.~D. Metzger$^{1,5}$\thanks{E-mail: bmetzger@astro.princeton.edu}, D.~Giannios$^{1}$, T.~A.~Thompson$^{2}$, N.~Bucciantini$^{3}$, E.~Quataert$^{4}$\\$^{1}$Department of Astrophysical Sciences, Peyton Hall, Princeton University, Princeton, NJ 08544, USA \\$^{2}$Department of Astronomy and Center for Cosmology $\&$ Astro-Particle Physics, The Ohio State University, Columbus, OH 43210, USA\\$^{3}$NORDITA, AlbaNova University Center, Roslagstullsbacken 23, SE 10691 Stockholm, Sweden\\$^{4}$Astronomy Department and Theoretical Astrophysics Center, 601 Campbell Hall, University of California, Berkeley, CA 94720, USA\\$^{5}$NASA Einstein Fellow}
\begin{document}
\date{Accepted . Received ; in original form }
\pagerange{\pageref{firstpage}--\pageref{lastpage}} \pubyear{????}
\maketitle
\label{firstpage}

\begin{abstract}
Long duration Gamma-Ray Bursts (GRBs) originate from the core collapse of massive stars, but the identity of the central engine remains elusive.  Previous work has shown that rapidly spinning, strongly magnetized proto-neutron stars (`millisecond proto-magnetars') produce outflows with energies, timescales, and magnetizations $\sigma_{0}$ (maximum Lorentz factor) that are consistent with those required to produce long duration GRBs.  Here we extend this work in order to construct a self-consistent model that directly connects the properties of the central engine to the observed prompt emission.  Just after the launch of the supernova shock, a wind heated by neutrinos is driven from the proto-magnetar.  The outflow is collimated into a bipolar jet by its interaction with the progenitor star.  As the magnetar cools, the wind becomes ultra-relativistic and Poynting-flux dominated ($\sigma_{0} \gg 1$) on a timescale comparable to that required for the jet to clear a cavity through the star.  Although the site and mechanism of the prompt emission are debated, we calculate the emission predicted by two models: magnetic dissipation and shocks.  

Magnetic reconnection may occur near the photosphere if the outflow develops an alternating field structure due to e.g.~magnetic instabilities or a misalignment between the magnetic and rotation axes.  Shocks may occur at larger radii because the Lorentz factor of the wind increases with time, such that the faster jet at late times collides with slower material released earlier.  Our results favor magnetic dissipation as the prompt emission mechanism, in part because it predicts a relatively constant `Band' spectral peak energy $E_{\rm peak}$ with time during the GRB.  The baryon loading of the jet decreases abruptly when the neutron star becomes transparent to neutrinos at $t = t_{\rm\nu-thin} \sim 10-100$ seconds.  Jets with ultra-high magnetization cannot effectively accelerate and dissipate their energy, which suggests this transition ends the prompt emission.  This correspondence may explain both the typical durations of long GRBs and the steep decay phase that follows.  Residual rotational or magnetic energy may continue to power late time flaring or afterglow emission, such as the X-ray plateau.  We quantify the emission predicted from proto-magnetars with a wide range of physical properties (initial rotation period, surface dipole field strength, and magnetic obliquity) and assess a variety of phenomena potentially related to magnetar birth, including low luminosity GRBs, very luminous GRBs, thermal-rich GRBs/X-ray Flashes, very luminous supernovae, and short duration GRBs with extended emission.  
\end{abstract}


\vspace{-0.7cm}
\section{Introduction}

\begin{figure*}
\resizebox{\hsize}{!}{\includegraphics[]{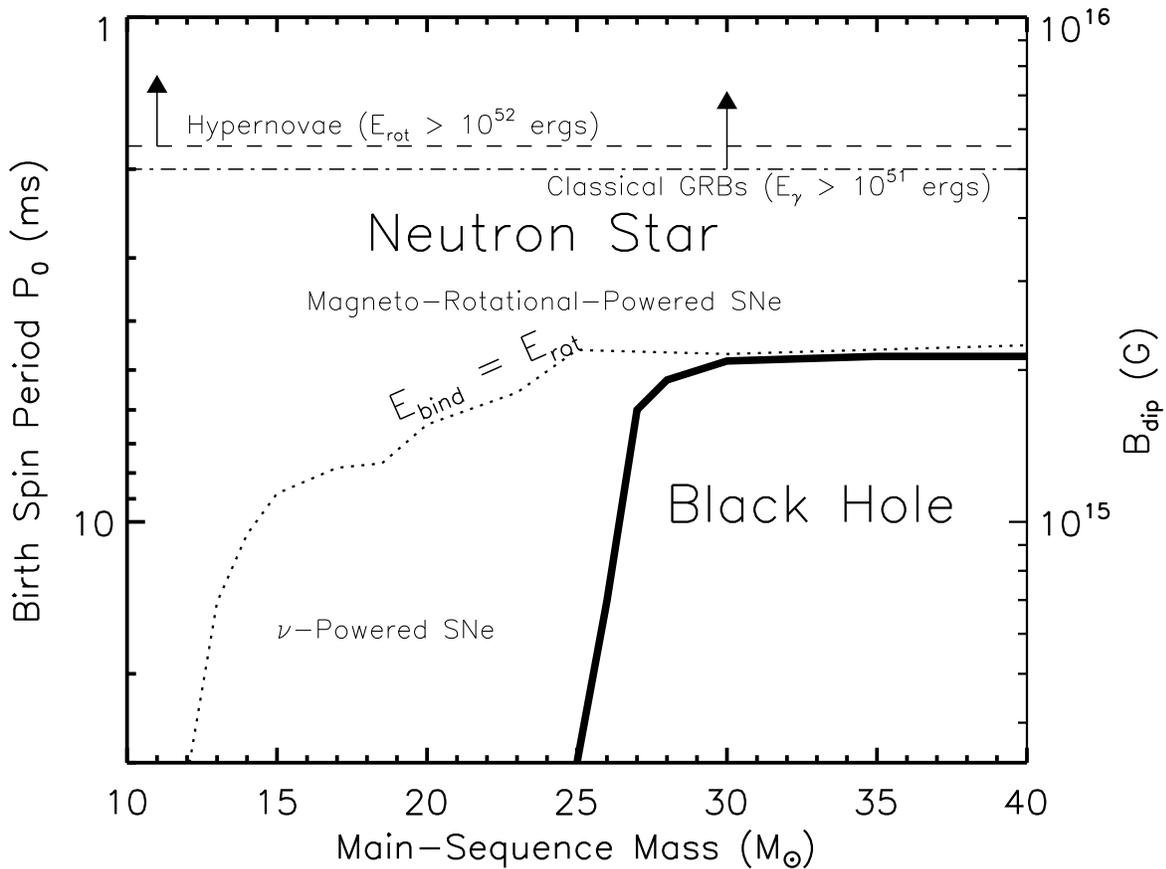}}
\caption{Schematic diagram of the regimes of neutron star versus black hole formation in core collapse SNe at sub-solar metallicities ({\it solid line}) in the space of main sequence mass and initial proto-NS spin period $P_{0}$, taking into account the possible effects of rapid rotation and strong magnetic fields.  The dotted line denotes the rotation rate above which the NS rotational energy $E_{\rm rot}$ (eq.~[\ref{eq:erot}]) exceeds the gravitational binding energy of the progenitor envelope.  The dashed line denotes the rotational energy $E_{\rm rot} = 10^{52}$ ergs sufficient to power a `hypernova'.  The right axis shows the magnetic field strength $B_{\rm dip}$ that would be generated if the magnetic energy in the dipole field is $\sim 0.1\%$ of $E_{\rm rot}$ (eq.~[\ref{eq:beq}]).  The dot-dashed line is the minimum rotation rate required for a magnetar with a field strength $B_{\rm dip}$ to produce a classical GRB with energy $E_{\gamma} > 10^{51}$ ergs, based on the model presented in $\S\ref{sec:GRB}$.}
\label{fig:diagram1}
\end{figure*}

Soon following the discovery of Gamma-Ray Bursts (GRBs; \citealt{Klebesadel+73}), there were possibly more theories for their origin than theorists \citep{Ruderman75}.  However, once GRBs were confirmed to originate from cosmological distances (e.g.~\citealt{Metzger+97}), the joint requirements of supernova-scale energies, short (millisecond) timescales, and relativistic speeds significantly narrowed the list of plausible central engines.  It is now generally accepted that GRBs result from the formation or catastrophic rearrangement of stellar-mass black holes (BHs) or neutron stars (NSs).  This conclusion has only been strengthened in recent years due to the much richer picture of the prompt and afterglow emission provided by the {\it Swift} and {\it Fermi} missions.  However, despite a wealth of new data, the identity of the central engine remains elusive. 

At least some long duration GRBs originate from the deaths of very massive stars \citep{Woosley&Bloom06}, as confirmed by their observed association with energetic core collapse supernovae (SNe) (e.g.~\citealt{Galama+98}; \citealt{Bloom+99}; \citealt{Stanek+03}; \citealt{Chornock+10}; \citealt{Starling+10}).  It nevertheless remains unsettled whether the central engine is a rapidly accreting BH (\citealt{Woosley93}; \citealt{MacFadyen&Woosley99}; \citealt{Nagataki+07}; \citealt{Barkov&Komissarov08}; \citealt{Lindner+10}) or a rapidly spinning, strongly magnetized NS (a `millisecond magnetar'; \citealt{Usov92}; \citealt{Thompson94}; \citealt{Blackman&Yi98}; \citealt{Wheeler+00}; \citealt{Zhang&Meszaros01}; \citealt{Thompson+04}; \citealt{Metzger+07}; \citealt{Bucciantini+07,Bucciantini+08,Bucciantini+09}).  Although much less is known about the origin of short duration GRBs, the properties of their host galaxies and their notable lack of an accompanying SN are consistent with an origin associated with the merger of NS-NS and NS-BH binaries (\citealt{Hjorth+05}; \citealt{Bloom+06}; \citealt{Berger+05}; see e.g.~\citealt{Berger10} for a recent review).  However, the unexpected discovery that many short GRBs are followed by an energetic X-ray `tail' lasting $\sim 100$ seconds has challenged basic predictions of the merger model (e.g.~\citealt{Gehrels+06}; \citealt{Gal-Yam+06}; \citealt{Perley+09}) and may hint at an alternative origin for some events, such as magnetar formation via the accretion-induced collapse (AIC) of a white dwarf \citep{Metzger+08b}.  

The large range in length scales and the complexity of the physics involved in producing a GRB have thus far prevented all steps in the phenomena from being studied in a single work.  Any attempt to construct a `first principles' model is hindered by uncertain intermediate steps relating the physics of the central engine to the properties of the relativistic jet and the gamma-ray emission mechanism.  Nevertheless, in this paper we argue that the magnetar model is uniquely predictive.  This allows us to construct a self-consistent model which can in principle be compared directly with observations.  Although we focus on magnetars formed via the core collapse of massive stars, we also apply our results to AIC ($\S\ref{sec:AIC}$).  Our primary conclusion is that a remarkable fraction of GRB properties find natural explanations within the proto-magnetar model.

\subsection{Black Hole vs.~Magnetar}  
\label{sec:BHvsNS}

In the original collapsar model, \citet{Woosley93} envisioned a `failed supernova,' in which the energy released by core collapse is insufficient to unbind the majority of the star, such that a black hole necessarily forms.  If the collapsing envelope has sufficient angular momentum, it accretes through a centrifugally-supported disk.  Energy released by accretion, or via the accretion-mediated extraction of the black hole's spin \citep{Blandford&Znajek77}, then powers a relativistic jet, which burrows through the star and ultimately powers the GRB at larger radii (\citealt{MacFadyen&Woosley99}; \citealt{Proga&Begelman03}; \citealt{Matzner03}; \citealt{Morsony+07}).  

The discovery that long GRBs are accompanied by hyper-energetic ($\sim 10^{52}$ erg) SNe propelled the collapsar model to the theoretical forefront.  However, it also proved, somewhat ironically, that GRB-SNe are far from the complete `failures' envisioned by \citet{Woosley93}.  Indeed, if the collapsar scenario is correct, then either (1) the BH forms promptly following stellar collapse and the explosion mechanism associated with GRB-SNe is fundamentally different than that associated with the death of normal (slower rotating) stars, which are instead powered by NS formation; or (2) a BH forms only after several seconds delay, due to the `fall-back' of material that remains gravitationally bound despite a successful and energetic SN (e.g.~\citealt{Chevalier93}; \citealt{Fryer99}; \citealt{Zhang+08}; \citealt{Moriya+10}).  

Modern core collapse simulations find that the shock produced at core bounce initially stalls due to neutrino and photo-dissociation losses (e.g.~\citealt{Rampp&Janka00}; \citealt{Liebendorfer+01}; \citealt{Thompson+03}).  It has long been thought that neutrino heating from the proto-NS may revive the shock, resulting in a successful explosion \citep{Bethe&Wilson85}.  Recent simulations suggest that the neutrino mechanism may work for low mass progenitors (e.g.~\citealt{Scheck+06}), but higher mass stars appear more difficult to explode.  Although multi-dimensional effects not captured by present simulations may be a crucial missing ingredient (e.g.~\citealt{Nordhaus+10}), neutrinos alone may well prove incapable of powering $\sim 10^{52}$ erg explosions.

GRB progenitors are, however, far from typical.  Essentially all central engine models require rapid rotation and a strong, large-scale magnetic field ($\gtrsim 10^{15}$ G; e.g.~\citealt{McKinney06}).  These ingredients may go hand-in-hand in core collapse because differential rotation provides a source of free energy to power field growth, via e.g. an $\alpha-\Omega$ dynamo in the convective proto-NS \citep{Duncan&Thompson92} or the magneto-rotational instability (MRI; e.g.~\citealt{Akiyama+03}; \citealt{Thompson+05}).  The crucial question then arises:  {\it Do SNe indeed fail and lead to BH formation if the progenitor core is rapidly rotating?} or stated more directly: {\it Are the requisite initial conditions for the collapsar model self-consistent?}

An additional energy reservoir (rotation) and means for extracting it (magnetic fields) make magneto-rotational effects a more promising way to produce hypernovae than neutrinos alone (e.g.~\citealt{LeBlanc&Wilson70}; \citealt{Symbalisty84}; \citealt{Ardeljan+05}).  Only recently, however, have simulations begun to capture the combined effects of MHD and neutrino heating (e.g.~\citealt{Burrows+07}).  

\citet{Dessart+08}, hereafter D08, calculate the collapse of a rotating 35$M_{\sun}$ ZAMS collapsar progenitor of \citet{Woosley&Heger06}, which they endow with a pre-collapse magnetic field that results in a $\sim 10^{15}$ G field strength when compressed to NS densities.  This reproduces the field strength, if not the field topology, expected from the saturated state of the MRI.  Soon after core bounce, a bipolar MHD-powered outflow develops from the proto-NS.  Although the explosion is not initially successful over all solid angles, matter continues to accrete through an equatorial disk.  By accreting angular momentum, the NS remains rapidly spinning, which in turn enhances the mass loss from higher latitudes due to magneto-centrifugal slinging (e.g.~\citealt{Thompson+04}; \citealt{Metzger+07}; see eq.~[\ref{eq:fcentmax}]).  Importantly, in the strongly magnetized model of D08, the wind mass loss rate eventually exceeds the accretion rate, such that for $t \gtrsim 300$ ms the NS mass begins {\it decreasing}.  Although D08 cannot address the possibility of later fall-back, and a different progenitor angular momentum profile could change the conclusion, their result is nonetheless suggestive: a core self-consistently endowed with the properties required to produce a GRB may not leave a BH at all.  The results of D08 highlight the fact that BH versus~NS formation may not be a function of progenitor mass and metallicity alone.  Delineating this dichotomy more definitively will, however, require addressing challenging theoretical issues, such as the precise mechanism responsible for amplifying the magnetic field (see \citealt{Spruit08} for a discussion).  

Figure \ref{fig:diagram1} is a schematic diagram of the possible effects of rapid rotation and strong magnetic fields on the regimes of NS versus BH formation as a function of main-sequence stellar mass $M_{\star}$ and the initial NS rotation period $P_{0}$.  The collapse of slowly rotating, low mass stars may result in a normal SN with kinetic energy $\sim 10^{51}$ ergs powered by neutrinos.  For higher mass stars, however, neutrino-powered explosions are less likely (or are accompanied by significant `fall-back' accretion) due to more massive, compact iron cores and higher envelope binding energies $E_{\rm bind}$.  For these reasons it has been argued that stars with $M_{\star} \gtrsim 25M_{\sun}$ leave BH remnants at the sub-solar metallicities that appear to characterize GRB progenitors (e.g.~\citealt{Fryer99}; \citealt{Heger+03}; \citealt{OConnor&Ott10}).  

Above the dashed line in Figure \ref{fig:diagram1}, however, the rotational energy $E_{\rm rot}$ of the proto-NS (eq.~[\ref{eq:erot}]) exceeds the binding energy of the stellar envelope, where
\begin{eqnarray}
E_{\rm rot} &\simeq& (1/2)I\Omega^{2} \nonumber \\ &\approx& 3\times 10^{52}{\rm ergs}\left(\frac{M_{\rm ns}}{1.4M_{\sun}}\right)\left(\frac{R_{\rm ns}}{12{\,\rm km}}\right)^{2}\left(\frac{P}{{\rm ms}}\right)^{-2},
\label{eq:erot}
\end{eqnarray}
and $I = (2/5)M_{\rm ns}R_{\rm ns}^{2}$, $M_{\rm ns}$, $R_{\rm ns}$, and $\Omega = 2\pi/P$ are the NS moment of inertia, mass, radius, and rotation rate, respectively.  We have defined $E_{\rm bind}$ exterior to $1.8 M_{\sun}$, as calculated by \citet{Dessart+10} from the stellar profiles of \citet{Woosley+02}.  Although the efficiency with which $E_{\rm rot}$ couples to the SN shock depends on uncertain details during the first few hundred milliseconds after core bounce, if $E_{\rm rot} > E_{\rm bind}$ then a NS remnant could in principle result, even for very massive stars.  The hypothetical boundary between NS and BH formation based on the above discussion is shown with a solid line in Figure \ref{fig:diagram1}.  We note that there is indeed evidence that some Galactic magnetars may have stellar progenitors with masses $\gtrsim 40M_{\sun}$ (\citealt{Muno+06}), although (consistent with Fig.~\ref{fig:diagram1}) this does not exclusively appear to be the case (\citealt{Davies+09}). 

If an MHD-powered SN does not leave a BH, then a rapidly spinning, strongly magnetized NS (a `proto-magnetar') may instead remain behind the outgoing SN shock.  The rotational energy $E_{\rm rot} \gtrsim 10^{52}$ ergs of a magnetar with $P_{0} \sim 1$ ms is more than sufficient to power most long GRBs.  However, not all of this energy is available to produce high energy emission; a fraction of $E_{\rm rot}$, for instance, is expended as the jet emerges from the star or is used to power an accompanying hypernova ({\it dashed line}; Fig.~\ref{fig:diagram1}).  The right axis in Figure \ref{fig:diagram1} shows the magnetic field strength $B_{\rm eq}$ that would be generated if the magnetic energy in the dipole field is $\sim 0.1\%$ of $E_{\rm rot}$ (eq.~[\ref{eq:beq}]).  A dot-dashed line shows the minimum rotation rate required to produce a classical GRB from a magnetar with a field strength $B_{\rm dip}$, based on the model presented in $\S\ref{sec:GRB}$.  The conditions for a hypernova and a GRB from a proto-magnetar are thus remarkably similar. 

\begin{figure*}
\resizebox{\hsize}{!}{\includegraphics[]{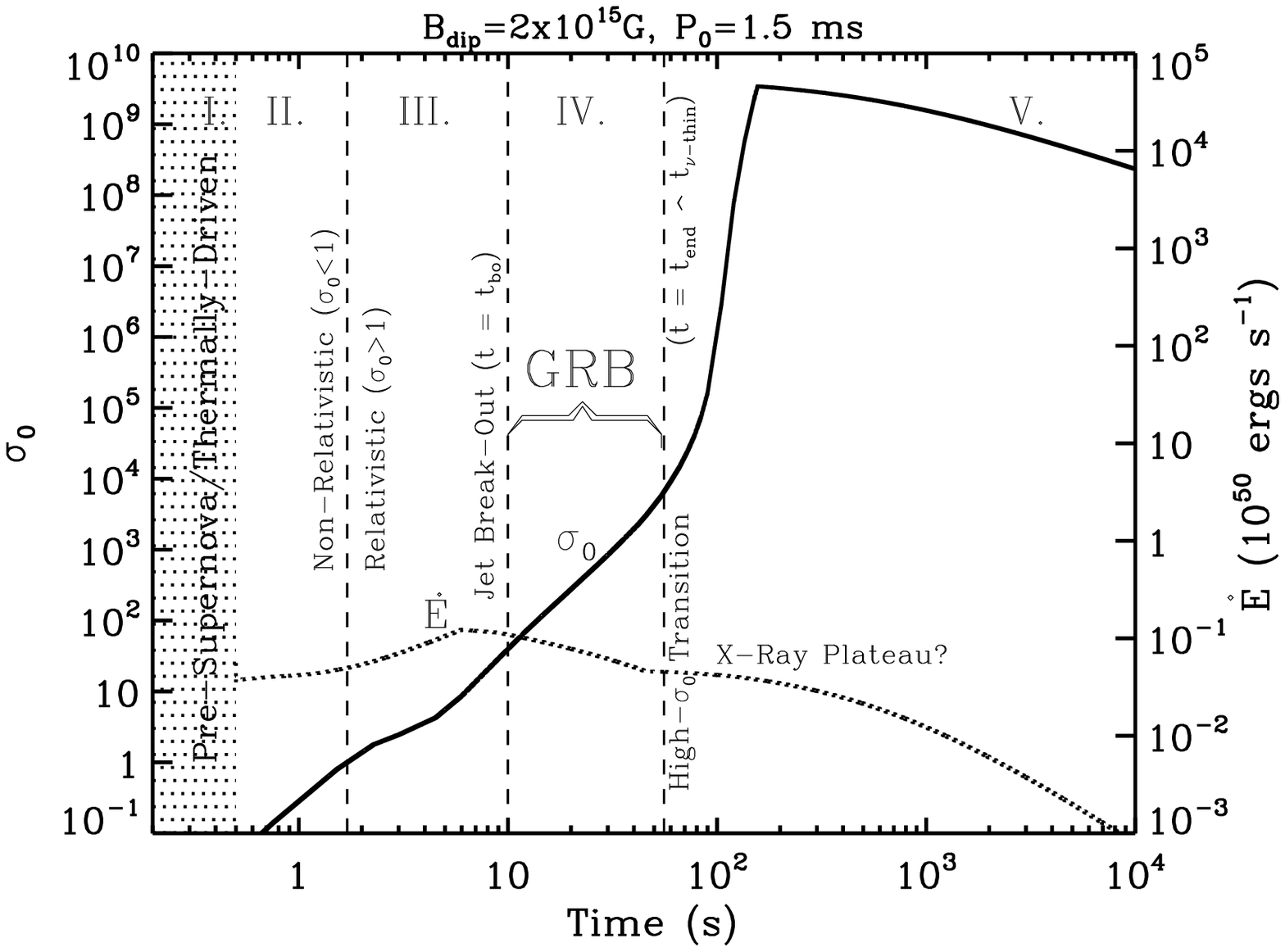}}
\caption{Wind power $\dot{E}$ ({\it right axis}) and magnetization $\sigma_{0}$ ({\it left axis}; eq.~[\ref{eq:sigma}]) of the proto-magnetar wind as a function of time since core bounce, calculated for a neutron star with mass $M_{\rm ns} = 1.4M_{\sun}$, initial spin period $P_{0} = 1.5$ ms, surface dipole field strength $B_{\rm dip} = 2\times 10^{15}$ G, and magnetic obliquity $\chi = \pi/2$.  Stages denoted I$.-$V.~are described in detail in $\S\ref{sec:stages}$.}
\label{fig:edotsig}
\end{figure*}
  

\subsection{Summary of the Magnetar Model and This Paper}

In this section we summarize the organization of the paper and orient the reader with a brief description of the model timeline (more details and references are provided in subsequent sections).

In $\S\ref{sec:windevo}$ we present calculations of the time-dependent properties of proto-magnetar winds and quantify the stages of the proto-magnetar model.  The basic picture is summarized by Figure \ref{fig:edotsig}, which shows the wind power $\dot{E}$ and magnetization $\sigma_{0}$ (maximum Lorentz factor) as a function of time following core bounce, calculated for a proto-magnetar with a surface dipole magnetic field strength $B_{\rm dip} = 2\times 10^{15}$ G, initial spin period $P_{0} = 1.5$ ms, and magnetic obliquity $\chi = \pi/2$.  Changes in the wind properties with time are driven largely by the increase in $\sigma_{0}(t)$ as the proto-NS cools.

Within the first few hundred milliseconds following core bounce, a successful SN shock is launched by neutrino heating or MHD forces (Stage I).  Soon after, a wind heated by neutrinos expands freely from the NS surface into the cavity evacuated by the outgoing shock.  The wind is initially non-relativistic ($\sigma_{0} \lesssim 1$) because the neutrino-driven mass loss rate is high (Stage II).  However, as the proto-NS cools, $\sigma_{0}$ increases to $\gtrsim$ 1 and the wind becomes relativistic (Stage III).   The wind is collimated by its interaction with the star into a bipolar jet, which breaches the stellar surface after $\sim 10$ seconds.  After jet break-out, the relativistic magnetar wind is directed through a relatively clear channel out of the star and the GRB commences (Stage IV; $\S\ref{sec:GRB}$).  Averaging over variability imposed by e.g.~interaction with the jet walls ($\S\ref{sec:variability}$), the time evolution of the power and mass-loading of the jet match those set by the magnetar wind at much smaller radii.  In $\S\ref{sec:stages}$ we provide a more quantitative description of the individual model stages described above using an extensive parameter study of wind models.

Although the site and mechanism of prompt GRB emission remain uncertain, in $\S\ref{sec:GRB}$ we calculate the light curves and spectra within two emission models.  Depending on the means and efficacy of the jet's acceleration ($\S\ref{sec:acceleration}$), GRB emission may be powered by the dissipation of the jet's Poynting flux directly (`magnetic dissipation'; $\S\ref{sec:magdiss}$) near or above the photosphere; and/or via `internal shocks' within the jet at larger radii\footnote{In this paper we define `internal shocks' as those resulting from the interaction between the magnetar jet and the accumulated (slower) shell of material released at earlier times.  This is in contrast to the standard internal shock model (e.g.~\citealt{Rees&Meszaros94}), which invokes the singular interaction between shells with similar properties released immediately after one another.  As we discuss in $\S\ref{sec:internalshocks}$, the former dominate the latter in the magnetar model because the mean Lorentz factor of the jet increases monotonically in time.} ($\S\ref{sec:internalshocks}$).  As Figure \ref{fig:edotsig} makes clear, self-interaction in the jet is inevitable because $\sigma_{0}-$and hence the jet speed$-$increase monotonically as the proto-NS cools.  

After $t\sim 30-100$ seconds, $\sigma_{0}$ increases even more rapidly as the proto-NS becomes transparent to neutrino emission.  Because magnetic dissipation and jet acceleration become ineffective when $\sigma_{0}$ is very large, this abrupt transition likely ends the prompt GRB.  In $\S\ref{sec:highsig}$ we address the possibility that residual rotational or magnetic energy may continue to power late time flaring or afterglow emission, such as the X-ray plateau.  In $\S\ref{sec:discussion}$ we discuss the implications of our results for the diversity of GRB-related phenomena, including very luminous GRBs ($\S\ref{sec:VLGRBs}$), low luminosity GRBs ($\S\ref{sec:LLGRBs}$), thermal-rich GRBs/X-ray Flashes ($\S\ref{sec:XRF}$), Galactic magnetars ($\S\ref{sec:galactic}$), very luminous supernova ($\S\ref{sec:choked}$), and magnetar formation via AIC ($\S\ref{sec:AIC}$).  We summarize our conclusions in $\S\ref{sec:conclusions}$.

\begin{figure}
\resizebox{\hsize}{!}{\includegraphics[]{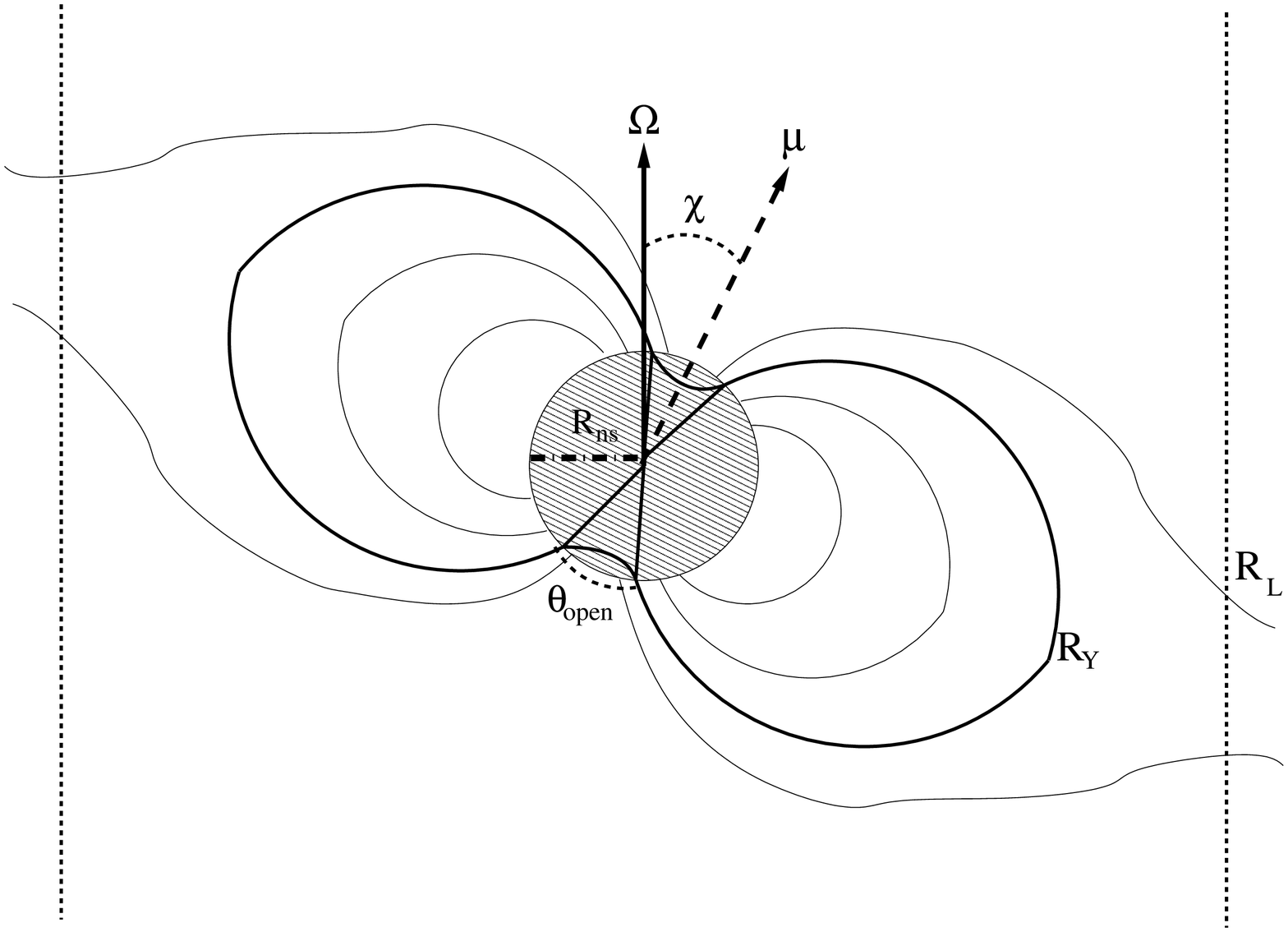}}
\caption{Geometry of magnetized proto-neutron star winds.  The neutron star radius $R_{\rm ns}$ is initially large ($\gtrsim 20$ km) following the launch of the supernova shock, but decreases to its final value $R_{\rm ns} \approx 12$ km in a few seconds (Fig.~\ref{fig:pons}).  The neutron star rotates at an angular velocity $\Omega = 2\pi/P$ about the vertical axis, where $P$ is the rotational period; the light cylinder radius is $R_{\rm L} = c/\Omega \simeq 50(P/{\rm ms})$ km.  The magnetic dipole moment $|\mu| = B_{\rm dip}R_{\rm ns}^{3}$ makes an angle $\chi$ with respect to the rotation axis.  The angle $\theta_{\rm open}$ defines the size of the open magnetosphere on the neutron star surface.  The magnetosphere is closed at angles $\theta > \theta_{\rm open}/2$ from the magnetic pole, while field lines with $\theta < \theta_{\rm open}/2$ form an `open' or `wind' zone along which matter may escape to infinity.  The size of the open zone affects both the spin-down rate and the mass loss rate from magnetized proto-neutron star winds.  The bundle of last closed field lines intersects the magnetic equator at the `Y' point radius $R_{\rm Y}$.  Ultra-relativistic, force-free winds ($\sigma_{0} \gg 1$) have $R_{\rm Y} \sim R_{\rm L}$, while less magnetized winds in general have $R_{\rm Y} < R_{\rm L}$ (see $\S\ref{sec:mdot}$ and Fig.~\ref{fig:radii}). }
\label{fig:nscartoon}
\end{figure}

\section{Proto-Magnetar Winds}
\label{sec:windevo}

In this section we present calculations of the time-dependent properties of magnetized proto-NS winds (\citealt{Thompson+04}; \citealt{Metzger+07}).  In $\S\ref{sec:windmodel}$ we summarize the model, which is similar to that presented in \citet{Metzger+07} but includes additional details not addressed in previous work.  Our results are presented in $\S\ref{sec:results}$.

\subsection{Evolutionary Wind Model}
\label{sec:windmodel}

\subsubsection{Model Description}


The two most important properties of the proto-magnetar wind are the mass loss rate $\dot{M}$ and the energy loss rate, or wind power, $\dot{E}$.  The wind power contains kinetic and magnetic (Poynting flux) components: $\dot{E} = \dot{E}_{\rm kin}$ + $\dot{E}_{\rm mag}$.  A related quantity, determined from $\dot{M}$ and $\dot{E}_{\rm mag}$, is the wind magnetization\footnote{Note that this definition may differ from that used elsewhere in the literature.  In particular, what we define as $\sigma_{0}$ is sometimes referred to as the `baryon loading' parameter (e.g.~\citealt{Drenkhahn&Spruit02}).}
\be
\sigma_{0} \equiv \frac{\phi^{2}\Omega^{2}}{\dot{M}c^{3}},
\label{eq:sigma}
\ee
where $\Omega$ is the NS rotation rate, $\phi \equiv B_{\rm r}r^{2}$ is the magnetic flux threading the open magnetosphere divided by 4$\pi$ steradians \citep{Michel69}, and $B_{\rm r} \sim$ the poloidal field strength.  As shown in Appendix \ref{sec:appendixA}, $\phi$ is directly related to the Poynting flux $\dot{E}_{\rm mag}$ (eqs.~[\ref{eq:edotmag}],[\ref{eq:phi}]).  The magnetization is important because it delineates non-relativistic ($\sigma_{0} \lesssim 1$) from relativistic ($\sigma_{0} \gtrsim 1$) outflows and affects the asymptotic partition between kinetic and magnetic energy in the wind.  In particular, in relativistic outflows most of the wind power resides in Poynting flux ($\dot{E}_{\rm mag} \gg \dot{E}_{\rm kin}$) at the fast magnetosonic surface.  The value of $\sigma_{0}$ in this case crucially affects the efficiency with which the jet may accelerate and dissipate its energy ($\S\ref{sec:acceleration}$) and is approximately equal to the outflow's maximum achievable Lorentz factor $\Gamma_{\rm max} \approx \dot{E}/\dot{M}c^{2} \simeq \sigma_{0}$.  

In Appendix \ref{sec:appendixA} we describe in detail how $\dot{E}$, $\dot{M}$, and $\sigma_{0}$ are determined in magnetized proto-NS winds.  To briefly summarize, mass loss during the first $t \sim 30-100$ seconds is caused by neutrino heating in the proto-NS atmosphere.  As a result, $\dot{M} \propto L_{\nu}^{5/3}\epsilon_{\nu}^{10/3}$ depends sensitively on the neutrino luminosity $L_{\nu}$ and the mean neutrino energy $\epsilon_{\nu}$ during the Kelvin-Helmholtz cooling phase (eq.~[\ref{eq:mdotnu}]).  In most cases we take $L_{\nu}(t)$ and $\epsilon_{\nu}(t)$ from the proto-NS cooling calculations of \citet{Pons+99} (see Fig.~\ref{fig:pons}), but modified by a `stretch factor' $\eta_{\rm s}$ (defined in eq.~[\ref{eq:stretch}]) that qualitatively accounts for the effects of rotation on the cooling evolution.
  
We assume that mass loss from the proto-NS occurs only from portions of the surface threaded by the open magnetic flux.  We assume a dipolar magnetosphere, bounded by the bundle of `last-closed' field lines which intersect the `Y' point radius in the magnetic equator (Figure \ref{fig:nscartoon} is an illustration of the relevant geometry).  We determine the dependence of the Y-point radius on the wind properties using results from the axisymmetric MHD simulations of \citet{Bucciantini+06}, which span the $\sigma_{0} < 1$ to $\sigma_{0} > 1$ transition.  Using numerical results from \citet{Metzger+08a}, we further account for the {\it enhancement} in $\dot{M}$ that occurs due to magneto-centrifugal forces in the heating region.  This effect is most important when the NS is rotating very rapidly ($P \lesssim 2$ ms) and the magnetic obliquity is large, such that the polar cap samples regions near the rotational equator.  After $t \equiv t_{\rm\nu-thin} \sim 30-100$ seconds, the proto-NS becomes transparent to neutrinos, which causes $L_{\nu}$ and $\epsilon_{\nu}$ to decrease sharply (Fig.~\ref{fig:pons}).  Once neutrino heating decreases sufficiently, other processes (e.g. $\gamma-B$ or $\gamma-\gamma$ pair production) likely take over as the dominant source of mass-loading (\citealt{Hibschman&Arons01}; \citealt{Thompson08}) and the wind composition may change from baryon- to pair-dominated.  Lacking a predictive model for $\dot{M}$ at late times, we assume that $\dot{M}$ scales with the \citet{Goldreich&Julian69} flux for a fixed value of the pair multiplicity $\mu_{-+} = 10^{6}$.  Our conclusions are fortunately insensitive to this choice (see $\S\ref{sec:highsig}$).  The full expression for $\dot{M}$ is given in equation (\ref{eq:mdot}).

Proto-magnetar winds are magnetically-driven throughout most of their evolution.  When the wind is non-relativistic, its speed at the fast surface is $v_{\infty} \approx \sigma_{0}^{1/3}c$, the wind power is $\dot{E} \propto \sigma_{0}^{2/3}\dot{M} \propto \dot{M}^{1/3}$ and $\dot{E}_{\rm mag} = 2\dot{E}_{\rm kin}$ \citep{Lamers&Cassinelli99}.  For relativistic winds $\dot{E} \propto \sigma_{0}\dot{M}$ is approximately independent of $\dot{M}$, and $\dot{E}_{\rm mag} \gg \dot{E}_{\rm kin}$ at the fast point.  Indeed, in the limit that $\sigma_{0} \gg 1$ we assume that $\dot{E}$ approaches the force-free spin-down rate \citep{Spitkovsky06}, which depends only on $\phi$ and $\Omega$.  Even for relatively large (but finite) values of $\sigma_{0}$, however, spin-down occurs more rapidly than in the force-free case because the `Y' point radius $R_{\rm Y}$ resides inside the light cylinder (see Fig.~\ref{fig:nscartoon}).  The full expression for $\dot{E}$ is given in equation (\ref{eq:edot}).     

\subsubsection{Spin-Down Evolution and Initial Conditions}

Proto-magnetar winds are magneto-rotationally powered throughout most of their evolution.  The NS thus loses angular momentum $J = I\Omega$ to the wind at the rate $\dot{J} = -\dot{E}/\Omega$.  Neglecting mass loss (a good approximation), the rotation rate $\Omega$ evolves according to
\begin{eqnarray}
\frac{\dot{\Omega}}{\Omega} = -\frac{2\dot{R_{\rm ns}}}{R_{\rm ns}}-\frac{2\dot{E}}{E_{\rm rot}},
\label{eq:omegadot}
\end{eqnarray}
where $E_{\rm rot}$ is the NS rotational energy (eq.~[\ref{eq:erot}]).  
In equation (\ref{eq:omegadot}) we neglect angular momentum losses due to gravitational waves, which become important if the NS is sufficiently aspherically distorted by its strong interior magnetic field (e.g.~\citealt{Cutler02}; \citealt{Arons03}; \citealt{Stella+05}; \citealt{Dallosso+09}).  This is a good approximation provided that either the magnetic obliquity is small or the interior magnetic field is less than $\sim 100$ times stronger than the outer dipole field.  We also neglect gravitational wave emission due to non-axisymmetric waves or instabilities (e.g.~$r$-modes; \citealt{Andersson98}), although these are implicitly taken into account through the maximum initial NS rotation rate that we consider (see below).  We also neglect the possibility of late-time accretion onto the proto-magnetar (e.g.~\citealt{Metzger+08b}; \citealt{Zhang&Dai09}), which could affect the spin-down evolution both through accretion torques and by altering the geometry of the magnetosphere.

Given $\dot{E}$ and $\dot{M}$ as a function of $\Omega$ and time, we solve equation (\ref{eq:omegadot}) to obtain $\Omega(t)$, $\dot{M}(t)$, $\dot{E}(t)$, and $\sigma_{0}(t)$.  A wind solution is thus fully specified by just four parameters: the NS mass $M_{\rm ns}$; the `initial' angular rotation rate $\Omega_{0} = 2\pi/P_{0}$; the surface dipole magnetic field strength $B_{\rm dip}$; and the inclination angle $\chi$ (`obliquity') between the magnetic and rotational axes (see Fig.~\ref{fig:nscartoon}).  Since the proto-NS is still contracting for several seconds following core bounce, $\Omega_{0}$ and $B_{\rm dip}$ are more precisely defined as the maximum values that {\it would be} achieved were the NS to contract at constant angular momentum $J \propto R_{\rm ns}^{2}M_{\rm ns}\Omega$ and magnetic flux\footnote{Note the distinction between the conserved dipole flux through the stellar interior $\Phi$ defined here and the open flux through the magnetosphere $\phi$ (eq.~[\ref{eq:sigma}]), which evolves in time.} $\Phi \propto B_{\rm dip}R_{\rm ns}^{2}$, respectively.  

If the magnetic field is amplified on a timescale comparable to the duration of the NS cooling epoch (e.g.~via linear field winding), the assumption of a fixed dipole flux may be a poor approximation.  On the other hand, if field growth occurs more rapidly via a convection-driven dynamo \citep{Duncan&Thompson92} or the dynamical-timescale MRI (e.g.~\citealt{Akiyama+03}; \citealt{Thompson+05}), then the field is probably established - and finds a MHD stable configuration \citep{Braithwaite&Spruit06} - in less than a few seconds (\citealt{Spruit08}).  In this case the assumption that $\Phi$ is fixed may be reasonable.

Given the uncertainty in the origin of magnetar fields, in general we allow both $P_{0}$ and $B_{\rm dip}$ to vary independently within their respective physical ranges ($P_{0} \gtrsim 1$ ms, $B_{\rm dip} \lesssim 3 \times 10^{16}$ G; see below).  However, if the magnetic field is in fact generated from the free energy available in differential rotation, then a relationship between $B_{\rm dip}$ and $P_{0}$ of the form
\be
B_{\rm dip} = 10^{16}{\,\rm G}\left(\frac{\epsilon_{\rm B}}{10^{-3}}\right)^{1/2}\left(\frac{R_{\rm ns}}{12{\,\rm km}}\right)^{-1/2}\left(\frac{P_{0}}{\,\rm ms}\right)^{-1}
\label{eq:beq}
\ee  
could result, where we have assumed that the magnetic energy in the dipole field ($\propto B_{\rm dip}^{2}R_{\rm ns}^{3}$) is a fraction $\epsilon_{\rm B}$ of the rotational energy $E_{\rm rot} \propto R_{\rm ns}^{2}P_{0}^{-2}$ (eq.~[\ref{eq:erot}]) and that the energy in differential rotation scales with $E_{\rm rot}$.  In our models we require that $P_{0} \gtrsim 1$ ms because this is the allowed range of stable proto-NS rotational periods (e.g.~\citealt{Strobel+99}).  This maximum rotation rate may be enforced in practice by the efficient loss of angular momentum incurred by very rapidly spinning NSs to MRI-generated turbulence or waves radiated by nonaxisymmetric instabilities (e.g.~\citealt{Thompson+05}; \citealt{Ott+05}; \citealt{Wheeler&Akiyama07}).  We furthermore only consider models with $B_{\rm dip} \lesssim 3\times 10^{16}$ G because although fields up to $\approx 3\times 10^{17}$ G are in principle possible if $\epsilon_{\rm B} \sim 1$, stable magnetic configurations generally require a total field strength which is larger than the dipole component by a factor $\gtrsim 10$ (e.g.~\citealt{Taylor73}; \citealt{Braithwaite09}).  In addition, our assumption that the magnetic field does not affect the neutrino-driven mass loss rate is invalid for $B_{\rm dip} \gtrsim 3\times 10^{16}$ G (see Appendix \ref{sec:appendixA}).  


\subsection{Results}
\label{sec:results}

\begin{figure}
\resizebox{\hsize}{!}{\includegraphics[]{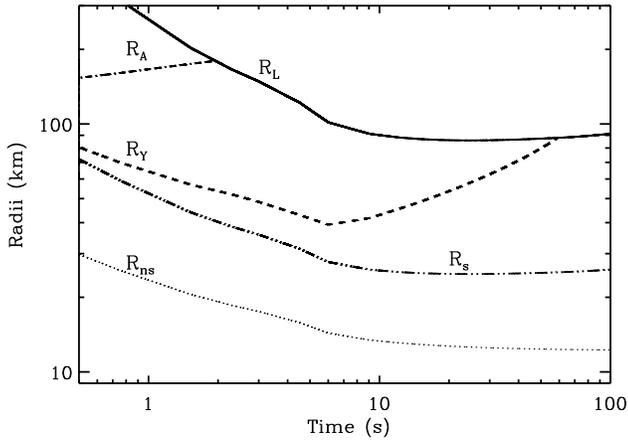}}
\caption{Time evolution of the light cylinder radius $R_{\rm L}$ ({\it solid line}), Alfven radius $R_{\rm A}$ ({\it dot-dashed line}; eq.~[\ref{eq:ra}]), `Y' point radius $R_{\rm Y}$ ({\it dashed line}), sonic radius $R_{\rm s}$ ({\it double dot-dashed line}), and neutron star radius $R_{\rm ns}$ (see Fig.~\ref{fig:pons}) for the solution shown in Figure \ref{fig:edotsig}.}
\label{fig:radii}
\end{figure}

\begin{figure}
\resizebox{\hsize}{!}{\includegraphics[]{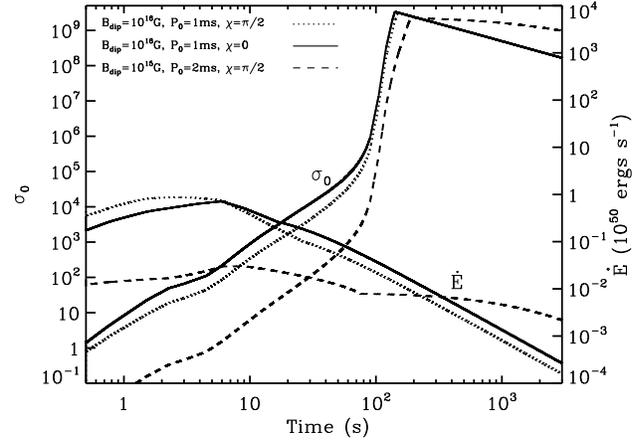}}
\caption{Same as Figure \ref{fig:edotsig}, but calculated for different proto-magnetar properties.  The first two models are for $P_{0} = 1$ ms and $B_{\rm dip} = 10^{16}$ G, and assume values of the magnetic obliquity $\chi = \pi/2$ ({\it dotted line}) and $\chi = 0$ ({\it solid line}), respectively.  The dashed line shows a lower spin-down case, calculated for $P_{0} = 2$ ms, $B_{\rm dip} = 10^{15}$ G, and $\chi = \pi/2$.  }
\label{fig:edotsig2}
\end{figure}

The results of our calculations are summarized in Figures $\ref{fig:edotsig}-\ref{fig:edotsig2}$ and Table \ref{table:winds}.  As already discussed, Figure \ref{fig:edotsig} shows the wind magnetization $\sigma_{0}(t)$ and power $\dot{E}(t)$ as a function of time since core bounce, calculated for $M_{\rm ns} = 1.4M_{\sun}$, $P_{0} = 1.5$ ms, $B_{\rm dip} = 2\times 10^{15}$ G, and $\chi = \pi/2$.  Figure \ref{fig:radii} shows the time evolution of several critical radii associated with this wind solution.

During the first few seconds, $\dot{E}$ rises because $\Omega$ and $B_{\rm dip}$ increase by angular momentum and magnetic flux conservation, respectively, as the proto-NS contracts to its final radius.  On longer timescales, $\dot{E}$ reaches a maximum and then decreases once the NS begins to spin down and the open magnetosphere shrinks.  The latter results because both the spin-down and the larger wind magnetization cause $R_{\rm Y}$ to increase (see Figs.~\ref{fig:nscartoon} and \ref{fig:radii}).  Figure \ref{fig:edotsig} also shows that $\sigma_{0}$ increases rapidly for the first $\sim 100$ seconds as the NS cools and the neutrino-driven mass loss rate decreases.  This results in several distinct stages in the wind evolution, which we denote by Roman numerals in Figure \ref{fig:edotsig} and are discussed individually in the next section.  At late times $\sigma_{0}$ plateaus and then begins decreasing once the wind mass loss rate reaches its minimum value proportional to the Goldreich-Julian flux (eq.~[\ref{eq:mdotGJ}]).  Once $\sigma_{0} \gg 1$ force-free spin-down obtains, such that $\dot{E}$ asymptotes at late times to the standard\footnote{Note, however, that the {\it measured} braking indices of Galactic pulsars generally differ from the force-free prediction (e.g.~\citealt{Livingstone+07}; see $\S\ref{sec:highsig}$).} force-free decay $\dot{E} \propto t^{-2}$.  

Figure \ref{fig:edotsig2} shows three additional wind models, calculated for different values of $B_{\rm dip}$, $P_{0}$, and $\chi$.  The models shown with solid and dotted lines correspond, respectively, to high spin-down cases with $B_{\rm dip} = 10^{16}$ G, $P_{0} = 1$ ms, calculated for different values of the magnetic obliquity $\chi = 0$ and $\pi/2$.  The third model shown with a dashed line is a lower spin-down case with $B_{\rm dip} = 10^{15}$ G, $P_{0} = 2$ ms, and $\chi = \pi/2$.  Although the evolution of $\dot{E}(t)$ and $\sigma_{0}(t)$ are qualitatively similar to the fiducial model in Figure~\ref{fig:edotsig}, differences are apparent.  Note that the higher(lower) spin-down models achieve larger(smaller) values of $\dot{E}$ and $\sigma_{0}$, except at late times.  Also note that at fixed $B_{\rm dip}$ and $P_{0}$, $\sigma_{0}$ is larger for the aligned rotator ($\chi = 0$) than in the oblique case ($\chi = \pi/2$) due to the enhanced mass loss in the latter case caused by centrifugal `slinging' (see eq.~[\ref{eq:fcentmax}] and surrounding discussion).   

Table \ref{table:winds} summarizes the results of several additional calculations, which explore the sensitivity of our results to variations in the proto-magnetar properties and in the adopted NS cooling model.  Our primary conclusion is that key observables are most sensitive to the dipole field $B_{\rm dip},$ rotation rate $P_{0}$, and obliquity $\chi$.  Plausible variations in the NS mass $M_{\rm ns}$, stretch parameter $\eta_{\rm s}$, and the cooling model, on the other hand, generally result in at most order unity differences.  For this reason we fix $M_{\rm ns} = 1.4M_{\sun}$ and $\eta_{\rm s} = 3$ in the sections to follow and confine our analysis to the 3D parameter space ($B_{\rm dip}, P_{0},\chi$).

\section{Stages of the Proto-Magnetar Model}
\label{sec:stages}
In this section we describe the stages of proto-magnetar wind evolution and quantify their relationship to GRB phenomenology.  Our discussion is guided closely by Figures \ref{fig:edotsig}$-$\ref{fig:edotsig2}. 

\subsubsection*{\bf I. Pre-Supernova/Thermally-Driven Wind}
\begin{center}
($\sigma_{0} \lesssim 10^{-3}$; t $\lesssim $ few$\times 100$ ms) \\
\end{center}

Simulations of core collapse fail to produce a prompt explosion, suggesting that the proto-NS continues to accrete for several hundred milliseconds before a delayed explosion occurs.  The proto-NS forms hot and its initial radius exceeds $\sim 30$ km.  Since magnetic forces are unlikely to be dynamically important yet, an explosion at this stage would be neutrino-driven \citep{Bethe&Wilson85}.  If this `standard' scenario applies, thermal pressure is initially responsible for accelerating the neutrino-heated wind into the cavity behind the outgoing SN shock (e.g.~\citealt{Burrows+95}; \citealt{Qian&Woosley96}; \citealt{Roberts+10}).

However, as already discussed, hypernovae are probably not powered by neutrinos alone.  For proto-magnetars the field is eventually amplified to a dynamically-relevant strength.  If this field mediates the transfer of a significant fraction of the rotational energy ($\gtrsim 10^{52}$ ergs) to the SN shock, the resulting explosion would indeed be hyper-energetic ($\S\ref{sec:BHvsNS}$).\footnote{Note also that the large temperatures behind the shock produced by such an energetic explosion will result in a large yield of $^{56}$Ni.}  For an MHD-powered SN, the neutrino wind is thus magnetically-driven from its onset.  The division between thermally- and magnetically-driven winds occurs at a critical magnetization $\sigma_{0} \sim 10^{-3}$, because above this value the asymptotic speed of a magnetically-driven wind $v_{\infty} = \sigma_{0}^{1/3}$c exceeds the speed $v_{\infty} \sim 0.1$ c obtained via thermal acceleration alone \citep{Metzger+07}.

\subsection*{\bf II. Magnetically-Driven, Non-Relativistic Wind}
\begin{center}
($10^{-3} \lesssim \sigma_{0} \lesssim 1$; few$\times 100$ ms $\lesssim t \lesssim$ few s)\\
\end{center}  

Regardless of whether the SN itself is powered by thermal or magnetic forces, the neutrino wind becomes magnetically-driven ($\sigma_{0} \gtrsim 10^{-3}$) less than a second later.  Because the neutrino luminosity $L_{\nu}$ is still large at these early times (Fig.~\ref{fig:pons}), the wind mass loss rate $\dot{M}$ remains high.  Though powerful at this stage, the outflow is thus still non-relativistic ($\sigma_{0} \lesssim 1$).  Non-relativistic magnetized winds are efficiently self-collimated by hoop stresses (e.g.~\citealt{Sakurai85}).  The proto-magnetar wind thus forms a bipolar jet, which catches up to the slower SN shock and begins boring a collimated cavity into the unshocked star.  

\begin{figure}
\resizebox{\hsize}{!}{\includegraphics[]{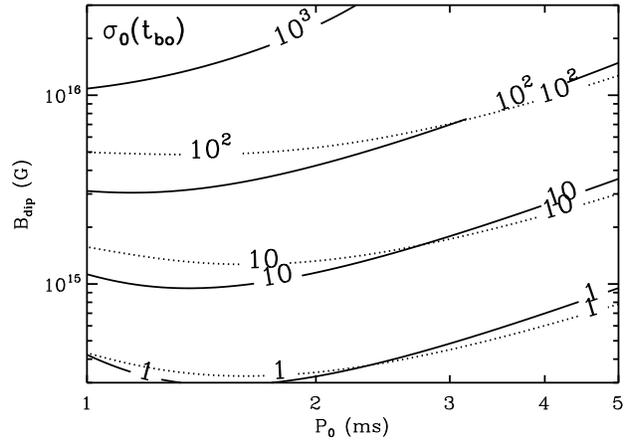}}
\caption{Contours of the wind magnetization at jet break-out $t = t_{\rm bo} = 10$ s, as a function of the magnetic field strength $B_{\rm dip}$ and initial rotation period $P_{0}$ of the magnetar.  Solid and dotted lines show calculations assuming magnetic obliquities $\chi = 0$ and $\chi = \pi/2$, respectively.}
\label{fig:sigbo}
\end{figure}

\subsubsection*{\bf III. Magnetically-Driven, Relativistic Wind (Pre-Breakout)}
\begin{center}
{($1 \lesssim \sigma_{0} \lesssim 10-100$; few s $\lesssim t \lesssim t_{\rm bo}$)} \\
\end{center}

As the NS continues to cool, $\sigma_{0}$ exceeds unity within a few seconds and the wind becomes relativistic.  Self-collimation fails in ultra-relativistic outflows (e.g.~\citealt{Bucciantini+06}).  The wind power thus becomes concentrated at low latitudes, where it collides with the slowly-expanding SN ejecta and forms a hot `proto-magnetar nebula' \citep{Bucciantini+07}.  As toroidal flux accumulates in the nebula, magnetic forces -- and the anisotropic thermal pressure they induce -- redirect the equatorial outflow towards the poles (\citealt{Begelman&Li92}; \citealt{Konigl&Granot02}; \citealt{Uzdensky&MacFadyen07}; \citealt{Bucciantini+07,Bucciantini+08,Bucciantini+09}; \citealt{Komissarov&Barkov07}).  Stellar confinement thus produces a mildly relativistic jet, which continues drilling a bipolar cavity where the earlier non-relativistic outflow left off. 

The jet propagates through the star at a significant fraction $\beta$ of the speed of light (e.g.~\citealt{MacFadyen&Woosley99}; \citealt{Aloy+00}; \citealt{Ramirez-Ruiz+02b}; \citealt{Zhang+03}; \citealt{Morsony+07}; \citealt{Bucciantini+08}), such that it `breaks out' of the stellar surface of radius $R_{\star}$ on a timescale 
\be
t_{\rm bo} \approx R_{\star}/\beta c \sim 7(R_{\star}/10^{11}{\,\rm cm})(\beta/0.5)^{-1} {\,\rm s}.
\label{eq:tbo}
\ee  
Although the precise value of $t_{\rm bo}$ will in general depend on both the properties of the jet and star, in what follows we assume a fixed value $t_{\rm bo} = 10$ seconds.  Although this is a reasonable estimate for moderately powerful jets, weaker jets could require significantly longer to reach the surface.  Below a critical jet power $\dot{E} \lesssim \dot{E}_{\rm min} \sim 10^{48}$ erg s$^{-1}$, both hydrodynamic (e.g.~\citealt{Woosley&Zhang07}) and MHD outflows \citep{Bucciantini+09} may fail to produce stable clean jets (e.g.~\citealt{Matzner03}) which may instead be `choked' inside the star, resulting in little direct electromagnetic radiation (see $\S\ref{sec:choked}$).  

\subsubsection*{\bf IV.~Magnetically-Driven, Relativistic Wind (GRB)}
\begin{center}
{($10-100 \lesssim \sigma_{0} \lesssim 10^{4}$; $t_{\rm bo} \lesssim t \lesssim t_{\rm end}$)}\\
\end{center}

After the jet breaches the stellar surface a relatively clean opening is soon established through the star (e.g.~\citealt{Morsony+07}).  Simulations suggest that after this point the power and mass loading of the jet reflect, in a time- and angle-averaged sense, the values of $\dot{E}(t)$ and $\dot{M}(t)$ set by the proto-magnetar wind at much smaller radii (e.g.~\citealt{Bucciantini+09}; \citealt{Morsony+10}).  

Figure $\ref{fig:sigbo}$ shows contours of the wind magnetization $\sigma_{0}$ at break-out ($t = t_{\rm bo} = 10$ s), as calculated using a grid of wind models spanning the physical range of magnetar parameters $B_{\rm dip}$ and $P_{0}$ for two values of the magnetic obliquity $\chi = 0,\pi/2$.  Note that high spin-down magnetars (upper left corner) produce outflows that are ultra-relativistic at break-out, i.e.~$\sigma_{0}|_{t_{\rm bo}} \gtrsim 10-100$.  

Over the next tens of seconds $\sigma_{0}$ increases from $\sigma_{0}|_{t_{\rm bo}}$ to $\gtrsim 10^{4}$ (Figs.~\ref{fig:edotsig} and \ref{fig:edotsig2}), resulting in ideal conditions for high energy emission.  Assuming that the wind is collimated into a jet with a half-opening angle $\theta_{\rm j}$, the `isotropic' jet luminosity $\dot{E}_{\rm iso}$ is larger than the wind power $\dot{E}$ by a factor $f_{\rm b}^{-1}$, where $f_{\rm b} \simeq \theta_{\rm j}^{2}/2$ is the beaming fraction \citep{Rhoads99}.  Using axisymmetric MHD simulations, \citet{Bucciantini+09} found $\theta_{\rm j} \sim 5-10^{\circ}$ for a magnetar with $B_{\rm dip} \sim 3\times 10^{15}$ G and $P_{0} \sim 1$ ms, values consistent with the typical opening angles inferred from GRB afterglow modeling (e.g.~\citealt{Frail+01}; \citealt{Bloom+03}).  

Although the more general dependence of $\theta_{\rm jet}$ on the properties of the magnetar and stellar progenitor has not yet been determined, some insight is provided directly from observations.  By combining the well-known correlation between the peak energy of the prompt emission spectrum $E_{\rm peak}$ and the isotropic energy $E_{\rm iso}$, $E_{\rm peak} \propto E_{\rm iso}^{0.4}$ \citep{Amati+02} with the correlation $E_{\rm peak} \propto E_{\gamma}^{0.7}$ between $E_{\rm peak}$ and the beaming-corrected energy $E_{\gamma} = f_{\rm b}E_{\rm iso}$ \citep{Ghirlanda+04}, we obtain the {\it empirical} relationship (cf.~\citealt{Nava+06})
\begin{eqnarray}
f_{\rm b} \approx 2\times 10^{-3}\left(\frac{E_{\gamma}}{10^{51}\,\rm ergs}\right)^{-3/4};\,\,
\theta_{\rm j} \approx 3.3^{\circ}\left(\frac{E_{\gamma}}{10^{51}\,\rm ergs}\right)^{-3/8}. \nonumber \\
\label{eq:fbcorrelation}
\end{eqnarray}
In what follows we assume for simplicity a fixed beaming fraction $f_{\rm b} = 2\times 10^{-3}$, but we return to an implication of the correlation $f_{\rm b} \propto E_{\gamma}^{-3/4}$ in $\S\ref{sec:magdiss}$.   

To produce high energy emission the jet must both accelerate to a high Lorentz factor $\Gamma_{\rm j} \sim \sigma_{0} \gg 1$ and dissipate much of its bulk energy internally.  Both of the emission models that we consider in $\S\ref{sec:GRB}$, magnetic dissipation and internal shocks, predict a characteristic emission radius where most dissipation occurs $R_{\gamma}= R_{\rm mag}$ and $R_{\gamma} = R_{\rm is}$, respectively, that increases with time.  Here $R_{\rm mag}$ and $R_{\rm is}$ are the radii at which magnetic dissipation peaks and internal shocks occur, respectively (see below).  Whether photons escape the emission region at a given epoch depends on the location of $R_{\gamma}$ with respect to the radius of the Thompson photosphere of the jet (e.g.~\citealt{Giannios06})
\be
R_{\rm ph} \simeq \frac{\dot{E}_{\rm iso}\kappa_{\rm es}}{8\pi c^{3}\sigma_{0}^{3}},
\label{eq:rph}
\ee
where $\kappa_{\rm es}$ is the Thomson opacity and we have assumed efficient acceleration, i.e.~$\Gamma_{\rm j} \approx \sigma_{0} \gg 1$ ($\S\ref{sec:acceleration}$).    

\begin{figure}
\resizebox{\hsize}{!}{\includegraphics[]{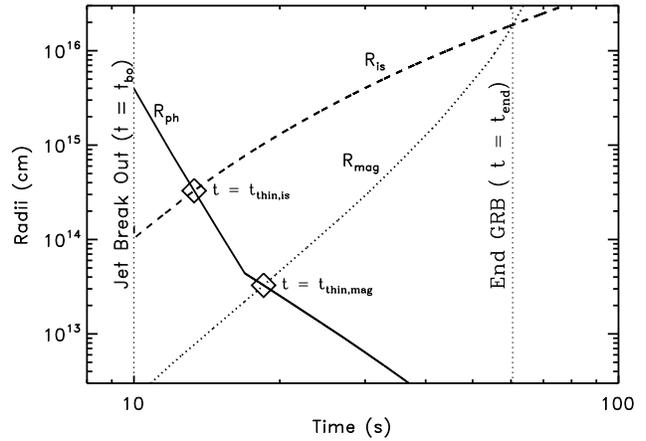}}
\caption{Photosphere radius $R_{\rm ph}$ ({\it solid line}; eq.~[\ref{eq:rph}]), internal shock radius $R_{\rm is}$ ({\it dashed line}; eq.~[\ref{eq:rsh}]), and the `saturation' radius at which magnetic dissipation peaks $R_{\rm mag}$ ({\it dotted line}; eq.~[\ref{eq:rsat}]) in the proto-magnetar jet as a function of time since core bounce, calculated for the model shown in Figure \ref{fig:edotsig}.  The jet breaks out of the star at the time $t = t_{\rm bo} = 10$ seconds.  At times $t_{\rm bo} \lesssim t \lesssim t_{\rm thin,mag}(t_{\rm thin,is})$ magnetic dissipation (internal shocks) occur below the photosphere and the resulting emission will be thermalized (Stage IVa).  By contrast, at times $t \gtrsim t_{\rm thin,mag}, t_{\rm thin,is}$ emission occurs in an optically-thin environment and may be non-thermal (Stage IVb).  The end of the GRB is defined as when $R_{\rm mag} = R_{\rm is}$ (Stage V).}
\label{fig:radiidis}
\end{figure}

\begin{figure}
\resizebox{\hsize}{!}{\includegraphics[]{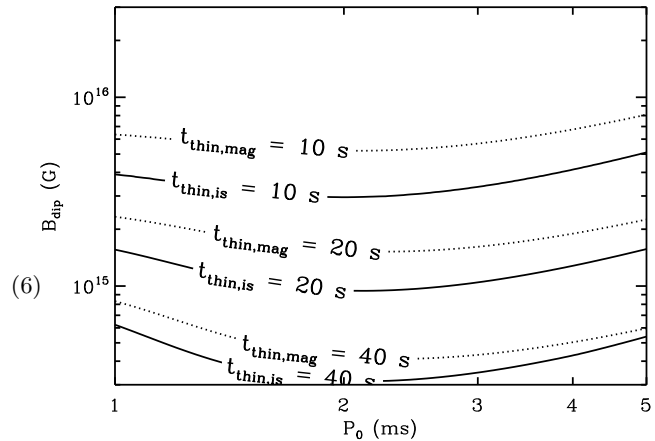}}
\caption{Contours of the time after core bounce $t_{\rm thin}$ when the jet becomes optically thin to emission at the magnetic dissipation radius $R_{\rm mag}$ ($t_{\rm thin,mag}$; {\it dotted line}) and the internal shock radius ($t_{\rm thin,is}$; {\it solid line}) as a function of magnetic dipole field strength $B_{\rm dip}$ and initial rotation period $P_{0}$, calculated for $\chi = \pi/2$.  Jets from lower field magnetars are optically thick at break-out (i.e. $t_{\rm thin} > t_{\rm bo} = 10$ s), potentially resulting in a short-lived phase of dim quasi-thermal emission (Stage IVa).  By contrast, jets from magnetars with stronger fields (upper diagram) have $t_{\rm thin} < t_{\rm bo}$ and may dissipate their energy in an optically-thin environment immediately after break-out (Stage IVb), thereby skipping Stage IVa entirely.}
\label{fig:tthin}
\end{figure}

\begin{figure}
\resizebox{\hsize}{!}{\includegraphics[]{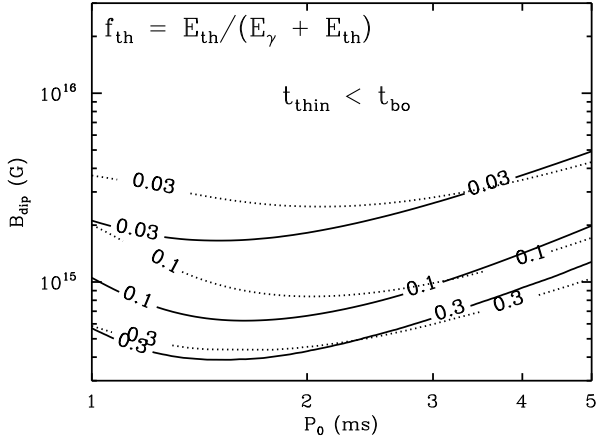}}
\caption{Contours of the fraction of energy released in thermal emission $f_{\rm th} \equiv E_{\rm th}/(E_{\gamma}+E_{\rm th})$ during the GRB phase, where $E_{\rm th}$ is defined in equation (\ref{eq:etherm}), as a function of surface dipole field $B_{\rm dip}$ and initial rotation rate $P_{0}$ for magnetic obliquities $\chi = \pi/2$ ({\it solid line}) and $\chi = 0$ ({\it dotted line}).}
\label{fig:thermfrac}
\end{figure}

\begin{flushleft}
{\bf IVa.~Quasi-Thermal, Photospheric Emission}
\end{flushleft}
\begin{center}
{($t_{\rm bo} \lesssim t \lesssim t_{\rm thin}$; $R_{\gamma} < R_{\rm ph}$)}
\end{center}

Figure \ref{fig:radiidis} shows the time evolution of the photosphere radius $R_{\rm ph}$ and the radii at which internal shocks ($R_{\rm is}$; eq.~[\ref{eq:rsh}]) and magnetic dissipation ($R_{\rm mag}$; eq.~[\ref{eq:rsat}]) occur, calculated for the fiducial model shown in Figure \ref{fig:edotsig}.  Just after break-out, magnetic dissipation and internal shocks occur below the photosphere, i.e.~$R_{\gamma} = \{R_{\rm mag},R_{\rm is}\} \ll R_{\rm ph}$, such that high energy emission will be partially thermalized and suppressed due to adiabatic losses.  At later times, jet dissipation occurs in an optically-thin environment ($R_{\gamma} > R_{\rm ph}$), such that brighter non-thermal emission\footnote{Throughout this paper we define `non-thermal' emission as a non-black body spectrum.  This does not necessarily imply that the radiating electrons have a non-thermal energy distribution.} is more likely.

Figure \ref{fig:tthin} shows contours of the time after core bounce at which $R_{\rm ph} = R_{\rm mag}$ ({\it dotted line}) and $R_{\rm ph} = R_{\rm is}$ ({\it solid line}), respectively, as a function of $B_{\rm dip}$ and $P_{0}$ for $\chi = \pi/2$.  Low field magnetars (lower diagram) produce jets that are optically thick at break-out (i.e. $t_{\rm thin} > t_{\rm bo} \approx 10$ s) and thus experience a phase of quasi-thermal photospheric emission, as in the fiducial model described above (Stage IVa).  In fact, if $t_{\rm thin}$ becomes comparable to the GRB duration itself (cf.~Fig.~\ref{fig:T90}), a thermal-rich sub-luminous GRB or X-ray Flash may result instead of a classical GRB ($\S\ref{sec:XRF}$).  By contrast, jets from strongly magnetized magnetars (upper diagram) dissipate their energy in an optically-thin environment just after jet break out, thereby skipping Stage IVa entirely. 

Figure \ref{fig:thermfrac} shows contours of the fraction of the energy released in thermal emission during the GRB phase $f_{\rm th} = E_{\rm th}/(E_{\gamma} + E_{\rm th})$.  Here $E_{\gamma}$ is the total {\it non-thermal} emission during the GRB (quantified in the next section) and $E_{\rm th}$ is the maximum thermal energy, which we estimate as (e.g.~\citealt{Meszaros&Rees00})
\be
E_{\rm th} = \int_{t_{\rm bo}}^{t_{\rm thin,is}}\epsilon_{\rm r}\dot{E}(R_{\rm ph}/R_{\rm is})^{-2/3}dt,
\label{eq:etherm}
\ee
where the factor $(R_{\rm ph}/R_{\rm is})^{-2/3}$ accounts for adiabatic losses, and we have (optimistically) assumed a radiative efficiency $\epsilon_{\rm r} = 0.5$ (eq.~[\ref{eq:epsilonr}]).  High field magnetars (upper diagram) produce little thermal emission $E_{\rm th} \approx 0$ because the jet is already optically thin at break out (i.e.~$t_{\rm thin,is} \lesssim t_{\rm bo}$; cf.~Fig.~\ref{fig:tthin}).  By contrast, somewhat lower-field magnetars (middle-left diagram) have $f_{\rm th} \gtrsim 0.1$ ($E_{\rm th} \sim 10^{48-50}$ ergs), consistent with measurements or upper-limits on quasi-thermal photospheric emission from GRBs (e.g.~\citealt{Meszaros&Rees00}; \citealt{Ramirez-Ruiz+02}; \citealt{Ryde05}; \citealt{Guiriec+10}).  

\begin{flushleft} 
{\bf IVb.~Main GRB Emission}
\end{flushleft}
\begin{center}
{\bf  ($t_{\rm thin} \lesssim t \lesssim t_{\rm end}$; $R_{\gamma} > R_{\rm ph}$)}
\end{center}

\begin{figure}
\resizebox{\hsize}{!}{\includegraphics[]{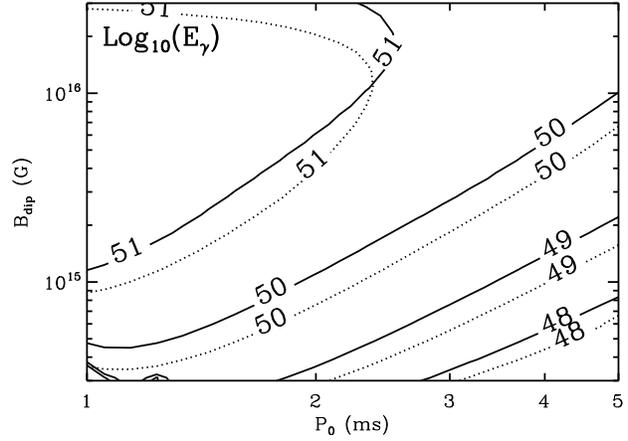}}
\caption{Contours of the maximum non-thermal gamma-ray emission $E_{\gamma}$ in ergs as a function of $B_{\rm dip}$ and $P_{0}$ for $\chi = 0$ ({\it solid line}) and $\chi = \pi/2$ ({\it dotted line}).  We calculate $E_{\gamma}$ as the total energy released by the magnetar in the time interval max[$t_{\rm bo},t_{\rm thin,is}] \lesssim t \lesssim t_{\rm end}$ times a factor $\epsilon_{\rm r} = 0.5$ to account for the maximum radiative efficiency.  Here $t_{\rm bo}$ is the time required for the jet to propagate through the star, $t_{\rm thin,is}$ is the time after which the outflow is optically thin at the internal shock radius (Fig.~\ref{fig:tthin}), and $t_{\rm end}$ is the end of the GRB, defined as when $R_{\rm mag} = R_{\rm is}$ (see Figs.~\ref{fig:radiidis} and \ref{fig:T90}).}
\label{fig:eGRB}
\end{figure}

From the time $t = $ min[$t_{\rm thin},t_{\rm bo}]$ until the GRB ends at $t = t_{\rm end}$ (which we define more precisely below), shocks or reconnection occur above the photosphere and non-thermal emission is likely.  Figure \ref{fig:eGRB} shows contours of the total energy released by the magnetar wind $E_{\gamma} \equiv \int\epsilon_{\rm r}\dot{E}dt$ integrated over the GRB duration as a function of $B_{\rm dip}$ and $P_{0}$ for $\chi = 0$ ({\it solid line}) and $\chi = \pi/2$ ({\it dotted line}), assuming a radiative efficiency $\epsilon_{\rm r} = 0.5$.  Note that $E_{\gamma} \gtrsim 10^{50-51}$ ergs across the entire range of high spin-down (`GRB capable') magnetars.  These values are consistent with the collimation-corrected energy released by GRBs in relativistic ejecta (e.g.~\citealt{Frail+01}; \citealt{Berger+03}; \citealt{Bloom+03}).   

\begin{figure}
\resizebox{\hsize}{!}{\includegraphics[]{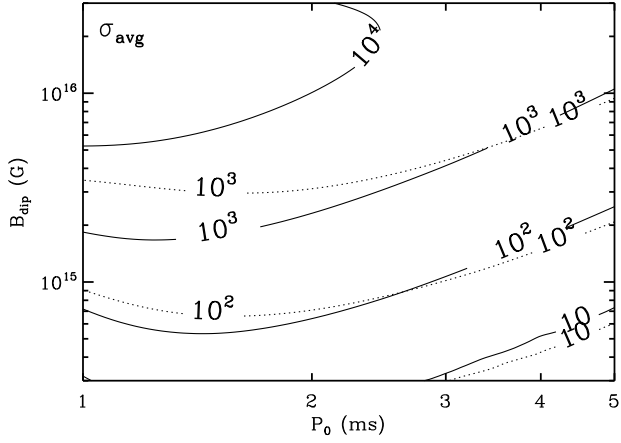}}
\caption{Contours of the (energy-weighted) average magnetization $\sigma_{\rm avg}$ as a function of $B_{\rm dip}$ and $P_{0}$, calculated for $\chi = 0$ ({\it solid line}) and $\chi = \pi/2$ ({\it dotted line}).  Note that in general $\sigma_{\rm avg}$ is smaller in the case of an oblique rotator ($\chi = \pi/2$) because of the enhanced mass loss due to centrifugal `slinging' (see eq.~[\ref{eq:fcentmax}] and surrounding discussion).}
\label{fig:meansig}
\end{figure}

\begin{figure}
\resizebox{\hsize}{!}{\includegraphics[]{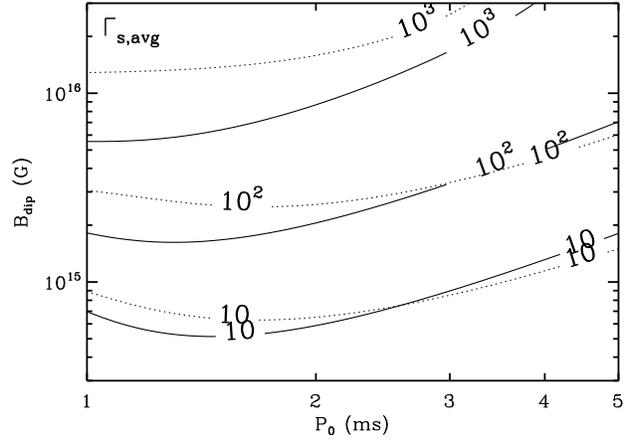}}
\caption{Contours of the (energy-weighted) average Lorentz factor $\Gamma_{\rm s,avg}$ of the bulk shell created by internal shocks, as a function of $B_{\rm dip}$ and $P_{0}$, calculated for $\chi = 0$ ({\it solid line}) and $\chi = \pi/2$ ({\it dotted line}).}
\label{fig:meangs}
\end{figure}

Figure \ref{fig:meansig} shows contours of the average (energy-weighted) magnetization of the jet, which we define as $\sigma_{\rm avg} \equiv \int\dot{E}\sigma_{0}dt/\int\dot{E}dt$ integrated over the duration of the GRB.  High spin-down magnetars (upper left diagram) achieve values $\sigma_{\rm avg} \approx \Gamma_{\rm max} \sim 10^{2}-10^{4}$ which are are consistent with observational constraints on the GRB Lorentz factors (i.e.~$\Gamma \gtrsim 100-1000$; e.g.~\citealt{Lithwick&Sari01}; \citealt{Zou&Piran10}; \citealt{Zou+10}).  We caution, however, that although $\sigma_{\rm avg}$ approximately equals the jets maximum instantaneous Lorentz factor, for internal shocks the Lorentz factor of the emitting material $\Gamma_{\rm s}$ is generally lower than $\sigma_{\rm avg}$ because the faster jet interacts with slower material released at earlier times ($\S\ref{sec:internalshocks}$).  In Figure \ref{fig:meangs} we show contours of the (energy-weighted) mean Lorentz factor $\Gamma_{\rm s,avg}$ of the bulk shell, from behind which internal shock emission originates.  Note that in general $\Gamma_{\rm s,avg}$ is a factor of a few times lower than $\sigma_{\rm avg}$.  A comparison of Figure \ref{fig:eGRB} with Figures \ref{fig:meansig} and \ref{fig:meangs} reveals a positive correlation between $E_{\gamma}$ and the mean magnetization/Lorentz factor.  We discuss this correlation and its implications further in $\S\ref{sec:magdiss}$.   

\subsubsection*{\bf{V.~Ultra High-$\sigma_{0}$ Phase (Post GRB)}}
\begin{center}
 ($R_{\rm mag} \gtrsim R_{\rm is}$; $t \gtrsim t_{\rm end}$).\\  
\end{center}

\begin{figure}
\resizebox{\hsize}{!}{\includegraphics[]{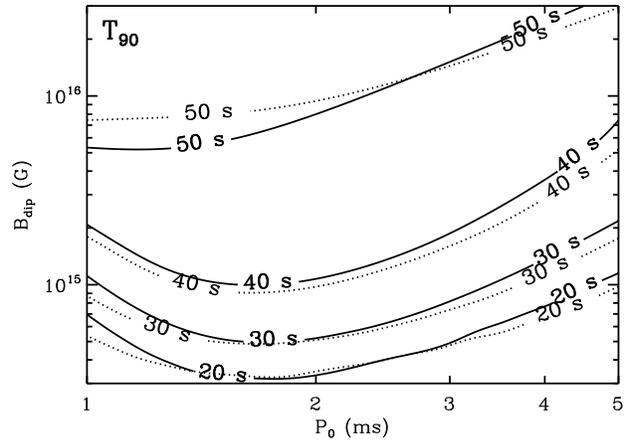}}
\caption{Duration $T_{90}$ in the time interval max[$t_{\rm bo},t_{\rm thin,is}] \lesssim t \lesssim t_{\rm end}$ during which 90$\%$ of the wind energy is released, calculated as a function of $B_{\rm dip}$ and $P_{0}$ for $\chi = 0$ ({\it solid line}) and $\chi = \pi/2$ ({\it dotted line}).  Here $t_{\rm bo}$ is the jet break-out time, $t_{\rm thin}$ is the time after internal shocks occur above the photosphere (Fig.~\ref{fig:tthin}), and $t_{\rm end}$ is the end of the prompt emission (i.e.~when $R_{\rm mag} = R_{\rm is}$; see Fig.~\ref{fig:radiidis}).}
\label{fig:T90}
\end{figure}

As $\sigma_{0}$ continues to increase, the jet becomes less and less effective at accelerating and dissipating its ordered energy ($\S\ref{sec:acceleration}$; e.g.~\citealt{Lyubarsky&Kirk01}).  A particularly abrupt jump in $\sigma_{0}$ occurs once the NS becomes transparent to neutrinos at $t = t_{\rm\nu-thin} \sim 30-100$ s (Fig.~\ref{fig:pons}), after which $\sigma_{0}$ rises to very large values $\gtrsim 10^{9}$.  This transition likely ends the prompt high energy emission.  Although the argument for why $t_{\rm end} \sim t_{\rm \nu-thin}$ is quite general, we can be concrete by defining $t_{\rm end}$ as the time after which the magnetic dissipation radius $R_{\rm mag}$ (eq.~[\ref{eq:rsat}]) exceeds the internal shock radius $R_{\rm is}$.  For $t \gtrsim t_{\rm end}$ the jet magnetization at the shock radius exceeds the critical value $\sigma \sim 0.1$ above which strong shocks are suppressed \citep{Kennel&Coroniti84}.  The association of $t_{\rm end}$ with $t_{\rm \nu-thin}$ explains both the typical duration of long GRBs $T_{\rm 90} \sim 10-100$ s and accounts for why the prompt $\sim$ MeV emission declines more rapidly at lates times $\gtrsim T_{\rm 90}$ ($\propto t^{-3}$) than the jet luminosity predicted by most central engine models (e.g.~\citealt{Tagliaferri+05}; \citealt{Barniol-Duran&Kumar09}).     

Figure \ref{fig:T90} shows contours of the rest-frame GRB duration $T_{\rm 90}$, defined as the time interval within max[$t_{\rm bo},t_{\rm thin,is}] \lesssim t \lesssim t_{\rm end}$ during which 90$\%$ of the wind energy is released.  Note that high spin-down (`GRB capable') magnetars have $T_{\rm 90} \sim 40-50$ seconds, similar to the average observed rest-frame duration of long GRBs.  A qualitatively similar, though somewhat shorter, $T_{\rm 90}$ distribution results if we assume that emission begins at $t_{\rm thin,mag}$ (magnetic dissipation) rather than $t_{\rm thin,is}$ (internal shocks).  The true predicted (rest frame) GRB duration distribution will of course be broader than suggested by Figure \ref{fig:T90} because we have not taken into account variations in the timescale for jet break-out $t_{\rm bo}$ (eq.~[\ref{eq:tbo}]) and neutrino transparency $t_{\rm \nu-thin}$, the latter of which depends on the NS mass and rotation rate.  Realistic variations in $t_{\rm bo}$, $\eta_{\rm s}$, and $M_{\rm ns}$ will undoubtedly broaden the rest-frame $T_{\rm 90}$ distribution by factors of a few as observed (see Table \ref{table:winds} for examples).

Except in the case of very luminous GRBs ($\S\ref{sec:VLGRBs}$), most of the magnetar's initial rotational energy remains when the prompt emission ends.  Though not released as gamma-rays, this residual energy may be dissipated at later times or larger radii and hence may contribute, for instance, to the GRB X-ray afterglow. In $\S\ref{sec:highsig}$ we discuss emission during the late-time high-$\sigma_{0}$ phase.

\section{Gamma-Ray Burst Emission}
\label{sec:GRB}
In this section we calculate the emission during the prompt phase ($t_{\rm bo} \lesssim t \lesssim t_{\rm end}$; Stage IV).  We begin with a discussion of the mechanisms for jet acceleration ($\S\ref{sec:acceleration}$) and variability ($\S\ref{sec:variability}$) and then present calculations of the gamma-ray emission produced by magnetic dissipation ($\S\ref{sec:magdiss}$) and internal shocks ($\S\ref{sec:internalshocks}$).

\subsection{Acceleration}
\label{sec:acceleration}

Energy carried by the relativistic wind is primarily in the magnetic field near the light cylinder radius $R_{\rm L} \sim 10^{7}$ cm.  Because GRBs originate from ultra-relativistic outflows (e.g.~\citealt{Lithwick&Sari01}), this magnetic energy must be transferred to kinetic energy prior to the radii $\sim 10^{12}-10^{16}$ cm at which the high energy emission occurs.  Unconfined, time-stationary Poynting-flux dominated outflows do not accelerate efficiently in ideal MHD (\citealt{Goldreich&Julian70}; \citealt{Beskin+98}; \citealt{Bogovalov&Tsinganos99}).  The Lorentz factor reached when acceleration slows near the fast magnetosonic surface $\Gamma_{\infty} \sim \sigma_{0}^{1/3}$ \citep{Goldreich&Julian70} is much less than the maximum possible value $\Gamma_{\rm max} \approx \sigma_{0}$.  Full acceleration to $\Gamma_{\infty} \sim \Gamma_{\rm max}$ therefore appears to require a combination of a differentially-collimated (non-monopolar) geometry, time variability, or violations of ideal MHD (see \citealt{Komissarov10} for a recent review).    

At small radii the wind is concentrated in the rotational equator.  On larger scales the outflow is redirected into a bipolar jet by its interaction with the star (\citealt{Komissarov&Barkov07}; \citealt{Bucciantini+09}).  Analytic (e.g.~\citealt{Vlahakis&Konigl01}; \citealt{Narayan+07}) and numerical (\citealt{Komissarov+07,Komissarov+09}; \citealt{Tchekhovskoy+09,Tchekhovskoy+10}) calculations show that if the jet is confined into a parabolic shape, additional acceleration is possible due to `equilibrium collimation'.  However, the maximum Lorentz factor that can be achieved in this manner is $\Gamma_{\infty} \sim 1/\theta_{\rm j} \sim 10$ (eq.~[\ref{eq:fbcorrelation}]) because only while $\Gamma\theta_{\rm j} \lesssim 1$ does the jet remain in lateral causal contact.  Although an additional boost of acceleration (by a factor $\lesssim$ 10) may occur as the jet emerges from the stellar surface (\citealt{Komissarov+09}; \citealt{Tchekhovskoy+09}), reaching $\Gamma_{\infty} \gtrsim 10^{2}$ and simultaneously achieving high conversion efficiency of magnetic to kinetic energy appears difficult via collimation alone.

A time-dependent flow can also produce acceleration.  In the so-called `astrophysical plasma gun' or `magnetic rocket' mechanism (\citealt{Contopoulos95}; \citealt{Granot+10}; \citealt{Lyutikov&Lister10}; \citealt{Lyutikov10}), a high-$\sigma_{0}$ magnetic pulse of finite width expands into a lower density medium (`vacuum'; see, however, \citealt{Levinson10}).  As the shell propagates, it `self-accelerates' via magnetic pressure gradients which develop as a rarefaction wave passes through the shell.  In this case $\Gamma$ increases $\propto r^{1/3}$ (e.g.~\citealt{Granot+10}), similar to the magnetic dissipation model described below.  Faster acceleration $\Gamma \propto r$ is possible in standard (high entropy) GRB fireball models (e.g.~\citealt{Goodman86}), but it remains unclear how the necessary thermalization would occur inside the star,\footnote{One possibility is if instabilities act within the jet to randomize the magnetic field, such that it behaves as a $\gamma = 4/3$ relativistic gas (\citealt{Heinz&Begelman00}; \citealt{Giannios&Spruit06}).} especially considering that reconnection may be slow in the collisional environment close to the central engine (\citealt{McKinney&Uzdensky10}).  Note that no ideal MHD model for jet acceleration accounts for the dissipation of energy responsible for powering the GRB, which must instead occur at larger radii after acceleration is complete.

An alternative possibility for jet acceleration is magnetic dissipation, i.e.~a break-down of ideal MHD (\citealt{Spruit+01}; \citealt{Drenkhahn02}; \citealt{Drenkhahn&Spruit02}).  One way this can occur is if the rotation and magnetic axes of the NS are misaligned ($\chi > 0$), such that the outflow develops an alternating or `striped' magnetic field geometry \citep{Coroniti90} on the scale of the light cylinder radius.  If this non-axisymmetric pattern is preserved when the flow is redirected along the polar jet, the resulting geometry is conducive to magnetic reconnection.  Magnetic dissipation occurs gradually from small radii up to the `saturation' radius $R_{\rm mag}$, beyond which reconnection is complete and the flow achieves its terminal Lorentz factor.  During this process, approximately half the Poynting flux is directly converted into kinetic energy (producing acceleration) and the other half is deposited into the internal (thermal) energy of the flow \citep{Drenkhahn&Spruit02}.  Acceleration and emission thus both result from the same physical mechanism.

\citet{Drenkhahn02} shows that the Lorentz factor of the jet as function of radius is given by
\begin{equation}
\Gamma_{\rm j} =
\left\{
\begin{array}{lr}
\sigma_{0}(r/R_{\rm mag})^{1/3}
, \qquad &
r<R_{\rm mag} \\
\sigma_{0}
, \qquad & 
r>R_{\rm mag} \\
\end{array}
\right.,
\label{eq:gamma_magdis}
\end{equation}
where the saturation radius is
\be
R_{\rm mag} = \frac{\pi c \sigma_{0}^{2}}{3\epsilon\Omega} = 5\times 10^{12}{\,\rm cm}\left(\frac{\sigma_{0}}{10^{2}}\right)^{2}\left(\frac{P}{\rm\,ms}\right)\left(\frac{\epsilon}{0.01}\right)^{-1} 
\label{eq:rsat}
\ee
and $\epsilon \lesssim 1$ parametrizes the reconnection speed $v_{\rm r} = \epsilon v_{\rm A}$, where $v_{\rm A} \simeq c$ is the Alfven speed.  In our calculations we assume that $\epsilon = 0.01$, independent of radius or jet properties.  This value is motivated by recent work finding a reconnection rate of this order due to secondary tearing instabilities in the current sheets (e.g.~\citealt{Uzdensky+10}), even in highly collisional environments, that characterizes the jet close to the central engine.  On the other hand, at larger radii (yet still well below the nominal saturation radius), reconnection may occur in the collisionless regime, such that faster reconnection is also likely (see e.g.~\citealt{Arons08}, \citealt{McKinney&Uzdensky10} for specific physical dissipation mechanisms).

\begin{figure*}
\begin{center}$
\begin{array}{cc}
\resizebox{\hsize}{!}{\includegraphics[width=8.5cm]{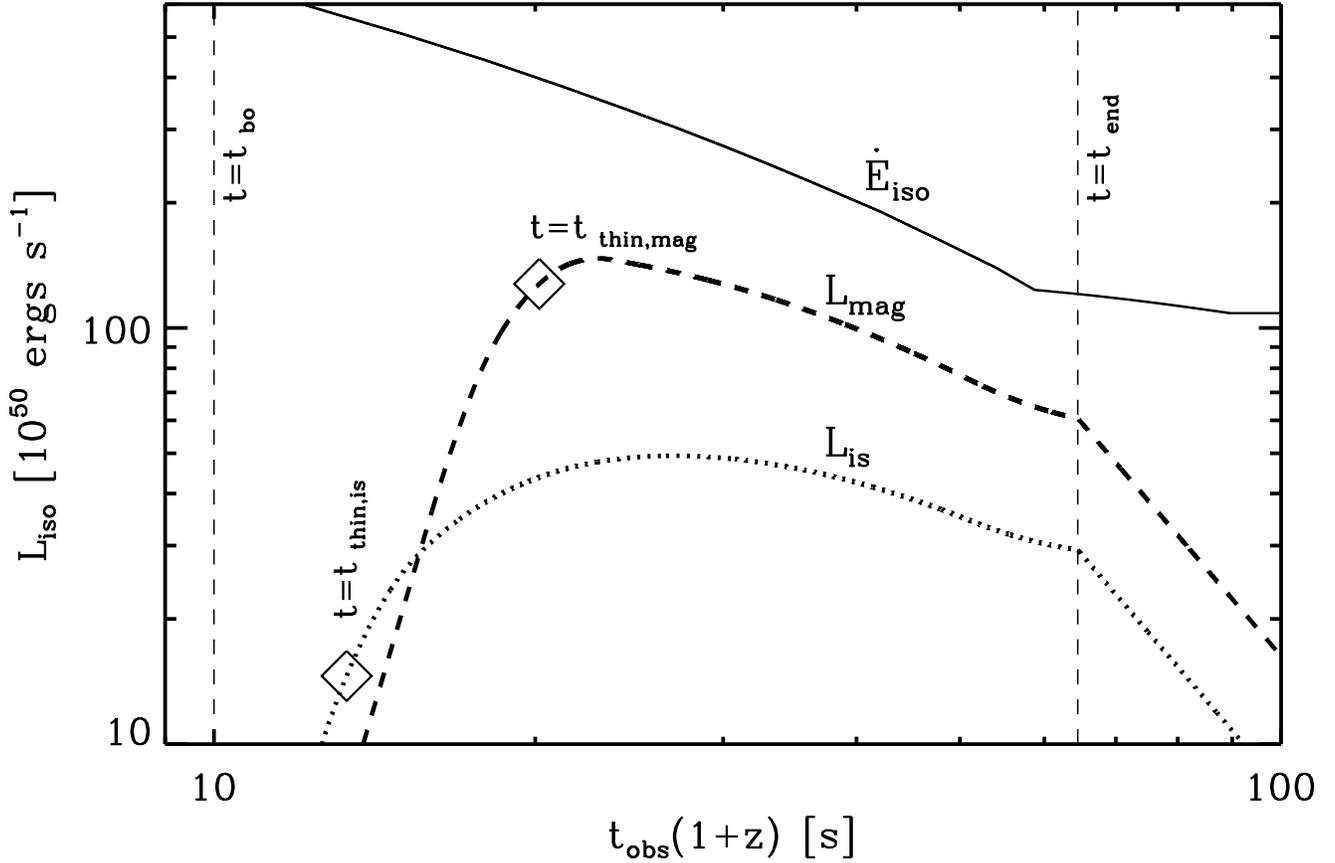}}
\end{array}$
\end{center}
\caption{Bolometric GRB luminosity due to magnetic dissipation ({\it dashed line}) and internal shocks ({\it dotted line}) as a function of observer time $t_{\rm obs}$, calculated for a proto-magnetar with $B_{\rm dip} = 2\times 10^{15}$ G, $P_{0} = 1.5$ ms, and $\chi = \pi/2$ (Fig.~\ref{fig:edotsig}).  For internal shocks we assume that $\epsilon_{\rm mag} = 0.5$ and $\epsilon_{e} = 1$ (see text for definitions).  The isotropic power of the jet $\dot{E}_{\rm iso}$ is shown for comparison with a solid line and is calculated assuming a beaming fraction $f_{\rm b} = 2\times 10^{-3}$.  The times when the jet becomes Thomson thin to emission from internal shocks and magnetic dissipation are marked with diamonds.  Although emission is suppressed at early times due to adiabatic losses, at times $t \gg t_{\rm thin}$ the radiative efficiency of both magnetic dissipation and shocks approaches $\sim 1/2$ (eq.~[\ref{eq:epsilonr}]).}  
\label{fig:luminosities}
\end{figure*}

\subsection{Variability}
\label{sec:variability}

Although GRBs are variable on timescales down to fractions of a millisecond (\citealt{Schaefer&Walker99}; \citealt{Walker+00}), most Fourier power is concentrated on a characteristic timescale $\sim 1$ second (\citealt{Beloborodov+98,Beloborodov+00}).  GRB variability may be related to the emission mechanism itself, or it may reflect real variations in the power and mass loading of the jet (e.g.~\citealt{MacFadyen&Woosley99}; \citealt{Aloy+00}; \citealt{Mizuta&Aloy09}; \citealt{Morsony+10}).

There are several potential sources of variability in proto-magnetar outflows.  Sporadic changes to the magnetosphere could modulate the magnetar wind properties on short ($\lesssim$ millisecond) timescales due to reconnection near the light cylinder \citep{Bucciantini+06} or on longer timescales due to neutrino heating in the closed zone \citep{Thompson03}.  Longer timescale variability could also be imposed on the outflow as it propagates to the stellar surface, due to instabilities associated with the termination shock(s) in the proto-magnetar nebula (\citealt{Bucciantini+09}; \citealt{Camus+09}) or at larger distances as the jet propagates through the stellar envelope \citep{Morsony+07,Morsony+10}.  The latter possibility is particularly promising because the sound crossing time across the jet near the stellar radius is in fact $\sim 1$ second (e.g.~\citealt{Morsony+10}; \citealt{Lazzati+10}) and might not evolve appreciably throughout the burst, a fact consistent with observations \citep{Ramirez-Ruiz&Fenimore99}.

The time-averaged wind properties calculated in $\S\ref{sec:windmodel}$ ($\dot{E}, \dot{M},$ and $\sigma_{0})$ do not account for any of the variability discussed above.  In fact, given the stochastic nature of GRB emission, it seems unlikely that any model will be capable of predicting the detailed light curve of individual bursts.  In our calculations below, we instead focus on predicting the {\it time-averaged} high energy emission over timescales of seconds or longer, which may be usefully compared with integrated GRB light curves and spectra (e.g.~\citealt{McBreen+02}).  We nevertheless emphasize that variability affects the observed emission differently depending on the emission model.  Magnetic dissipation, for instance, occurs at relatively small radii, such that variability is directly encoded in the emitted radiation.  Variability from internal shocks instead manifests indirectly through the effects of subsequent collisions at larger radii.

\subsection{Emission from Magnetic Dissipation}
\label{sec:magdiss}

\citet{Drenkhahn&Spruit02} make specific predictions for the rate that magnetic energy is dissipated with radius.  However, reconnection can in principle energize particles in a variety of ways.  Reconnection can lead to plasma heating and acceleration in localized regions\footnote{This is the approach adopted by \citet{Lyutikov&Blandford03}.  However, because the mechanisms responsible for particle acceleration in magnetic reconnection are uncertain, it is difficult to make concrete predictions for the resulting GRB emission in this case.} (e.g.~current layers).  Alternatively, reconnection may drive bulk motions in the jet that excite Alfvenic turbulence (e.g.~\citealt{Thompson94}), which cascades to small scales and heats larger volumes in the plasma.  We follow the model of \citet{Giannios06,Giannios08}, who assumes that the dissipated energy heats the plasma smoothly throughout the flow (slow heating model; see \citealt{Ghisellini&Celotti99}; \citealt{Stern&Poutanen04}).  Similar qualitative conclusions would, however, result from any model that invokes localized modest particle acceleration close to the photosphere (e.g.~\citealt{Lazzati&Begelman10}).

\citet{Giannios08} shows that energy dissipated at large Thomson optical depths is thermalized, such that a portion emerges through the photosphere with a peak at $\sim$ MeV energies (cf.~\citealt{Goodman86}; \citealt{Meszaros&Rees00}; \citealt{Ramirez-Ruiz05}; \citealt{Peer+06}; \citealt{Giannios06}; \citealt{Beloborodov10}).  At times when $R_{\rm mag} \gtrsim R_{\rm ph}$ most of the Poynting flux is dissipated near or above the photosphere and the equilibrium temperature of the electrons exceeds the radiation temperature.  Inverse Compton scattering of the photons advected outwards with the flow then results in power-law emission with a flat spectral slope $E\cdot L_{E}\propto E^0$ above the thermal peak.  Larger radii in the flow are heated to yet higher temperatures, resulting in an additional component of synchrotron and synchrotron-self-Compton emission at lower frequencies (i.e.~optical, UV, and X-ray bands).  This softens the spectrum below the MeV peak close to the observed $E\cdot L_{E}\propto E^1$ value.  

Figure \ref{fig:luminosities} shows the bolometric (isotropic) luminosity due to magnetic dissipation $L_{\rm mag}$, calculated for the fiducial model shown in Figure \ref{fig:edotsig}.  At late times $t \gtrsim t_{\rm thin,mag} \approx 20$ s, magnetic dissipation occurs above the photosphere ($R_{\rm mag} \gtrsim R_{\rm ph}$) and $L_{\rm mag} = \dot{E}_{\rm iso}/2$ \citep{Drenkhahn&Spruit02}.  At early times $t \lesssim t_{\rm thin,mag}$ when $R_{\rm mag} \lesssim R_{\rm ph}$, by contrast, $L_{\rm mag}$ is suppressed below $\dot{E}_{\rm iso}$ by an additional factor $\sim 0.4(R_{\rm mag}/R_{\rm ph})^{2/3}$ due to adiabatic losses incurred between the dissipation radii and the photosphere.  

\begin{figure}
\begin{center}$
\begin{array}{cc}
\includegraphics[width=8.5cm]{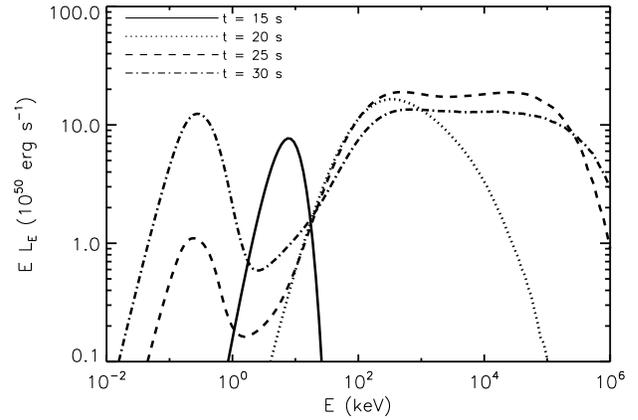}
\end{array}$
\end{center}
\caption{Spectral energy distributions $E\cdot L_{\rm E}$ of the magnetic dissipation model calculated at several times, $t = 15$ s ({\it solid}), $t = 20$ s ({\it dotted}), $t = 25$ s ({\it dashed}), and $t = 30$ s ({\it dot-dashed}), for the same model shown in Figures \ref{fig:edotsig}, \ref{fig:luminosities}, and \ref{fig:epeak}.}  
\label{fig:magdis2}
\end{figure}

Figure \ref{fig:magdis2} shows snapshots of the high energy spectrum, calculated at the times $t = 15$, 20, 25, and 30 seconds.  The spectrum at $t = 15$ s corresponds to an early epoch when $R_{\rm ph} \gg R_{\rm mag}$ and the dissipated energy is thermalized; the spectrum in this case is approximately Planckian with temperature $T \simeq 2$ keV.  Due to its low luminosity and X-ray peak, such a component of early thermal emission may be challenging to detect in an actual GRB (Fig.~\ref{fig:thermfrac}).  At later times $t \gtrsim t_{\rm thin,mag} \approx 20$ s, by contrast, dissipation peaks near or above the photosphere.  This results in a spectral peak at energy $E_{\rm peak} \sim 10^{2}$ keV (described below) and a nonthermal Comptonized tail that extends to increasingly higher energies as $\sigma_{0}$ rises and the outflow becomes cleaner with time.  Also note the component of synchrotron emission at softer X-ray/UV wavelengths, which increases in relative importance to the Comptonized gamma-rays as the jet magnetization increases and dissipation peaks at larger radii.   

\begin{figure}
\begin{center}$
\begin{array}{cc}
\includegraphics[width=8.5cm]{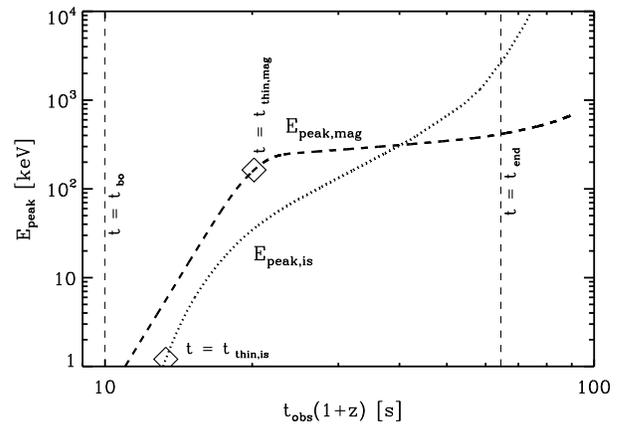}
\end{array}$
\end{center}
\caption{Peak spectral energy (or break energy) $E_{\rm peak}$ as a function of observer time $t_{\rm obs}$ in magnetic dissipation ({\it dashed line}) and synchrotron internal shock ({\it dotted line}) models, calculated for the model shown in Figures \ref{fig:edotsig} and \ref{fig:luminosities}}  
\label{fig:epeak}
\end{figure}

The magnetic dissipation model predicts a spectral energy peak $E_{\rm peak}$ (or break\footnote{The $E\cdot L_{E}$ spectrum above the break may (depending on parameters) be slowly rising.  In this case the $\sim$ MeV `peak' is formally a break.}) similar to the observed Band spectrum peak $\sim$ few hundred keV and which is relatively insensitive to the jet properties.  \citet{Giannios&Spruit07} show that to good approximation
\begin{eqnarray}
E_{\rm peak,mag} \simeq 270{\,\rm keV}\left(\frac{\dot{E}_{\rm iso}}{10^{52}\,{\rm ergs\,s^{-1}}}\right)^{0.11}\left(\frac{\epsilon\Omega}{10^{2}}\right)^{0.33}\left(\frac{\sigma_{0}}{10^{2}}\right)^{0.2} \nonumber \\
\label{eq:epeakMD}
\end{eqnarray}
for $R_{\rm mag} \gtrsim R_{\rm ph}$.

Figure \ref{fig:epeak} shows $E_{\rm peak}$ as a function of time for the fiducial model shown in Figures \ref{fig:luminosities} and \ref{fig:magdis2}.  Note that although $E_{\rm peak}$ rises rapidly at times $t \lesssim t_{\rm thin,mag}$ (when the luminosity is highly suppressed), $E_{\rm peak}$ is relatively constant during the GRB itself, increasing from $\sim 200$ keV to $\sim 400$ keV between $t = t_{\rm thin,mag}$ and $t = t_{\rm end}$.  This slow evolution results from the weak dependence of $E_{\rm peak}$ on $\dot{E}_{\rm iso}(t)$ and $\sigma_{0}(t)$ in equation (\ref{eq:epeakMD}).  A rising value of $E_{\rm peak}$ at first seems in conflict with the observation that GRBs are usually inferred to spectrally `soften' throughout their duration.  This behaviour may, however, still be consistent with spectral evolution predicted by magnetic dissipation if the synchrotron emission at lower frequencies begins to contaminate the soft X-ray bands at late times ($E_{\rm peak,mag}$ refers to the spectral peak of the Inverse-Compton emission; see Fig.~\ref{fig:magdis2}).

\begin{figure*}
\centering
\subfigure[$\chi = 0$]{
\includegraphics[width=0.48\textwidth]{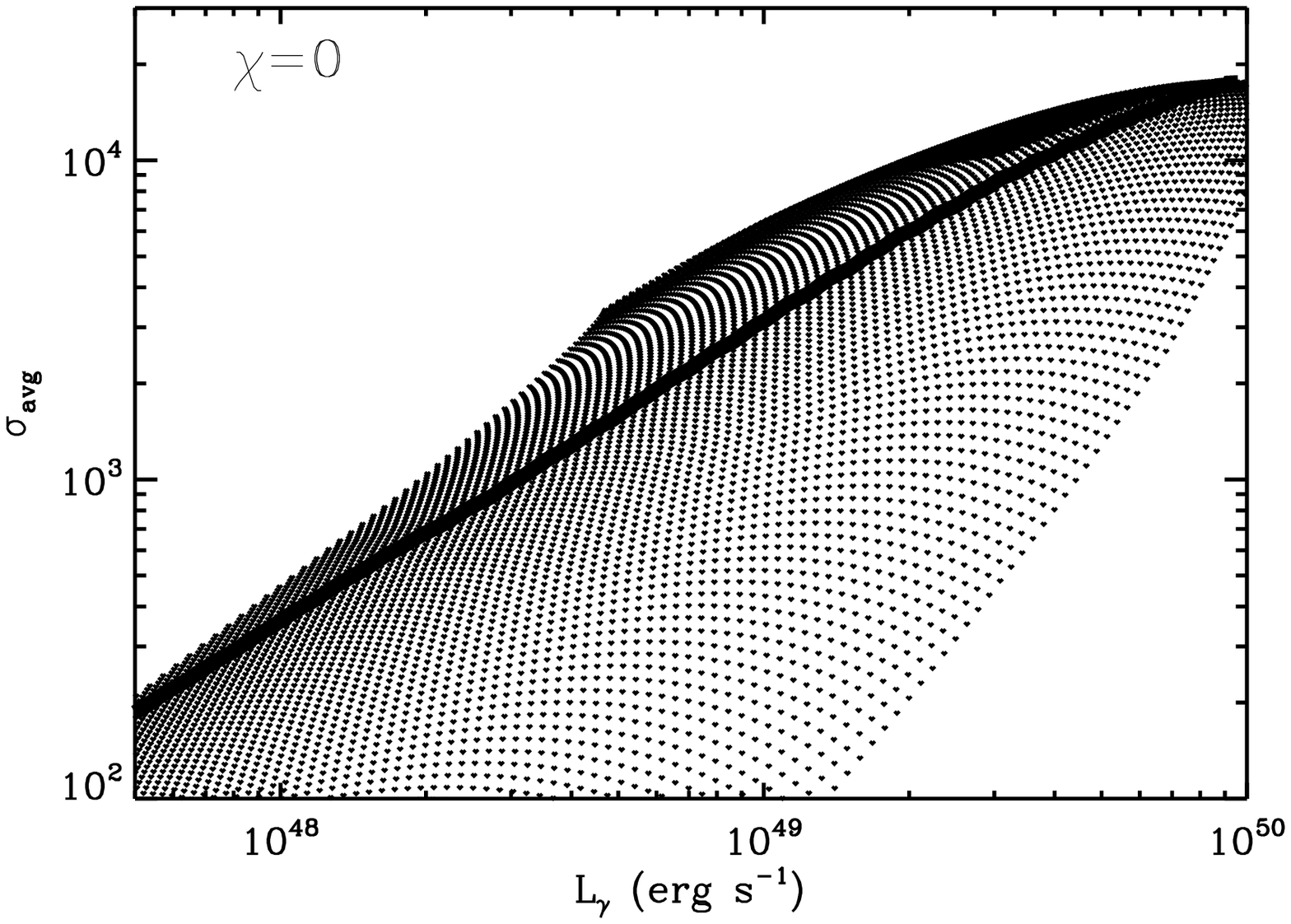}
}
\subfigure[$\chi = \pi/2$]{
\includegraphics[width=0.48\textwidth]{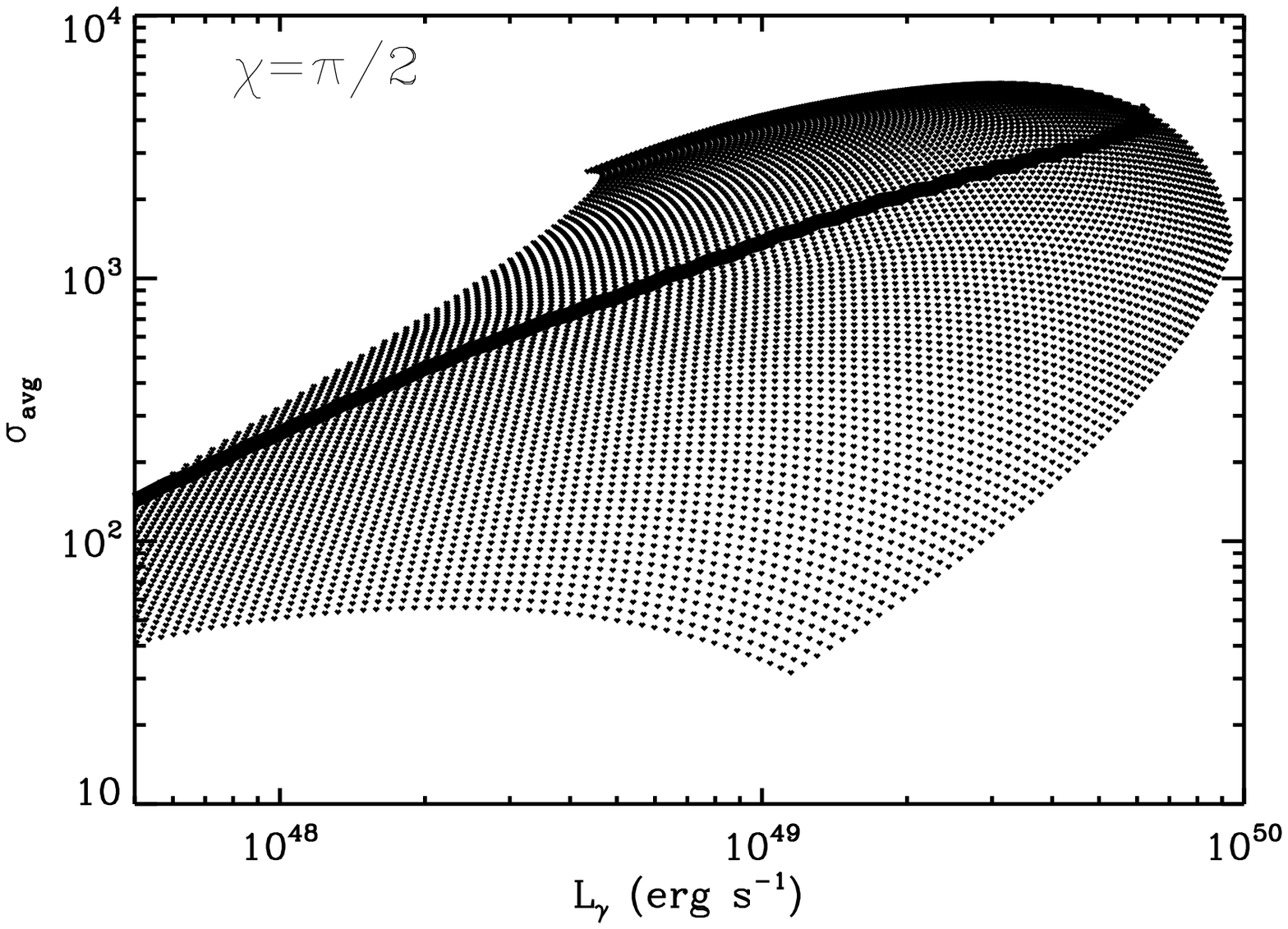}
}
\caption{Average magnetization ${\sigma}_{\rm avg}$ during the GRB versus the average GRB luminosity $L_{\gamma} \equiv E_{\gamma}/T_{\rm 90}$.  Each point represents a model calculated within the range of initial spin periods 1 ms $\lesssim P_{0} \lesssim$ 5 ms and surface dipole fields $3\times 10^{14}$ G $\lesssim B_{\rm dip} \lesssim 3\times 10^{16}$ G.  The left and right panels show calculations performed assuming the magnetic obliquity $\chi = 0$ and $\chi = \pi/2$, respectively.  A solid line connects solutions lying along the one-parameter family $B_{\rm dip} \propto P_{0}^{-1}$ defined by equation (\ref{eq:beq}) assuming $\epsilon_{\rm B} = 10^{-3}$.}
\label{fig:meansiglum}
\end{figure*}

We now consider the implications of equation (\ref{eq:epeakMD}) for the population of magnetar-powered GRBs as a whole.  Figure \ref{fig:meansiglum} shows a scatter plot of the average magnetization $\sigma_{\rm avg}$ during the GRB (Fig.~\ref{fig:meansig}) as a function of the average GRB luminosity $L_{\gamma} \equiv E_{\gamma}/T_{\rm 90}$ (Figs.~\ref{fig:eGRB} and \ref{fig:T90}), where we have included data points from all models within the range of magnetar parameters explored previously (1 ms $\lesssim P_{0} \lesssim$ 5 ms; $3\times 10^{14}$ G $\lesssim B_{\rm dip} \lesssim 3\times 10^{16}$ G; $\chi = 0$ and $\chi = \pi/2$, respectively).  Magnetars lying on the one-parameter family $B_{\rm dip} \propto P_{0}^{-1}$ defined by equation (\ref{eq:beq}) for $\epsilon_{B} = 10^{-3}$ are connected with a solid line.  

Figure \ref{fig:meansiglum} shows that the magnetar model predicts, with large scatter, a positive correlation between $\sigma_{\rm avg}$ and $L_{\gamma}$.  In particular, for the one-parameter family of solutions we find that $\sigma_{\rm avg} \propto L_{\gamma}^{\alpha}$, where $\alpha \simeq 0.5-1$, depending on $L_{\gamma}$ and $\chi$.  Assuming that the GRB duration, radiative efficiency $\epsilon_{\rm r}$, and beaming fraction $f_{\rm b}$ are similar from burst to burst, this correlation implies that $E_{\gamma} \propto \sigma_{\rm avg}^{1/\alpha}$.  From equation (\ref{eq:epeakMD}) this in turn leads to the prediction that $E_{\rm peak} \propto E_{\gamma}^{0.2-0.4}$ for $\epsilon\Omega \sim$ constant.  Note that this is close to the \citet{Amati+02} relationship $E_{\rm peak} \propto E_{\gamma}^{0.4}$.  If one furthermore drops the assumption that $f_{\rm b}$ is constant and instead assumes $f_{\rm b} \propto E_{\gamma}^{-0.75}$, as motivated by the combined Amati and Ghirlanda relations (eq.~[\ref{eq:fbcorrelation}]), one finds $E_{\rm peak} \propto E_{\gamma}^{0.3-0.5}$, resulting in even better agreement with observations.  A qualitatively similar correlation is predicted between $E_{\gamma}$ and the peak jet power, consistent with the related `Yonetoku' relation (\citealt{Yonetoku+04}; cf.~\citealt{Wei&Gao03}).  We emphasize that both the normalization and the slope of the Amati/Yonetoku correlations are reproduced if we assume a reconnection rate $\epsilon = 10^{-2}$ favored by recent work \citep{Uzdensky+10}.  

\subsection{Emission from Internal Shocks}
\label{sec:internalshocks}

If the acceleration of the jet is efficient ($\S\ref{sec:acceleration}$), then a significant fraction of the Poynting flux is converted into kinetic energy.  The kinetic luminosity and Lorentz factor of the outflow in this case are given by $L_{\rm j}(t) \simeq (1-\epsilon_{\rm mag})\dot{E}(t)$ and $\Gamma_{\rm j}(t) \simeq (1-\epsilon_{\rm mag})\sigma_{0}(t)$, respectively, where $\epsilon_{\rm mag} \lesssim 0.5$ is the fraction of the power radiated during the acceleration phase, due to magnetic dissipation ($\S\ref{sec:magdiss}$).  In what follows we assume $\epsilon_{\rm mag} = 0.5$, although $\epsilon_{\rm mag} = 0$ would be appropriate if the magnetic energy that is dissipated is not radiated away or if acceleration is achieved by another mechanism.

Because $\sigma_{0} \sim \Gamma_{\rm j}$ increases monotonically during the GRB (Fig.~\ref{fig:edotsig}), slower material is released prior to faster material.  Strong shocks will occur once the faster material catches up provided that the residual magnetization of the jet is $\lesssim 0.1$ (\citealt{Kennel&Coroniti84}; \citealt{Mimica+09}).  This scenario is similar to the standard internal shock model for GRB emission (e.g.~\citealt{Rees&Meszaros94}; \citealt{Kobayashi+97}; \citealt{Daigne&Mochkovitch98}) with a few key differences to be discussed below.   

Immediately after the jet breaks out of the star, the fast and slow ejecta have comparable energies and speeds.  With time, however, the slow material released early accumulates into a common `bulk' shell, which we characterize by its total rest mass $M_{\rm s} = \int_{t_{\rm bo}}^{t}\dot{M}_{\rm j}dt$, energy $E_{\rm s}$, mean velocity $\beta_{\rm s}$, and mean Lorentz factor $\Gamma_{\rm s} \equiv E_{\rm s}/M_{\rm s}c^{2}$, where $\dot{M}_{\rm j} = L_{\rm j}/\Gamma_{\rm j}c^{2}$ (see Fig.~\ref{fig:meangs}).  At most times the jet's self-interaction is well described as a collision between the fast, variable jet and a slower (yet still ultra-relativistic) shell.  We model this interaction using a one-dimensional kinematic model, as described in Appendix \ref{sec:isappendix}.  Although this approach neglects the effects of pressure forces and the true multi-dimensional geometry (e.g.~\citealt{Zhang&MacFadyen09}), it provides a reasonable first approximation to the full hydrodynamic problem \citep{Daigne&Mochkovitch98, Daigne&Mochkovitch00, Daigne&Mochkovitch03}.    

Figure \ref{fig:luminosities} shows the (average) bolometric luminosity $L_{\rm is} = \epsilon_{\rm r}L_{\rm j}$ from shocks as a function of observer time $t_{\rm obs}(1+z)$, calculated assuming that the fraction of the shock's energy imparted to electrons is $\epsilon_{\rm e} \approx 1$.  As in the case of magnetic dissipation, at times $t \lesssim t_{\rm thin,is}$ we suppress $L_{\rm is}$ by an additional factor $\sim (R_{\rm ph}/R_{\rm is})^{-2/3}$ to account for adiabatic losses when shocks occur below the photosphere, where $R_{\rm is}$ is the internal shock radius given in eq.~[\ref{eq:rsh}].  Note that although $L_{\rm j}$ decreases by a factor of $\approx 6$ throughout the burst, $L_{\rm is}$ changes by only a factor of a few.  Indeed, both magnetic dissipation and shock models predict that the bolometric luminosity should be relatively constant in time, a result in agreement with the approximately linear slope of cumulative GRB light curves (e.g.~\citealt{McBreen+02}). 

In Appendix $\ref{sec:isappendix}$ we calculate the peak energy $E_{\rm peak,is}$ of the synchrotron spectrum as a function of the jet and shell properties, assuming that a fraction $\zeta_{e}$ of electrons are accelerated and that a fraction $\epsilon_{\rm B}$ of the shock energy goes into generating the magnetic field (see eq.~[\ref{eq:epeakis}] and surrounding discussion).  

Figure \ref{fig:epeak} shows the evolution of $E_{\rm peak,is}$ during the GRB for the fiducial model, calculated assuming $\epsilon_{e} \approx 1$, $\epsilon_{B} = 0.1$, and $\zeta_{e} = 0.3$.  These microphysical parameters are chosen {\it ad hoc} such that $E_{\rm peak}$ attains a value $\sim 10^{2}$ keV at peak luminosity characteristic of observed GRB spectra.  Even after this fine tuning, however, two problems remain for the internal shock model.  First, Figure \ref{fig:epeak} shows that $E_{\rm peak,is} $ increases by over three orders of magnitude during the burst, in contradiction with the relatively constant (or decreasing) peak energy measured during actual bursts.  Although both $\Gamma_{\rm s}$ and $t_{\rm j}$ increase with time, $E_{\rm peak,is} \propto t^{-1}\Gamma_{\rm j}^{2}\Gamma_{\rm s}^{-4}$ increases because the jet Lorentz factor $\Gamma_{\rm j}$ increases even more rapidly (see eq.~[\ref{eq:epeakis}]).  Although a slowly-evolving peak energy could in principle be recovered by invoking e.g.~time-dependent microphysical parameters, fine tuning appears unavoidable (e.g.~\citealt{Zhang&Meszaros02}).  A second problem is that the variability timescale produced by subsequent internal collisions $\delta t_{\rm var} \propto R_{\rm sh}/2\Gamma_{\rm s}^{2} \propto t$ is predicted to increase linearly with time, again contrary to observations suggesting that $\delta t_{\rm var}$ evolves weakly during the burst \citep{Ramirez-Ruiz&Fenimore99}.\footnote{Note that this problem does not arise in the standard internal shock model because $\Gamma_{\rm j}$ is assumed to vary randomly throughout the burst (e.g.~\citealt{Beloborodov+00}), rather than to systematically increase as predicted by the magnetar model.}  We conclude that synchrotron emission from internal shocks appears disfavored as the source of prompt emission from proto-magnetars.  

\begin{figure*}
\resizebox{\hsize}{!}{\includegraphics[]{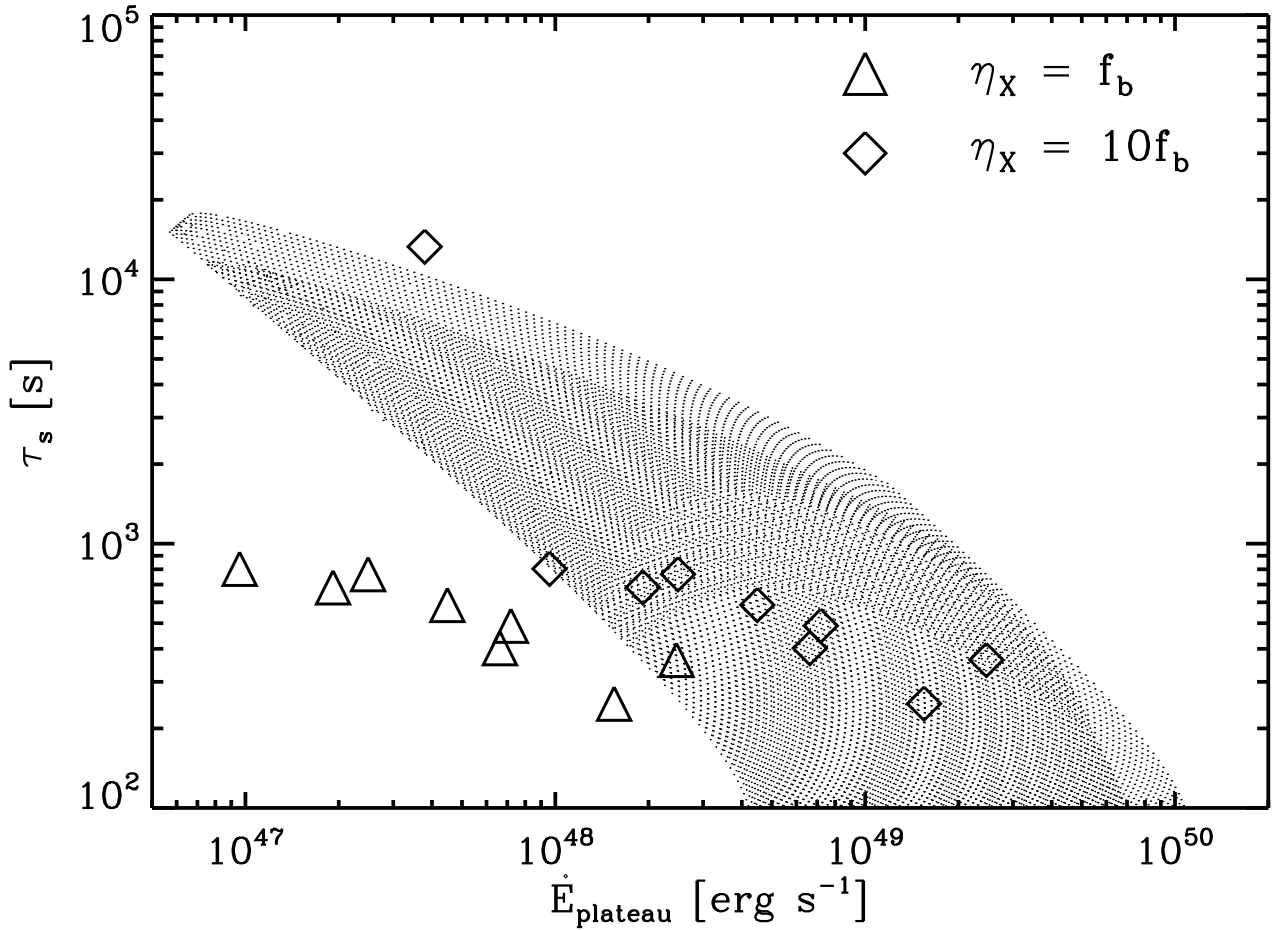}}
\caption{Scatter plot of the wind power at the beginning of the plateau-like, high-$\sigma_{0}$ phase $\dot{E}_{\rm plateau} \equiv \dot{E}|_{t_{\rm end}}$ as a function of the spin-down timescale $\tau_{\rm s}|_{t_{\rm end}}$.  Each point represents a model calculated within the range of initial spin periods 1 ms $\le P_{0} \le 5$ ms and surface dipole fields $3\times 10^{14} \le B_{\rm dip} \le 3\times 10^{16}$ G; results are shown for both magnetic obliquities $\chi = 0$ and $\chi = \pi/2$.  We also show for comparison the luminosities and end times $t_{\rm end,X}$ of the sample of plateaus from \citet{Lyons+10}, which show a steep decline in flux at times $t \gtrsim t_{\rm end,X}$.  Triangles and diamonds show the luminosities calculated assuming that the ratio between (observed) isotropic X-ray luminosity and wind power is equal to, or is a factor of 10 larger than, respectively, the gamma-ray beaming fraction (eq.~[\ref{eq:fbcorrelation}]). }
\label{fig:tauedot}
\end{figure*}

\section{Late-Time Emission}
\label{sec:highsig}

When $\sigma_{0}$ becomes very large at late times ($\gtrsim 10^{5}$; Fig.~\ref{fig:edotsig}), $R_{\rm mag}$ becomes so large that, even if reconnection occurs at the speed of light in the co-moving frame ($\epsilon \sim 1$), no efficient acceleration or dissipation occurs before the outflow begins to interact with itself or the external ISM.  This is the `causality limit' of \citet{Lyubarsky&Kirk01}.  Without acceleration, shocks cannot occur; and without efficient reconnection, there can be no dissipation-powered emission.  As we argued in $\S\ref{sec:stages}$, this transition ends the phase of prompt internal emission.  Similar physics occurs in the wind from the Crab Pulsar, for which the very high initial magnetization may prevent internal dissipation prior to the wind termination shock at $R \sim 10^{17}$ cm (\citealt{Kennel&Coroniti84}; \citealt{Lyubarsky&Kirk01}).  Emission from the nebula may be the result of forced reconnection at the termination shock itself \citep{Lyubarsky03, Lyubarsky05} or from dissipation in a striped wind (\citealt{Coroniti90}) if the pair multiplicity is higher than is commonly assumed (e.g.~\citealt{Arons08}).  Although internal dissipation in proto-magnetar winds is unlikely at $t > t_{\rm end}$, forced reconnection at large radii is a potential source of late-time emission in this case as well (see below). 

When the prompt emission ends a significant fraction of the magnetar's initial rotational energy remains to be released in other forms.  Since the beginning of the {\it Swift} mission, evidence has accumulated that GRB central engines are indeed active at late times, from minutes to $\gtrsim$ hours following the burst.  The X-ray afterglow in particular shows a complex evolution, including a `plateau' phase in the light curve which is not predicted by the standard forward shock model (\citealt{Nousek+06}; \citealt{Willingale+07}).  Superimposed on the smoother afterglow are large amplitude X-ray flares (\citealt{Piro+05}; \citealt{Burrows+05, BurrowsD+07}; \citealt{Chincarini+07}; \citealt{Chincarini+10}), which share many properties with the prompt GRB emission \citep{Margutti+10b} and also appear to result from late-time central engine activity (\citealt{Lazzati&Perna07}; \citealt{Margutti+10}).  

Although the magnetic dissipation or internal shocks responsible for the prompt emission become ineffective when $\sigma_{0}$ is very large, spin-down luminosity can in principle power late-time emission in other ways.  Indeed, a spin-down origin for the X-ray plateau is suggested by the `plateau-like' evolution of the late-time wind power $\dot{E}(t)$ illustrated in Figures \ref{fig:edotsig} and \ref{fig:edotsig2} (e.g.~\citealt{Zhang+06}).  Spin-down can in principle power X-ray emission either indirectly by refreshing the forward shock (e.g.~\citealt{Granot&Kumar06}; \citealt{Dall'Osso+10}) or directly (`internally') by e.g.~forced reconnection at the forward shell (e.g.~\citealt{Lyubarsky03, Lyubarsky05}; \citealt{Thompson06}; \citealt{Zhang+11}) or by upscattering forward shock photons \citep{Panaitescu08}.  Internal emission appears favored in at least some cases due to the very steep decay observed in the X-ray flux following the plateau (e.g.~GRB 070110; \citealt{Troja+07}; \citealt{Lyons+10}, hereafter L10; \citealt{Rowlinson+10}).  

Figure \ref{fig:tauedot} shows a scatter plot of the wind power evaluated at the beginning of the plateau-like high-$\sigma_{0}$ phase $\dot{E}_{\rm plateau} \equiv \dot{E}|_{t_{\rm end}}$ as a function of the spin-down timescale $\tau_{\rm s}|_{t_{\rm end}}$, calculated for models spanning the usual range of magnetar parameters (1 ms $\lesssim P_{0} \lesssim$ 5 ms; $3\times 10^{14}$ G $\lesssim B_{\rm dip} \lesssim 3\times 10^{16}$ G; $\chi = 0,\pi/2$).  Force-free spin down is characterized by $\dot{E}_{\rm plateau} \propto B_{\rm dip}^{2}P_{\rm 0,p}^{-4}$ and $\tau_{\rm s} \propto B_{\rm dip}^{-2}P_{\rm 0,p}^{2}$, such that $\tau_{\rm s} \propto \dot{E}_{\rm plateau}P_{\rm 0,p}^{-2}$.  The vertical scatter in Figure \ref{fig:tauedot} therefore results entirely from the distribution in `initial' rotational periods $P_{\rm 0,p} \equiv P|_{t_{\rm end}}$ following the GRB.  

L10 measure the isotropic X-ray luminosities $L_{\rm X,iso}$ and end times $t_{\rm end,X}$ of the plateau phase for a subset of GRBs that show a steep decline in their X-ray flux at times $t > t_{\rm end,X}$.  In Figure \ref{fig:tauedot} we overplot $t_{\rm end,X}$ and the luminosity from L10 $L_{\rm X} = L_{\rm X,iso}\eta_{\rm X}^{-1}$ corrected by a factor $\eta_{\rm X} = f_{\rm b,X}\epsilon_{\rm r,X}^{-1}$ that accounts for both the X-ray beaming fraction $f_{\rm b,X}$ and the efficiency that spin-down power is converted into X-ray luminosity $\epsilon_{\rm r,X}$.  We show two cases, in which $\eta_{\rm X}$ equals, or is a factor $\simeq 10$ times larger than, the gamma-ray beaming fraction $f_{\rm b}$ (which we estimate using equation (\ref{eq:fbcorrelation}) and the measured isotropic GRB energies).  Note that because $t_{\rm end,X}$ is a lower limit on $\tau_{\rm s}$, figure \ref{fig:tauedot} shows that all of the plateaus measured by L10 are consistent with being powered by magnetar spin-down for $\eta_{\rm X} \lesssim 10f_{\rm b}$.  If $t_{\rm end}$ is instead interpreted as the spin-down time itself,\footnote{As would be the case if spin-down triggers an abrupt end to the emission due to e.g.~the delayed formation of a black hole from a rotationally-supported hyper-massive NS (e.g.~\citealt{Baumgarte+00}).} our results indicate that either (1) the jet opening angle during the plateau phase is a few times larger than during the GRB itself, i.e.~$f_{\rm b,X} \gg f_{\rm b}$ and/or (2) the fraction of the spin-down power escaping through the jet and radiated in X-rays is $\ll 1$.  Although it is natural to expect that the radiative efficiency may be low when $\sigma_{0}$ is very large at late times, too low of an efficiency may be inconsistent with afterglow energetics.  It is also possible that a fraction of the late-time spin-down energy is instead transferred to the supernova shock, although numerical simulations of the interaction of the wind with the star suggest this need not be the case during the GRB itself \citep{Bucciantini+09}.  
    
Late-time magnetar activity could also produce X-ray flaring.  \citet{Margutti+10} find that the average flare luminosity decreases as $L_{\rm flare} \propto t^{-\alpha}$ where $\alpha = 2.7$ (cf.~\citealt{Lazzati+08}, \citealt{Margutti+10}).  Although standard force-free spin-down predicts $\alpha = 2$ at times $\gg \tau_{\rm s}$, steeper decays are inferred from the {\it measured} braking indices $n$ of some pulsars (e.g.~$\alpha = 4/(n-1) \simeq 2.42$ for PSR J1846-0258 with $n = 2.65$; \citealt{Livingstone+07}).  If prompt emission is indeed suppressed at late times by the high magnetization of the jet, periodic enhancements in the jet's mass-loading could temporarily `revive' prompt-like internal emission, resulting in flaring.  Temporarily enhanced mass loss could result, for instance, from currents driven by a sudden rearrangement of the magnetosphere, analogous to Galactic magnetar flares \citep{Thompson&Beloborodov05}.  Indeed, X-ray flares could also be powered by the release of magnetic energy itself, which is $\gtrsim 10^{49}-10^{50}$ ergs for typical values of the interior field strength $B \sim 10^{16}-10^{17}$ G.  \citet{Giannios10} recently proposed searching for such `super-flares' in nearby Galaxies, which could in principle be observed even long after the GRB, possibly in coincidence with a relic radio afterglow.

\section{Discussion - A Diversity of Phenomena}
\label{sec:discussion}

\begin{figure*}
\resizebox{\hsize}{!}{\includegraphics[]{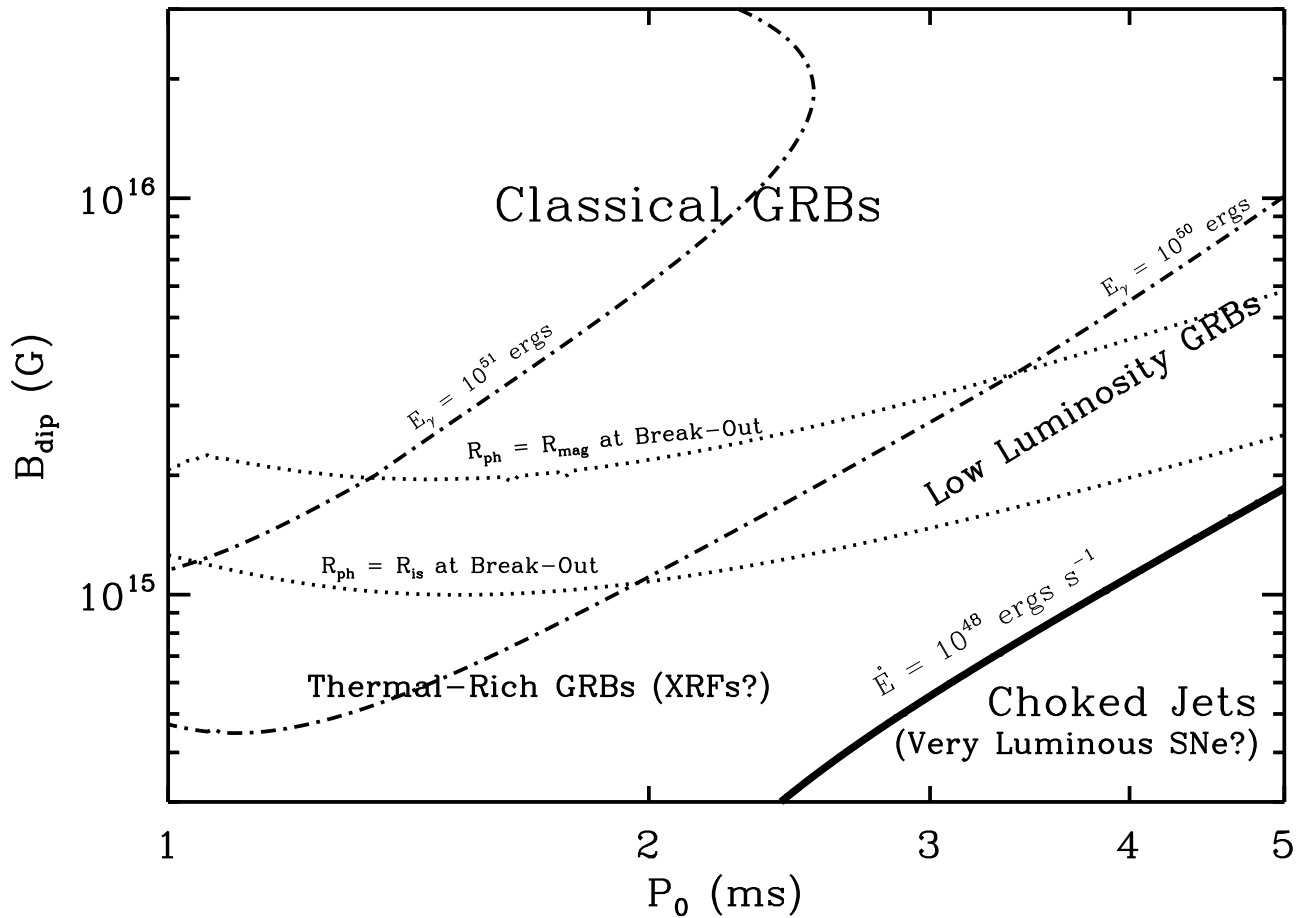}}
\caption{Regimes of high energy phenomena produced by magnetar birth in core collapse supernovae, as a function of the magnetic dipole field strength $B_{\rm dip}$ and initial rotation period $P_{0}$, calculated for an aligned rotator ($\chi = 0$).}
\label{fig:regimes}
\end{figure*}

Magnetars may form with a variety of properties (and under a variety of conditions) which, in turn, manifests as a diversity of high energy phenomena.  Figure \ref{fig:regimes} summarizes the possible observable signatures of magnetar birth as a function of the dipole field strength $B_{\rm dip}$ and birth period $P_{0}$.  Although the plot shown is for an aligned rotator ($\chi = 0$) qualitatively similar results apply to the oblique case as well.

\subsection{Classical GRBs}
\label{sec:classical}

Magnetars in the upper left hand quadrant of Figure \ref{fig:regimes} produce `classical GRBs' because (1) above the dotted lines the high energy emission is almost exclusively non-thermal because the relativistic jet dissipates its energy$-$through reconnection or shocks$-$above the photosphere beginning just after stellar break-out; (2) magnetars to the left of the dot-dashed line produce GRBs with energies $E_{\gamma} \gtrsim 10^{50}$ ergs (see Fig.~\ref{fig:eGRB}); (3) magnetars in this regime produce outflow with average magnetization $\sigma_{\rm avg} \sim 10^{2}-10^{3}$, consistent with the inferred Lorentz factors of long GRBs (Figs.~\ref{fig:meansig},\ref{fig:meangs}).  Note that the initial rotational energies of magnetars in this parameter regime are $\gtrsim 3\times 10^{51}$ ergs ($P_{0} \lesssim 3$ ms), implying that the requirements for a classical GRB and a hyper-energetic SN are remarkably similar (Fig.~\ref{fig:diagram1}). 

\subsection{Very Luminous GRBs}
\label{sec:VLGRBs}

Magnetars in the extreme upper left corner of Figure \ref{fig:regimes} produce classical GRBs with energies $E_{\gamma} \sim 10^{52}$ ergs which are comparable to the total rotational energy available (eq.~[\ref{eq:erot}]).  Evidence has recently grown for a class of `Very Luminous GRBs' (VLGRBs; e.g.~\citealt{Cenko+10a,Cenko+10b}), which includes several {\it Fermi} bursts such as GRB 080916C with an isotropic energy $E_{\rm \gamma,iso} \approx 8\times 10^{54}$ ergs (e.g.~\citealt{Abdo+09}).  The observation that energetic {\it Fermi} bursts appear to be distinguished by larger inferred Lorentz factors\footnote{Note, however, that the lower limit constraints on $\Gamma$ derived for {\it Fermi} bursts become weaker if the $\sim$ GeV and $\sim$ MeV photons originate from different radii (e.g.~\citealt{Zou+10}).} $\Gamma \gtrsim 10^{3}$ than is estimated for more typical GRBs is consistent with the correlation predicted by the magnetar model between the average GRB luminosity and jet magnetization $\sigma_{\rm avg}$ (maximum Lorentz factor), as shown in Figure~\ref{fig:meansiglum}.  Many extremely energetic GRBs, such as GRB 990123 (e.g.~\citealt{Kulkarni+99}) and 080319B \citep{Bloom+09}, are also distinguished by bright optical emission coincident with the GRB.  The synchrotron emission predicted by the magnetic dissipation model at optical-UV wavelengths contributes an especially large fraction of the total radiated energy in bursts with large magnetization $\sigma_{\rm avg}$ (see the late-time spectra in Fig.~\ref{fig:magdis2}).

At present, the properties of VLGRBs appear consistent with resulting from magnetars with extreme, but physically reasonable, properties.  However, measurements of the total energy in relativistic ejecta $E_{\rm tot} = E_{\gamma} + E_{\rm k}$ (where $E_{\rm k}$ is the kinetic energy) could constrain$-$or even rule out$-$the magnetar model as the central engine if $E_{\rm tot}$ were found to exceed the maximum rotational energy $\sim E_{\rm rot}(P_{0} \approx 1{\,\rm ms}) \sim 3\times 10^{52}$ ergs.  Although efforts are presently under way to determine $E_{\rm tot}$ for a sample of well-studied bursts \citep{Cenko+10a, Cenko+10b}, the results of these studies are hindered at present by simplifying assumptions in the afterglow modeling and jet structure, which may lead to systematic overestimates in $E_{\rm tot}$ (e.g.~\citealt{Zhang&MacFadyen09}; \citealt{vanEerten+10}).  Nevertheless, VLGRBs provide important probes of the most extreme central engine properties.

\subsection{Low Luminosity GRBs}
\label{sec:LLGRBs}

Magnetars to the right of the dot-dashed line in Figure \ref{fig:regimes} produce GRBs with energies $\lesssim 10^{50}$ ergs which may contribute to the class of so-called `low luminosity GRBs' (LLGRBs; e.g.~\citealt{Bloom+03b}; \citealt{Soderberg+04}; \citealt{Cobb+06}; \citealt{Liang+07}; \citealt{Kaneko+07}).  LLGRBs are distinguished from classical GRBs by their lower energies, simple gamma-ray light curves (generally a single pulse), longer durations, and higher local rates (e.g.~\citealt{Coward05}; \citealt{Le&Dermer07}; \citealt{Liang+07}).  Because large angular momentum is probably rare in core collapse supernovae, LLGRB-producing magnetars with weaker fields and/or slower rotation may indeed be formed more commonly than the magnetars responsible for classical GRBs.  

\subsection{Thermal-Rich GRBs and X-Ray Flashes}
\label{sec:XRF}

Magnetars below the dotted lines in Figure \ref{fig:regimes} produce jets that dissipate a significant fraction of their energy under optically thick conditions after breaking through the star (i.e.~they pass through Stage IVa described in $\S\ref{sec:stages}$) and produce jets with lower Lorentz factors than classical GRBs, i.e.~$\sigma_{\rm avg} \lesssim 10^{2}$.  We speculate that proto-magnetars in this regime may produce X-ray-rich GRBs or X-ray Flashes (XRFs; \citealt{Heise+01}; \citealt{Mazzali+06}) because they are accompanied by lower-frequency, quasi-thermal emission with an energy comparable to, or somewhat lower than, the non-thermal GRB emission itself (Fig.~\ref{fig:thermfrac}).  Although XRFs share many properties with long GRBs, such as an association with massive star formation (e.g.~\citealt{Bloom+03c}; \citealt{Soderberg+04b,Soderberg+07}), they may be distinguished from GRBs by their ability to couple a significant energy to highly relativistic material (e.g.~\citealt{Soderberg+04b}).  This is consistent with the fact that magnetars in the lower portions of Figure \ref{fig:regimes} indeed radiate a smaller fraction of their total energy during the GRB (as compared to the radiatively-inefficient high-$\sigma_{0}$ phase; $\S\ref{sec:highsig}$) than magnetars in the classical GRB regime.

\subsection{Choked Jets and Very Luminous Supernovae}
\label{sec:choked}

Magnetars in the lower right hand corner of Figure \ref{fig:regimes} produce jets with peak isotropic luminosities $\lesssim 10^{48}$ ergs s$^{-1}$.  Low power jet may be unstable \citep{Bucciantini+09} or take longer to propagate through the star than the duration of the GRB (e.g.~\citealt{Meszaros&Waxman01}; \citealt{Waxman&Meszaros03}).  Magnetars in this regime may thus produce `choked' jets with little direct electromagnetic radiation (although they could still be a source of high energy cosmic rays or neutrinos; e.g.~\citealt{Waxman95}; \citealt{Vietri95}; \citealt{Ando&Beacom05}; \citealt{Murase+09}).  

A number of core-collapse SNe have been recently discovered that are unusually bright and/or optically-energetic (e.g.~\citealt{Ofek+07}; \citealt{Smith+07}; \citealt{Smith+08}; \citealt{Quimby+07}; \citealt{Rest+09}; \citealt{Gal-Yam+07}; \citealt{Quimby+09}).  Proposed explanations for these events, collectively known as very luminous SNe (VLSNe), include pair-instability SNe (\citealt{Barkat+67}); interaction of the supernova shock with dense circumstellar material (e.g.~\citealt{Gal-Yam+07}; \citealt{Smith+07,Smith+08}; \citealt{Metzger10}); and the injection of late-time rotational energy from a rapidly-spinning magnetar (\citealt{Kasen&Bildsten10}; \citealt{Woosley10}).  In order to energize the supernova ejecta on the $\sim$ days-weeks timescales relevant for powering VLSNe, \citet{Kasen&Bildsten10} conclude that a magnetar with $B_{\rm dip} \sim 5\times 10^{14}$ G must possess an initial rotation period $P_{0} \sim 2-20$ ms.  This nominally places VLSNe-producing magnetars in the `choked jet' regime.  We note, however, that in order to explain VLSNe, the initially Poynting flux-dominated magnetar wind must thermalize its energy behind the SN shock, instead of escaping in a jet \citep{Bucciantini+09}, which might still be able to propagate through the star on the longer timescales of relevance for VLSNe.

\subsection{Galactic Magnetars}
\label{sec:galactic}

If known Galactic magnetars were born with magnetic fields similar to their current observed strengths $B_{\rm dip} \sim 10^{14}-10^{15}$ G (e.g.~\citealt{Kouveliotou+98}) and as fast rotators, then Figure \ref{fig:regimes} suggests that their formation was accompanied by a thermal-rich GRB/XRF or choked jet, depending on their initial rotational period.  Slower rotation, corresponding to a choked jet, may be likely in the majority of cases because Galactic magnetars are formed in $\sim 10\%$ of core collapse SN \citep{Woods&Thompson06}, yet only a small fraction of envelope-stripped SN are accompanied by relativistic ejecta \citep{Soderberg+06}.  Furthermore, the SN remnants of Galactic magnetars do not show evidence for hyper-energetic SN explosions (e.g.~\citealt{Vink&Kuiper06}; see, however, \citealt{Horvath&Allen10}).

\subsection{Accretion-Induced Collapse}
\label{sec:AIC}

This paper has focused on the core collapse of massive stars, but magnetars may also form via the accretion-induced collapse (AIC) of white dwarfs (WD, e.g.~\citealt{Nomoto+79}; \citealt{Usov92}).  Although AIC is probably intrinsically rarer than standard core collapse (e.g.~\citealt{Fryer+99}), millisecond magnetars may be a more common byproduct of AIC because the WD is spun up considerably as it accretes up to the Chandrasekhar mass.

A distinguishing characteristic of AIC is the lack of a massive overlying stellar envelope.  However, AIC does not produce a vacuum around the magnetar.  A small quantity of mass $\sim 10^{-3}-10^{-1}M_{\sun}$ is ejected during the supernova explosion itself \citep{Dessart+06} and in the early, mildly-relativistic phase of the neutrino wind (Stage II).  If the collapsing white dwarf furthermore has sufficient angular momentum, an accretion disk forms around the neutron star (\citealt{Michel87}; \citealt{Dessart+06}).  As this disk accretes onto the NS on a timescale $\lesssim 1$ s, outflows from the disk powered by nuclear recombination eject $\gtrsim 10^{-2}M_{\sun}$ in Nickel-rich material \citep{Metzger+09}.\footnote{This implies that although AIC is not accompanied by a bright supernova, it may produce a dimmer $\sim$ day-long transient (\citealt{Metzger+09}; \citealt{Fryer+09}; \citealt{Darbha+10}).}

Because the proto-magnetar is surrounded by a modest `sheath' of material, its relativistic wind from the magnetar may be collimated into a bipolar jet, analogous to the standard core collapse case.  Because of the lower inertia of this surrounding mass, however, collimation may be less effective and the opening angle of any `jet'-like structure may be considerably larger.  If this speculation is correct, it would imply a larger beaming fraction $f_{\rm b}$, lower isotropic luminosity, and softer spectral peak (e.g.~eq.~[\ref{eq:epeakMD}]) than in the core collapse case.  Perhaps equally important, the fact that the jet is no longer required to escape the star in order to produce high energy emission may `select' for magnetars with lower fields and/or slower rotation (and, hence, lower spin-down luminosities, lower Lorentz factors, and softer spectra) than in the core collapse case.
 
One of the biggest mysteries associated with short-duration GRBs is that $\gtrsim 1/4$ are followed by a `tail' of emission (usually soft X-rays) starting $\sim 10$ seconds after the GRB and lasting for $\sim 30-100$ seconds (\citealt{Norris&Bonnell06}; \citealt{Gehrels+06}; \citealt{Gal-Yam+06}; \citealt{Perley+09}; \citealt{Norris+10}).  Although the large inferred energies and durations of the tails are difficult to explain in NS merger models (e.g.~\citealt{Metzger+10}), their properties are similar to the prompt emission expected from magnetar birth via AIC.  \citet{Metzger+08b} proposed an AIC model for `short GRBs with extended emission', in which the short GRB is powered by the accretion of the disk onto the NS as described above and the subsequent `tail' is powered by the (wider-angle) proto-magnetar wind.  This model is consistent with the host galaxy demographics, and the lack of a bright supernova, associated with short GRBs (e.g.~\citealt{Bloom+06}; \citealt{Berger+05}; \citealt{Hjorth+05}; \citealt{Perley+09}; \citealt{Berger10}; \citealt{Fong+10}).  We note that an analogous model invoking a long-lived magnetar remnant that survives a NS-NS merger could also in principle explain the late-time X-ray activity.  Such a possibility is supported by the recent discovery of a $\approx 2M_{\sun}$ pulsar by \citet{Demorest+10}, which demonstrates that the high density equation of state is relatively stiff.

\section{Conclusions}
\label{sec:conclusions}

In this paper we take the first steps towards developing the millisecond proto-magnetar model into a quantitative theory for gamma-ray bursts.  Using detailed evolutionary models of magnetar spin-down, we explore a wide range of magnetar properties and calculate the prompt emission predicted by magnetic dissipation and internal shock models.  Although the picture we construct may not be accurate in all details, it serves as a `proof of principle' that the basic concepts can be constructed into a self-consistent model.  Our work also provides a baseline for future improvements, as will be necessitated in particular by advances in our understanding of the origin of prompt GRB emission.  

Several theoretical uncertainties remain that should be addressed with future work.  These include a more detailed understanding of the effects of rotation and convection on the cooling evolution of the proto-neutron star, and the  effects of strong magnetic fields on the neutrino-driven mass loss rate.  Although most of our results are at least qualitatively robust to these uncertainties, predictions for the GRB duration (and how it correlates with other observables) is in particular sensitive to the time of neutrino transparency.  The mass loss rate from the proto-NS (and, hence, the wind magnetization) also depends on fraction of the magnetosphere open to outflows, which depends on the poorly-understood sources of dissipation near the Y-point.  Future studies would also be aided by a more detailed understanding of the dependence of the jet properties (e.g. break-out time and opening angle) on the properties of the proto-magnetar and the stellar progenitor.  The source of the rapid rotation and strong magnetic fields required to produce millisecond magnetars also remains a major uncertainty.  However, we note that black hole models place similar, if not more extreme, constraints on the progenitor rotation and the large-scale magnetic field of the central engine (e.g.~\citealt{McKinney06}).

Our primary conclusion is that a surprisingly large fraction of GRB properties can be explained by the magnetar model.  These include:
\begin{itemize}
\item{{\bf Energy.}  Magnetars with properties in the `classical GRB' regime in Figure~\ref{fig:regimes} radiate $E_{\gamma} \sim 10^{50}-10^{52}$ ergs during the GRB phase, consistent with the beaming-corrected gamma-ray energies inferred from afterglow modeling.  Magnetars with stronger(weaker) magnetic fields and/or shorter(longer) initial periods may produce very luminous(low luminosity) GRBs.}
\item{{\bf Lorentz Factor.}  Magnetars in the `classical GRB' regime produce jets with average and instantaneous magnetizations $\sigma_{0} \gtrsim 10^{2}-10^{3}$ (Fig.~\ref{fig:meansig}) which are remarkably similar to the typical Lorentz factors inferred from GRB observations (cf.~Fig.~\ref{fig:meangs}).  The baryon loading of the jet is not fine-tuned or put in by hand, but instead results naturally from the physics of neutrino heating above the proto-magnetar surface.  This is contrast to black hole models, for which current predictions for $\Gamma$ depend on the uncertain rate at which baryons diffuse into an otherwise clean jet (e.g.~\citealt{Levinson&Eichler03}; \citealt{McKinney05}).  The magnetar model predicts that $\sigma_{0}$ (and probably $\Gamma$) increases monotonically with time during the burst.  Among other things, this implies that any thermal emission present will be strongest at early times and will decrease in relative strength as the outflow becomes cleaner with time (Fig.~\ref{fig:magdis2}). }
\item{{\bf Duration.}  The GRB begins once the jet breaks out of the star and becomes optically thin at the internal shock or magnetic dissipation radius.  The GRB ends once the jet magnetization increases sufficiently that jet acceleration and dissipation become ineffective.  Because the latter generally occurs when the NS becomes transparent to neutrinos at $t = t_{\rm\nu-thin} \sim 10-100$ s (Fig.~\ref{fig:pons}), the magnetar model naturally explains the typical durations of long GRBs.}
\item{{\bf Steep Decay Phase.}  The abrupt onset of the high-$\sigma_{0}$ transition at $t \approx t_{\rm \nu-thin}$ (Fig.~\ref{fig:edotsig}) explains why GRB prompt emission decreases rapidly after the prompt emission ends (the `steep decay' phase; e.g.~\citealt{Tagliaferri+05}).}
\item{{\bf Association with Hypernova.}  It is natural to associate energetic, MHD-powered supernovae with magnetar birth.  If the magnetar model is correct, all long GRBs formed from the core collapse of massive stars should be accompanied by an energetic (and possibly hyper-energetic) supernova.  Magnetars formed via AIC, by contrast, may produce long GRBs not accompanied by a bright SN.  This is a promising explanation for the $\sim 100$ second X-ray tails observed following some short GRBs ($\S\ref{sec:AIC}$) and explains why they resemble long GRBs in many of their properties.}
\item{{\bf High Lorentz Factors$\leftrightarrow$Energetic Bursts.} The magnetar model predicts a positive correlation (with significant scatter) between the (energy-weighted) average magnetization $\sigma_{\rm avg}$ of the jet and the (beaming-corrected) GRB luminosity/energy (Fig.~\ref{fig:meansiglum}).  This is consistent with the fact that energetic {\it Fermi} bursts appear to have the largest Lorentz factors.} 
\item{{\bf High Radiative Efficiency.} Both magnetic dissipation and internal shocks may occur in proto-magnetar winds, resulting in the prompt high-energy emission.  Both models predict maximum radiative efficiencies $\epsilon_{\rm r} \sim 30-50\%$, consistent with the high values of $\epsilon_{\rm r}$ inferred from afterglow modeling (e.g.~\citealt{Panaitescu&Kumar01}; \citealt{Zhang+07}; \citealt{Fan&Piran06}).}
\item{{\bf Amati-Yonetoku Relation.}  Our spectral modelling favors magnetic dissipation over internal shocks as the prompt emission mechanism, in part because magnetic dissipation predicts a relatively constant spectral energy peak $E_{\rm peak}$ as a function of time (Fig.~\ref{fig:epeak}).  Strong internal shocks may be suppressed by the residual magnetization of the ejecta or if the toroidal field geometry is not conducive to particle acceleration (e.g.~\citealt{Sironi&Spitkovsky10}).  In combination with the predicted $\sigma_{\rm avg}-L_{\gamma}$ correlation (Fig.~\ref{fig:meansiglum}), the magnetic dissipation model reproduces both the slope and normalization of the observed Amati-Yonetoku correlations.}
\item{{\bf Late-Time Emission.} Although we expect that prompt internal emission becomes ineffective when $\sigma_{0}$ becomes very large at late times, the plateau X-ray afterglow phase may also be powered by magnetar spin-down, as proposed by previous authors and suggested by Figure \ref{fig:edotsig}.  The predicted correlation between the plateau luminosity and duration (Fig.~\ref{fig:tauedot}) is consistent with the sample of `internal' plateaus studied by \citet{Lyons+10}.  Late-time X-ray flaring may be powered by residual rotational or magnetic energy. 
}
\end{itemize}


\begin{table*}
\begin{center}
\vspace{0.05 in}\caption{Properties of Proto-Magnetar Winds}
\label{table:winds}
\resizebox{18cm}{!}{
\begin{tabular}{cccccccccccccc}
\hline
\hline
\multicolumn{1}{c}{$B_{\rm dip}^{(a)}$} &
\multicolumn{1}{c}{$P_{0}$[P$|_{t=0}$]$^{(b)}$} &
\multicolumn{1}{c}{$M_{\rm ns}$} &
\multicolumn{1}{c}{$\chi^{(c)}$} &
\multicolumn{1}{c}{$\eta_{\rm s}^{(d)}$} &
\multicolumn{1}{c}{$\sigma_{\rm bo}^{(e)}$} &
\multicolumn{1}{c}{$\sigma_{\rm avg}^{(f)}$} &
\multicolumn{1}{c}{$\Gamma_{\rm s,avg}^{(g)}$} &
\multicolumn{1}{c}{$T_{90}^{(h)}$} &
\multicolumn{1}{c}{$t_{\rm end}^{(i)}$} &
\multicolumn{1}{c}{$E_{\rm th}^{(j)}$} &
\multicolumn{1}{c}{$E_{\gamma}^{(k)}$} &
\multicolumn{1}{c}{$\dot{E}|t_{\rm end}^{(l)}$} & 
\multicolumn{1}{c}{$\tau_{\rm s}|_{t_{\rm end}}^{(m)}$}
\\
(G) & (ms) & ($M_{\sun}$) & (rad) & - & - & - & - & (s) & (s) & ($10^{50}$ ergs) & ($10^{50}$ ergs) & ($10^{50}$ erg s$^{-1}$) & (s) \\
\hline 
\\
$2\times 10^{15}$ & 1.5[4.3] & 1.4 & $\pi/2$ & 3 & 22 & 570 & 68 & 47 & 60 & 1.6 & 24 & 0.25 & 270 \\
$2\times 10^{15}$ & 1.5[4.3] & 1.2 & $\pi/2$ & 3 & 25 & 500 & 74 & 38 & 51 & 1.3 & 19 & 0.23 & 240 \\
$2\times 10^{15}$ & 1.5[4.3] & 2.0 & $\pi/2$ & 3 & 14 & 760 & 51 & 80 & 96 & 1.8 & 35 & 0.24 & 400 \\
$2\times 10^{15}$ & 1.5[4.3] & 1.4 & 0       & 3 & 40 & 1200 & 140 & 46 & 56 & 0 & 14 & 0.19 & 430 \\
$2\times 10^{15}$ & 1.5[4.3] & 1.2 & 0       & 3 & 54 & 1300 & 180 & 37 & 47 & 0 & 11 & 0.18 & 380 \\
$2\times 10^{15}$ & 1.5[4.3] & 2.0 & 0       & 3 & 17 & 1100 & 70 & 72 & 85 & 0.7 & 19 & 0.17 & 660 \\
$2\times 10^{15}$ & 1.5[4.3] & 1.4 & $\pi/2$ & 1 & 37 & 360 & 93 & 18 & 29 & 0.7 & 12 & 0.28 & 240 \\
$2\times 10^{15}$ & 1.5[4.3] & 1.2 & $\pi/2$ & 1 & 46 & 340 & 110 & 13 & 23 & 0 & 7.7 & 0.27 & 210 \\
$2\times 10^{15}$ & 1.5[4.3] & 2.0 & $\pi/2$ & 1 & 21 & 450 & 60 & 41 & 54 & 1.9 & 22 & 0.25 & 390 \\
$2\times 10^{15}$ & 1.5[4.3] & 1.4 & 0       & 1 & 70 & 750 & 200 & 18 & 28 & 0 & 6.5 & 0.20 & 420 \\
$2\times 10^{15}$ & 1.5[4.3] & 1.2 & 0       & 1 & 100 & 830 & 280 & 12 & 22 & 0 & 4.0 & 0.19 & 370 \\
$2\times 10^{15}$ & 1.5[4.3] & 2.0 & 0       & 1 & 28 & 670 & 87 & 40 & 52 & 0.69 & 14 & 0.17 & 660 \\
$        10^{16}$ & 1[2.8]   & 1.4 & $\pi/2$ & 3 & 340 & 5100 & 740 & 54 & 64 & 0 & 61 & 0.30 & 50 \\
$        10^{16}$ & 1[2.8]   & 1.4 & 0       & 3 & 890 & $2.0\times 10^{4}$ & 2200 & 61 & 71 & 0 & 110 & 0.50 & 53 \\
$        10^{16}$ & 1[2.8]   & 1.4 & $\pi/2$ & 1 & 490 & 4400 & 1100 & 19 & 29 & 0 & 22 & 0.60 & 32 \\
$        10^{16}$ & 1[2.8]   & 1.4 & 0       & 1 & 1200 & $1.4\times 10^{4}$& 3000 & 19 & 29 & 0 & 43 & 1.40 & 33 \\
$5\times 10^{14}$ & 2[5.7]   & 1.4 & $\pi/2$ & 3 & 1.9 & 55 & 6.4 & 30 & 60 & 0.15 & 0.86 & 0.020 & 2600 \\
$5\times 10^{14}$ & 2[5.7]   & 1.4 & 0       & 3 & 2.2 & 62 & 7.5 & 30 & 55 & 0.063 & 0.43 & 0.010 & 5100 \\
$5\times 10^{14}$ & 2[5.7]   & 1.4 & $\pi/2$ & 1 & 3.3 & 31 & 7.8 & 7.6 & 28 & 0.083 & 0.24 & 0.023 & 2300 \\
$5\times 10^{14}$ & 2[5.7]   & 1.4 & $0$ & 1 & 4.0 & 38 & 9.7 & 9.2 & 27 & 0.040 & 0.15 & 0.011 & 4700 \\
$^{\dagger}2\times 10^{15}$ & 1.5[4.3] & 1.28 & $\pi/2$ & 3 & 36 & 350 & 91 & 14 & 25 & 0.94 & 12 & 0.39 & 190 \\ 
$^{\dagger}2\times 10^{15}$ & 1.5[4.3] & 1.28 & 0       & 3 & 81 & 910 & 230 & 14 & 24 & 0 & 5.9 & 0.26 & 330 \\ 
$^{\dagger}        10^{16}$ & 1[2.8]   & 1.28 & $\pi/2$       & 3 & 590 & 4300 & 1300 & 13 & 23 & 0 & 32 & 1.3 & 21 \\
$^{\dagger}        10^{16}$ & 1[2.8]   & 1.28 & 0 & 3 & 1800 & $1.9\times 10^{4}$ & 4400 & 14 & 24 & 0 & 61 & 2.7 & 21 \\
$^{\dagger}5\times 10^{14}$ & 2[5.7]   & 1.28 & $\pi/2$ & 3 & 3.4 & 34 & 8.1 & 5.7 & 25 & 0.072 & 0.21 & 0.027 & 1900 \\
$^{\dagger}5\times 10^{14}$ & 2[5.7]   & 1.28 & 0       & 3 & 4.6 & 47 & 11 & 7.4 & 24 & 0.036 & 0.14 & 0.012 & 4200 \\

\hline
\end{tabular}
}
\end{center}
{\small
$^{(a)}$Surface dipole magnetic field strength following NS contraction.
$^{(b)}$Spin period if NS were to contract to final radius with fixed angular momentum [Actual initial spin period at t=0]. 
$^{(c)}$Magnetic obliquity (see Fig.~\ref{fig:nscartoon}).
$^{(d)}$`Stretch' correction applied to neutrino luminosities and energies to account for the effects of rotation (see eq.~[\ref{eq:stretch}]).
$^{(e)}$Wind magnetization when the jet breaks out of the stellar surface at $t = t_{\rm bo} = 10$ s (eq.~[\ref{eq:tbo}]; Fig.~\ref{fig:sigbo}).
$^{(f)}$Energy-weighted average magnetization between jet break-out and the end of the prompt emission, $t_{\rm bo} \lesssim t \lesssim t_{\rm end}$ (Fig.~\ref{fig:meansig}).
$^{(f)}$Energy-weighted average Lorentz factor of the bulk shell produced by internal shocks (Fig.~[\ref{fig:meangs}]).
$^{(h)}$Duration of the prompt GRB emission $T_{90} \equiv t_{\rm end} - t_{\rm thin,is}$ (Fig.~\ref{fig:T90}).
$^{(i)}$Time after core bounce when the prompt GRB emission ends, defined as the point when the `saturation radius' $r_{\rm mag}$ (eq.~[\ref{eq:rsat}]) exceeds the internal shock radius $r_{\rm sh}$ (eq.~[\ref{eq:rsh}]).  This transition generally occurs simultaneous with the transition of the proto-NS to neutrino transparency (see Fig.~\ref{fig:edotsig}). 
$^{(j)}$Maximum `thermal' energy produced by the jet (eq.~[\ref{eq:etherm}]). 
$^{(k)}$Maximum GRB energy, defined as the rotational energy released in the time interval $min[t_{\rm bo},t_{\rm thin,is}] < t < t_{\rm end}$ (see Fig.~\ref{fig:eGRB}).
$^{(l)}$Wind power at $t \simeq t_{\rm end}$. 
$^{(m)}$Dipole spin-down timescale at $t = t_{\rm end}$. 
$^{\dagger}$Calculated using the neutrino cooling calculations of \citet{Hudepohl+10}.
\\

}
\end{table*}

\section*{Acknowledgments}

We thank J.~Pons for providing the NS cooling curves used in our calculations.  We thank L.~Roberts, P.~Kumar, J.~McKinney, R.~Margutti, J.~Arons, P.~O'Brien, and A.~Spitkovsky for helpful discussions and information.  BDM is supported by NASA through Einstein Postdoctoral Fellowship grant number PF9-00065 awarded by the Chandra X-ray Center, which is operated by the Smithsonian Astrophysical Observatory for NASA under contract NAS8-03060.  DG is supported by the Lyman Spitzer, Jr.~Fellowship award by the Department of Astrophysical Sciences at Princeton University.  TAT is supported in part by an Alfred P.~Sloan Foundation Fellowship.  NB is supported by a Fellowship grant from NORDITA.

\begin{appendix}

\section{Proto-Magnetar Wind Properties}

\label{sec:appendixA}

In this appendix we describe how to calculate the power $\dot{E}$ ($\S\ref{sec:edot}$) and mass loss rate $\dot{M}$ ($\S\ref{sec:mdot}$) of proto-magnetar winds. 

\subsection{Energy Loss Rate $\dot{E}$}
\label{sec:edot}

The winds from millisecond proto-magnetars are accelerated primarily by magnetic forces rather than by thermal pressure.  At large radii the wind power can be divided into components of kinetic energy and magnetic Poynting flux, viz.~$\dot{E} = \dot{E}_{\rm kin} + \dot{E}_{\rm mag}$.  

The kinetic luminosity of the wind is $\dot{E}_{\rm kin} = (\Gamma_{\infty}-1)\dot{M}c^{2}$, where $\Gamma_{\infty} \equiv (1-v_{\infty}^{2}/c^{2})^{-1/2}$ and $v_{\infty}$ are the asymptotic Lorentz factor and velocity of the outflow, respectively.  Nonrelativistic outflows have magnetization $\sigma_{0} \lesssim 1$ and reach an asymptotic speed $v_{\infty} \approx c\sigma_{0}^{1/3}$, resulting in a kinetic luminosity $\dot{E}_{\rm kin} \propto (1/2)\dot{M}v_{\infty}^{2}$ \citep{Lamers&Cassinelli99}.  Relativistic outflows have $\sigma_{0} \gtrsim 1$ and achieve $\Gamma_{\infty} \approx \sigma_{0}^{1/3}$ near the fast magnetosonic surface, beyond which acceleration effectively ceases \citep{Goldreich&Julian70}.  This weak 1/3 power embodies the classical problem that (unconfined, time-stationary) high-$\sigma_{0}$ winds accelerate inefficiently in ideal MHD (\citealt{Kennel&Coroniti84}; see $\S\ref{sec:acceleration}$ for further discussion).

The magnetosphere is completely open outside of the Alfven radius $R_{A}$.  The Poynting flux $\dot{E}_{\rm mag}$ is thus related to the open magnetic flux $\phi$, and hence to the magnetization $\sigma_{0}$ (eq.~[\ref{eq:sigma}]), via the relationship
\begin{eqnarray}
\dot{E}_{\rm mag} \simeq \left(\int\left.\frac{c}{4\pi}|{\mathop E \limits^{\rightarrow}}\times {\mathop B \limits^{\rightarrow}}|_{r}\times r^{2}d\Omega\right)\right|_{R_{A}} \approx \frac{2}{3}\left(\left.v_{\phi}B_{\phi}^{2}r^{2}\right)\right|_{R_{A}} \nonumber \\
\approx \frac{2\phi^{2}\Omega}{3R_{A}} \approx \frac{2}{3}
\left\{
\begin{array}{lr}
\dot{M}c^{2}\sigma_{0}^{2/3}
, \qquad &
\sigma_{0} \ll 1 \\
\dot{M}c^{2}\sigma_{0}
, \qquad &
\sigma_{0} \gg 1 \\
\end{array}
\right.,
\label{eq:edotmag}
\end{eqnarray}
where ${\mathop E \limits^{\rightarrow}} = -({\mathop v \limits^{\rightarrow}}/c)\times{\mathop B \limits^{\rightarrow}} $ is the electric field and the factor 2/3 accounts for the angular integration.  The equalities in the second line follow because (1) the outflow co-rotates with the star, such that $v_{\phi} \sim \Omega r$ out to radii $\sim R_{A}$, where $v_{\phi}$ is the toroidal velocity; and (2) near $R_{A}$ the poloidal field begins to bend back appreciably due to the fluid's inertia.  The toroidal magnetic field strength $B_{\phi}$ thus becomes comparable to the poloidal field $\sim B_{\rm r}$ at $r \sim R_{A}$, such that $\phi \equiv B_{\rm r}r^{2}|_{R_{A}} \simeq B_{\phi}|_{R_{A}}R_{A}^{2}$ (the equality is exact in the case of force-free winds).  In writing the third and fourth lines we have made use of the fact that 
\begin{equation}
R_{A} =
\left\{
\begin{array}{lr}
v_{\infty}/\Omega \sim R_{\rm L}\sigma_{0}^{1/3}
, \qquad &
\sigma_{0} \ll 1 \\
R_{\rm L} = c/\Omega
\qquad & 
\sigma_{0} \gg 1 \\
\end{array}
\right..
\label{eq:ra}
\end{equation}

The open magnetic flux $\phi$ of a rotating dipole with a surface magnetic field strength $B_{\rm dip}$ is given by
\be
\phi \simeq f_{\rm open}B_{\rm dip}R_{\rm ns}^{2},
\label{eq:phi}
\ee
where 
\begin{eqnarray}
f_{\rm open} &&\simeq (1/2\pi)\int_{0}^{2\pi}d\phi \int_{0}^{\theta_{\rm open}/2}\sin\theta\,d\theta \nonumber \\
&=&1-\cos\left(\frac{\theta_{\rm open}}{2}\right) {\mathop \simeq \limits_{R_{\rm Y} \gg R_{\rm ns}}}  \frac{\theta_{\rm open}^{2}}{8} \approx (1+\sin^{2}\chi)^{1/2}\frac{R_{\rm ns}}{2R_{\rm Y}} \nonumber \\
\label{eq:fopen}
\end{eqnarray}
is the fraction of the NS surface threaded by open field.  Here $\theta_{\rm open} \simeq 2\sin^{-1}[(R_{\rm ns}/R_{\rm Y})^{1/2}]$ is the opening angle of the polar cap, corrected by a factor $(1+\sin^{2}\chi)^{1/2}$ to account for the larger cap size of an oblique rotator (e.g.~\citealt{Cheng+00}; \citealt{Bai&Spitkovsky10}); $R_{\rm ns}$ is the NS radius; $R_{\rm Y}$ is the `Y' point radius where the close zone ends in the magnetic equatorial plane; and the second equality holds in the small-cap limit $\theta_{\rm open} \ll 1$ ($R_{\rm Y} \gg R_{\rm ns}$).  See Figure \ref{fig:nscartoon} for an illustration of the relevant geometry.

Just after core bounce, thermal pressure may dominate above the NS surface and the entire magnetosphere may open into a `split-monopole' configuration with $f_{\rm open} \sim 1$.  As the NS contracts, cools, and spins up, however, its magnetic field is amplified and magnetic pressure eventually comes to dominate.  This produces a `closed' or `dead' zone at low magnetic latitudes from which a steady-state wind cannot escape (i.e.~$R_{\rm Y} > R_{\rm ns}$).  In the limit of a force-free wind ($\sigma_{0} \gg 1$) the radius of the Y-point likely extends close to the radius of the light cylinder, but in general $R_{\rm Y}$ is $\le R_{\rm L}$ for less magnetized (finite-$\sigma_{0}$) winds.  Following \citet{Metzger+07}, we assume that $R_{\rm Y}/R_{\rm L} = \min[0.3\sigma_{0}^{0.15},1]$ for $R_{\rm Y} > R_{\rm ns}$, based on an empirical fit to the axisymmetric relativistic MHD simulations of \citet{Bucciantini+06}, which span the non-relativistic to relativistic transition.  The values of $R_{\rm Y}$ that we adopt are similar to those we estimate by applying the toy model of \citet{Mestel&Spruit87} to the proto-magnetar context.  Determining the detailed time-dependence of $R_{\rm Y}$ will, however, ultimately require incorporating a self-consistent, physical model for the resistivity in the magnetosphere and equatorial current sheet. 

Combining our results, the total wind power is given by
\begin{eqnarray}
\dot{E} &=& \dot{E}_{\rm mag} + \dot{E}_{\rm kin} \nonumber \\
&\simeq& \left\{
\begin{array}{lr}
\dot{M}c^{2}\sigma_{0}^{2/3}
, &
\sigma_{0} \ll 1 \\
 (2/3)\dot{M}c^{2}\sigma_{0}
, &
\sigma_{0} \gg 1 \\
\end{array}
\right., \nonumber \\
\label{eq:edot}
\end{eqnarray}
where the magnetization (eq.~[\ref{eq:sigma}]) can now be written 
\begin{eqnarray}
\sigma_{0} &\simeq& \frac{B_{\rm dip}^{2}R_{\rm ns}^{4}\Omega^{2}f_{\rm open}^{2}}{\dot{M}c^{3}} \nonumber \\ &{\mathop \simeq \limits_{R_{\rm Y} \gg R_{\rm ns}}}&  \frac{B_{\rm dip}^{2}R_{\rm ns}^{6}\Omega^{4}(1+\sin^{2}\chi)}{4\dot{M}c^{5}}\left(\frac{R_{\rm Y}}{R_{\rm L}}\right)^{-2}. 
\label{eq:sig0}
\end{eqnarray}
Note that in the nonrelativistic case the kinetic and magnetic contributions to the total power are similar ($\dot{E}_{\rm mag} = 2\dot{E}_{\rm kin}$), while in the relativistic case the outflow is Poynting dominated since $\dot{E}_{\rm mag}/\dot{E}_{\rm kin} \sim \sigma_{0}^{2/3} \gg 1$.  

\subsection{Mass Loss Rate $\dot{M}$}
\label{sec:mdot}

Mass loss from the proto-NS results from neutrino heating in the atmosphere just above the NS surface.  The dominant heating and cooling processes are the charged-current reactions
\be 
\nu_{e} + n \leftrightarrow e^{-} + p\,\,\,\,\,{\rm and}\,\,\,\,\,\bar{\nu}_{e} + p \leftrightarrow e^{+} + n
\label{eq:chargedcurrent}
\ee
For unmagnetized winds, the mass loss rate is well-approximated by the analytic expression \citep{Qian&Woosley96}
\begin{eqnarray}
\dot{M}_{\nu} = 5\times 10^{-5}M_{\sun}{\,\rm s^{-1}}\left(\frac{L_{\nu}}{10^{52}{\,\rm ergs\,s^{-1}}}\right)^{5/3}\times \nonumber \\\left(\frac{\epsilon_{\nu}}{10{\,\rm MeV}}\right)^{10/3}\left(\frac{M_{\rm ns}}{1.4M_{\sun}}\right)^{-2}\left(\frac{R_{\rm ns}}{10{\,\rm km}}\right)^{5/3}(1+\epsilon_{\rm es})^{5/3}.
\label{eq:mdotnu}
\end{eqnarray}
Although both electron neutrinos and antineutrinos contribute to the heating, for simplicity we combine their contributions into a single product of the neutrino luminosity $L_{\nu}$ and mean energy $\epsilon_{\nu}$, defined by 
\be 
L_{\nu}\epsilon_{\nu}^{2} \equiv L_{\nu_{e}}|\epsilon_{\nu_{e}}|^{2} + L_{\bar{\nu}_{e}}|\epsilon_{\bar{\nu}_{e}}|^{2},
\label{eq:lnu}
\ee 
where the $|...|$ represent an appropriate average over the neutrino absorption cross sections.  The normalization adopted in equation (\ref{eq:mdotnu}) includes both this averaging and a general relativistic correction \citep{Thompson+01}. The parameter
\be
\epsilon_{\rm es} \equiv 0.2\left(\frac{M_{\rm ns}}{1.4M_{\sun}}\right)\left(\frac{R_{\rm ns}}{10{\,\rm km}}\right)^{-1}\left(\frac{\epsilon_{\nu}}{10{\,\rm MeV}}\right)^{-1}
\ee
is a correction $\lesssim 1$ for the additional heating due to inelastic electron scattering (see \citealt{Qian&Woosley96}, eq.~50).    

\begin{figure*}
\begin{center}$
\begin{array}{cc}
\includegraphics[width=8.5cm]{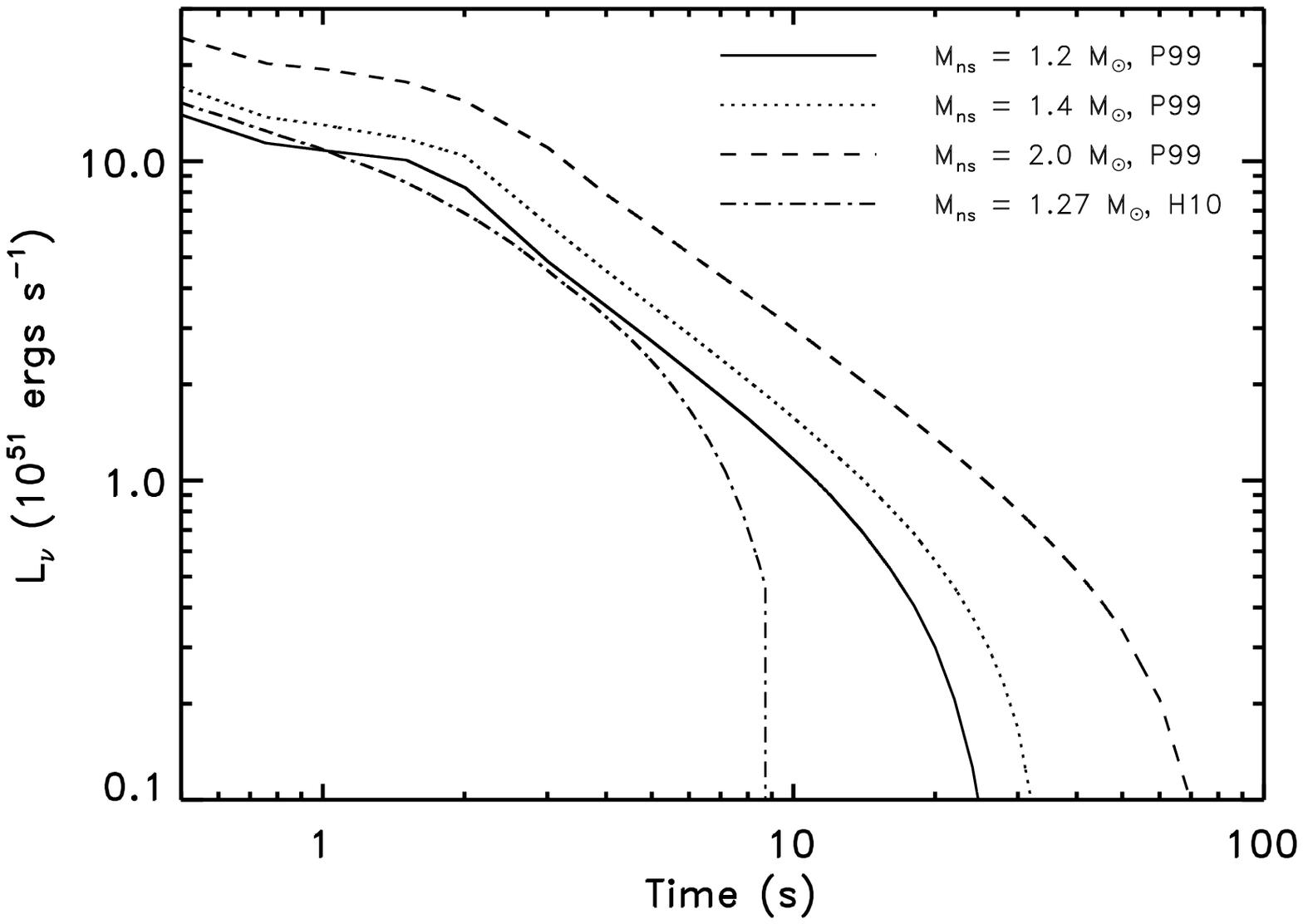} &
\includegraphics[width=8.5cm]{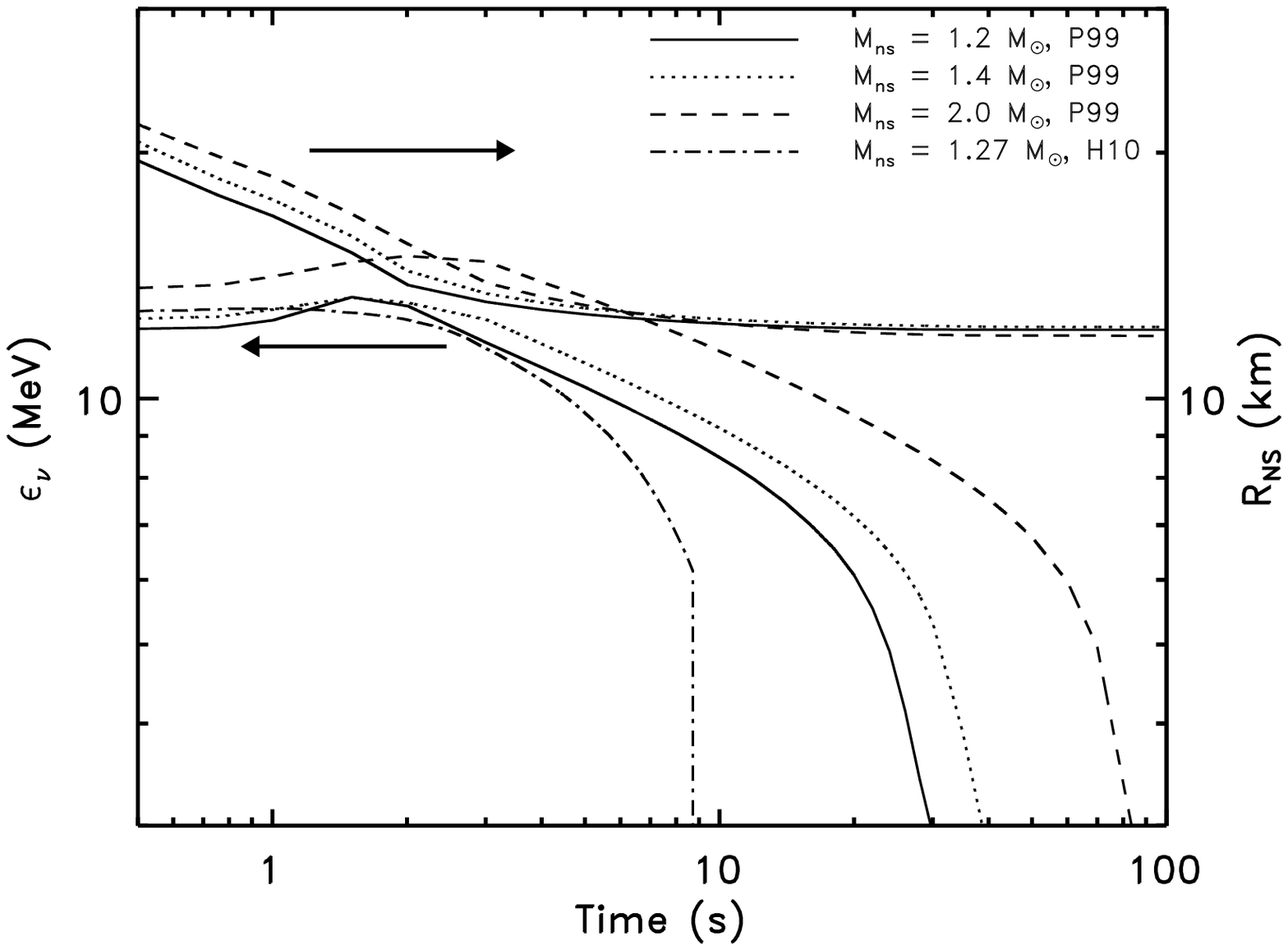}
\end{array}$
\end{center}
\caption{{\it \bf--Left Panel:} Weighted neutrino luminosity $L_{\nu}$ (defined in eq.~[\ref{eq:lnu}]) as a function of time after core bounce.  The first three models are from \citet{Pons+99}, calculated for NS masses $M_{\rm ns} = 1.2M_{\sun}$ ({\it solid line}), 1.4$M_{\sun}$ ({\it dotted line}), and $2.0M_{\sun}$ ({\it dashed line}).  The last model is from \citet{Hudepohl+10} ($M_{\rm ns} = 1.27M_{\sun}$; {\it dot-dashed line}). {\it \bf Right Panel:} Mean electron anti-neutrino energy $\epsilon_{\nu}$ ({\it left axis}) and NS radius ({\it right axis}) as a function of time after core bounce, for the same models shown in the left panel (the radius from H10 is not shown).}  
\label{fig:pons}
\end{figure*}

In most calculations we use $L_{\nu}(t), \epsilon_{\nu}(t),$ and $R_{\rm ns}(t)$ from \citet{Pons+99}, hereafter P99, who calculate the deleptonization and cooling evolution of non-rotating proto-NSs (cf.~\citealt{Burrows&Lattimer86}).  Examples of $L_{\nu}(t), \epsilon_{\nu}(t),$ and $R_{\rm ns}(t)$ are shown in Figure \ref{fig:pons} for different NS masses.  Note that for $t\gtrsim 1$ s, $L_{\nu}$ and $\epsilon_{\nu}$ decrease relatively gradually as a power-law until a time $t_{\rm \nu-thin} \sim 10-60$ seconds, after which $L_{\nu}$ and $\epsilon_{\nu}$ plummet as the proto-NS becomes transparent to neutrinos.  As we show in $\S\ref{sec:stages}$, $t_{\rm \nu-thin}$ determines the GRB duration in the proto-magnetar model.  

Since $\dot{M}_{\nu}$ depends sensitively on $L_{\nu}$ and $\epsilon_{\nu}$ we briefly discuss the limitations and the uncertainties in the calculations of P99.  First, although portions of the proto-NS are convectively unstable during its early cooling evolution (\citealt{Burrows&Fryxell93}; \citealt{Keil+96}), convective transport is not accounted for by P99.  The primary effect of convection is to increase the cooling rate and hence to speed up the temporal evolution of the neutrino luminosity (L.~Roberts, private communication).  P99 find that the rate at which $L_{\nu}$ and $\epsilon_{\nu}$ decrease at late times, and hence the precise value of $t_{\rm \nu-thin}$, also depends sensitively on the high density equation of state, which is uncertain.  In order to explore the sensitivity of our results to uncertainties in $L_{\nu}$ and $\epsilon_{\nu}$, we also calculate models using neutrino luminosities and energies from the recent proto-NS cooling calculations of \citet{Hudepohl+10}, hereafter H10 (L.~Roberts, private communication), which follow a successful electron-capture supernova \citep{Kitaura+06}.  This calculation, which includes improvements in the neutrino opacities over previous work, is shown for comparison in Figure \ref{fig:pons}.  The primary difference between the cooling curves of P99 and H10 is the significantly faster late-time evolution found by H10.

Finally, neither P99 or H10 include the effects of magnetic fields or rotation, yet this paper focuses on proto-magnetars rotating at a significant fraction of their break-up speed.  Rapid rotation decreases the interior temperature of the NS, which slows its cooling evolution.  Using one-dimensional rotating core collapse calculations, \citet{Thompson+05} find that $L_{\nu}$ and $\epsilon_{\nu}$ are reduced by factors of $\sim 0.5$ and $\sim 0.8$, respectively in their fastest rotating model at $t \approx 600$ ms following core bounce, compared to an otherwise equivalent nonrotating case.  Ideally the effects of rotation on $L_{\nu}$ and $\epsilon_{\nu}$ should be calculated self-consistently.  Lacking such a model, however, we account for rotational effects qualitatively by introducing a `stretch' parameter $\eta_{s}$, which modifies the cooling evolution from the non-rotating case ($\Omega = 0$) as follows:
\begin{eqnarray}
L_{\rm \nu} \rightarrow L_{\rm \nu}|_{\Omega=0}\eta_{s}^{-1};\,\, t \rightarrow t|_{\Omega=0}\eta_{s};\,\, \epsilon_{\rm \nu} \rightarrow \epsilon_{\rm \nu}|_{\Omega=0}\eta_{s}^{-1/4},
\label{eq:stretch}
\end{eqnarray}
where a value $\eta_{s} \sim $ few is motivated by the calculations of \citet{Thompson+05} for millisecond rotators.  Note that this simple parametrization preserves the total energy radiated in electron neutrinos and increases the time of neutrino transparency $t_{\rm \nu-thin} \propto \eta_{s}$.  Although we expect $\eta_{\rm s}$ to be an increasing function of $\Omega$, in our calculations we fix $\eta_{\rm s} = 3$ for lack of a predictive model.  We also neglect differences in the neutrino radiation field with latitude caused by rapid rotation (e.g.~\citealt{Brandt+10}), which if properly included would impart the total wind mass loss rate with an additional dependence on the magnetic obliquity $\chi$

A strong magnetic field modifies $\dot{M}$ from the standard expression in equation (\ref{eq:mdotnu}) in three ways.  First, $\dot{M}$ is reduced by a factor $f_{\rm open}$ (eq.~[\ref{eq:fopen}]) since only the open fraction of the surface contributes to the outflow.  Second, $\dot{M}$ is enhanced by a factor $f_{\rm cent}$ due to centrifugal `slinging.' This occurs when rotation is sufficiently rapid and the magnetic field is sufficiently strong that centrifugal forces increase the scale height in the heating region \citep{Thompson+04}.  By fitting the numerical results of \citet{Metzger+08a} we find that the {\it maximum} centrifugal enhancement to $\dot{M}$ (obtained in the strong-field limit of strict co-rotation described below) is well-approximated by the functional form (for $P \gtrsim 1{\,\rm ms}$)
\be
f_{\rm cent,max} \simeq \exp[(P_{c}/P)^{\beta}]
\label{eq:fcentmax}
\ee
for a value $\beta \simeq 1.5$, where 
\be
P_{c} \simeq 2.1 \sin{\alpha}\left(\frac{R_{\rm ns}}{10{\,\rm km}}\right)^{3/2}\left(\frac{M_{\rm ns}}{1.4M_{\sun}}\right)^{-1/2}\,{\rm\,ms},
\ee
where $\alpha \approx$ max[$\theta_{\rm open}/2,\chi$] is a typical angle from the rotational axis sampled by the open zone.  The normalization we adopt for $P_{c}$ is determined by fitting the numerical results of \citet{Metzger+08a}, which were calculated for equatorial field lines ($\alpha = \pi/2$).  The scaling of $P_{c}$ with mass, radius, and $\alpha$, however, are chosen based on the theoretical expectation that $P_{c} \propto R_{\perp}/c_{\rm s}$ \citep{Thompson+04}, where $R_{\perp} \sim R_{\rm ns}\sin\alpha$ and $c_{\rm s}$ are the centrifugal `lever arm' and sound speed in the gain region, respectively.  The latter is proportional to the NS escape speed $\propto (M/R_{\rm ns})^{1/2}$ (see \citealt{Qian&Woosley96}, eq.~45).  

Although $f_{\rm cent,max}$ is the maximum enhancement of $\dot{M}$, it obtains only if the magnetic field is sufficiently strong that the outflow co-rotates with the star to a location outside the sonic radius $R_{\rm s}$.  This requires $R_{A} \gtrsim R_{\rm s} = \left(\frac{GM}{\Omega^{2}}\right)^{1/3}$ \citep{Lamers&Cassinelli99}.  Using the numerical results of \citet{Metzger+08a}, we find that a satisfactory interpolation of the mass loss enhancement between the centrifugal ($R_{\rm A} \gg R_{\rm s}$) and non-centrifugal ($R_{\rm A} \ll R_{\rm s}$) regimes is given by $f_{\rm cent} = f_{\rm cent,max}(1-\exp[-R_{\rm A}/R_{\rm s}]) + \exp[-R_{\rm A}/R_{\rm s}]$.  Note that relativistic outflows are necessarily in the centrifugal regime because for $\sigma_{0} \gg 1$, $R_{A} \sim R_{\rm L} = c/\Omega > R_{\rm s}$ (eq.~[\ref{eq:ra}]).  

Finally, a strong magnetic field changes $\dot{M}_{\nu}$ (eq.~[\ref{eq:mdotnu}]) by altering the neutrino heating and cooling rates in the proto-NS atmosphere (e.g.~\citealt{Duan&Qian04}; \citealt{Riquelme+05}).  The most important effect is that the electrons and positrons participating in the charged-particle reactions (eq.~[\ref{eq:chargedcurrent}]) are restricted into discrete Landau levels \citep{Duan&Qian04}.  In this paper we neglect these effects because we estimate that the corrections to $\dot{M}_{\nu}$ are relatively minor for surface field strengths $B_{\rm dip} \lesssim 3\times 10^{16}$ G.\footnote{Note that $\dot{M}_{\nu}$ depends most sensitively on the heating and cooling rates in the gain region, which is typically located a few kilometers above the NS surface, where the magnetic field strength $B\propto r^{-3}$ is a factor of a few weaker than the surface field strength.}  However, more detailed future work should address the dependence of $\dot{M}_{\nu}$ on $B_{\rm dip}$.    

Once the NS becomes transparent to neutrinos at late times, $L_{\nu}$ and $\epsilon_{\nu}$ decrease rapidly (Fig.~\ref{fig:pons}) and the mass loss rate decreases abruptly.  Neutrino heating only determines $\dot{M}$ so long as the magnetosphere is sufficiently dense that vacuum electric fields do not develop.  This assumption breaks down, however, once $\dot{M}$ decreases to near the critical \citet{Goldreich&Julian69} (GJ69) mass loss rate.  However, because in actual pulsar winds $\dot{M}$ exceeds the GJ69 value, we instead assume that the minimum mass loss rate in the pair-dominated regime is given by a multiple of GJ69 rate, viz.~
\begin{eqnarray}
\dot{M}_{\rm GJ} &\equiv& \mu_{-+} m_{\rm e}(I/e) \sim 3\times 10^{-15}M_{\sun}{\rm s}^{-1}\times \nonumber \\
&& \left(\frac{\mu_{-+}}{10^{6}}\right)\left(\frac{B_{\rm dip}}{10^{15}{\,\rm G}}\right)\left(\frac{P}{\,\rm ms}\right)^{-2}\left(\frac{R_{\rm ns}}{\,\rm 10\,km}\right)^{3},
\label{eq:mdotGJ}
\end{eqnarray}
where $I \equiv 4\pi R_{\rm L}^{2}\eta_{\rm GJ}|_{\rm R_{\rm L}}c$, $m_{\rm e}$ and $e$ are the electron mass and charge, and $\eta_{\rm GJ} \approx (\Omega B/2\pi c)$ is the GJ69 charge density, evaluated at the light cylinder.  The multiplicity $\mu_{-+}$ of positron/electrons produced by magnetospheric acceleration is uncertain, especially in the case of magnetars \citep{Thompson08}.  Lacking a predictive model, in our calculations we fix the multiplicity at a value $\mu_{-+} = 10^{6}$ which is consistent with estimates based on detailed synchrotron emission models of pulsar wind nebulae (e.g.~\citealt{Bucciantini+10}).  Although the late-time wind magnetization depends sensitively on the multiplicity ($\sigma_{0}|_{t \gg t_{\rm \nu-thin}} \propto 1/\mu_{-+}$), most of our conclusions regarding late-time emission ($\S\ref{sec:highsig}$) are insensitive to this choice.        

To summarize, the mass loss rate is given by
\begin{equation}
\dot{M} =
\left\{
\begin{array}{lr}
\dot{M}_{\nu}f_{\rm open}
, \qquad &
(\chi + \theta_{\rm open}/2) \ll \pi/2 \\
\dot{M}_{\nu}f_{\rm open}f_{\rm cent}
, \qquad &
(\chi + \theta_{\rm open}/2) \gtrsim \pi/2 \\
\dot{M}_{\rm GJ}
, \qquad &
\dot{M} \le \dot{M}_{\rm GJ} 
\end{array}
\right.,
\label{eq:mdot}
\end{equation}
where
\be
f_{\rm cent} = f_{\rm cent,max}(1-\exp[-R_{\rm A}/R_{\rm s}]) + \exp[-R_{\rm A}/R_{\rm s}]
\ee
and
\be
\frac{R_{\rm A}}{R_{\rm s}} \simeq \left(\frac{\sigma_{0}c^{3}}{GM\Omega}\right)^{1/3}
\ee

\section{Internal Shock Emission}
\label{sec:isappendix}

In this Appendix we derive the luminosity and synchrotron spectral peak of internal shock emission from proto-magnetar winds as described in $\S\ref{sec:internalshocks}$.      

In a small interval $dt$ centered around the time $t_{\rm j}$, the jet releases a mass $dM_{\rm j} = \dot{M}_{\rm j}|_{t_{\rm j}}dt$ with energy $dE_{\rm j} = \dot{E}_{\rm j}|_{t_{\rm j}}dt$ and Lorentz factor $\Gamma_{\rm j}|_{t_{\rm j}}$.  The collision between this fast shell and the bulk shell occurs at the time $t_{\rm sh}$ and radius $R_{\rm sh}$ defined by the condition
\be
R_{\rm sh} = \int_{t_{\rm bo}}^{t_{\rm sh}}\beta_{\rm s}dt = (t_{\rm sh}-t_{\rm j})\beta_{\rm j},
\ee
where $\beta_{\rm j} = (1-\Gamma_{\rm j}^{-2})^{1/2}$.  If $\Gamma_{\rm j},\Gamma_{\rm s} \gg 1$ (usually valid at all times) and $\Gamma_{\rm j} \gg \Gamma_{\rm s}$ (valid at late times) one finds that  
\be
t_{\rm sh} {\mathop \simeq \limits_{\Gamma_{\rm j} \gg \Gamma_{\rm s}}}   t_{\rm bo} + 2\Gamma_{\rm s}^{2}t_{\rm j}\left(1 + \frac{\Gamma_{\rm s}^{2}}{\Gamma^{2}}\right);
\label{eq:tsh}
\ee
\be
R_{\rm sh} {\mathop \simeq \limits_{\Gamma_{\rm j} \gg \Gamma_{\rm s}}} 2\Gamma_{\rm s}^{2}ct_{\rm j}\left(1-\frac{1}{2\Gamma_{\rm s}^{2}}\right).
\label{eq:rsh}
\ee
Because both $\Gamma_{\rm s}$ and $t_{\rm j}$ increase with time, $R_{\rm sh}$ also increases as the GRB proceeds.  Neglecting cosmological dilation, emission from the shock is received by a distant observer at the time 
\be
t_{\rm obs} = t_{\rm sh} - R_{\rm sh}/c + t_{\rm bo} {\mathop \simeq \limits_{\Gamma_{\rm j} \gg \Gamma_{\rm s}}}  t_{\rm j} + t_{\rm bo}, 
\ee
where in the second equality we have used equations (\ref{eq:tsh}) and (\ref{eq:rsh}).  This shows explicitly how relativistic effects conspire to produce emission on timescales matching those set by the central engine.  
When the fast shell becomes fully incorporated into the bulk, the bulk Lorentz factor increases by an amount $d\Gamma_{\rm s}$ determined by conservation of momentum,
\be
\frac{d\Gamma_{\rm s}}{\Gamma_{\rm s}} \simeq \frac{dM_{\rm j}}{2M_{\rm s}}\left(\frac{\Gamma_{\rm j}}{\Gamma_{\rm s}} - \frac{\Gamma_{\rm s}}{\Gamma_{\rm j}}\right),
\ee
resulting in the release of thermal energy 
\begin{eqnarray}
dE_{\rm sh} &=& M_{\rm s}\Gamma_{\rm s}c^{2} + dM_{\rm j}\Gamma_{\rm j}c^{2} - (dM_{\rm j} + M_{\rm s})(\Gamma_{s} + d\Gamma_{\rm s}) \nonumber \\
&\approx& \frac{dM_{\rm j}c^{2}}{2}\left(\Gamma_{\rm j} + \frac{\Gamma_{s}^{2}}{\Gamma_{\rm j}} - 2\Gamma_{s}\right)
\end{eqnarray}
If a fraction $\epsilon_{\rm e}$ of $dE_{\rm sh}$ is imparted to the electrons, which then radiate it away on the expansion timescale, the total radiative efficiency is given by
\be
\epsilon_{\rm r} \equiv \frac{dE_{\rm sh}\epsilon_{e}}{dM\Gamma_{\rm j}c^{2}} = \frac{\epsilon_{\rm e}}{2}\left(1 + \frac{\Gamma_{\rm s}^{2}}{\Gamma_{\rm j}^{2}} - 2\frac{\Gamma_{\rm s}}{\Gamma_{\rm j}}\right).
\label{eq:epsilonr}
\ee
At early times $\Gamma_{\rm j} \gtrsim \Gamma_{\rm s}$ and $\epsilon_{\rm r} \ll 1$, but at later times when $\Gamma_{\rm j} \gg \Gamma_{\rm s}$, $\epsilon_{\rm r}$ reaches a maximum value $\approx \epsilon_{\rm e}/2$.
 
We now consider how the jet and shell interact in greater detail in order to evaluate the peak energy of the resulting synchrotron emission.  In the frame of the bulk shell (hereafter denoted by a tilde), the jet velocity and Lorentz factor are given by
\be
\tilde{\beta}_{\rm j} \simeq \frac{1 - \Gamma_{\rm s}^{2}/\Gamma_{\rm j}^{2}}{1 + \Gamma_{\rm s}^{2}/\Gamma_{\rm j}^{2}};\,\,\,\,\,\,\tilde{\Gamma}_{\rm j} \equiv (1-\tilde{\beta}_{\rm j}^{2})^{-1/2} {\mathop \simeq \limits_{\Gamma_{\rm j} \gg \Gamma_{\rm s}}} \Gamma_{\rm j}/2\Gamma_{\rm s}.
\label{eq:betajetprime}
\ee
Note that unlike in the standard internal shock scenario, the relative Lorentz factor between the shells $\tilde{\Gamma}_{\rm j}$ is $\gg 1$ at late times when $\Gamma_{\rm j} \gg \Gamma_{\rm s}$.  This allows for high radiative efficiency.

If the shocked gas is relativistically hot, the (rest frame) post-shock number and energy densities are given, respectively, by \citep{Blandford&McKee76}
\be
n_{\rm sh} = (4\tilde{\Gamma}_{\rm j} + 3)n_{\rm j};\,\,\,\epsilon_{\rm sh} = (\tilde{\Gamma}_{\rm j}-1)n_{\rm sh}m_{\rm p}c^{2},
\ee
 where 
\be
n_{\rm j} = \frac{L_{\rm j,iso}}{4\pi\Gamma_{\rm j}^{2}R_{\rm sh}^{2}m_{\rm p}c^{3}}
\ee
is the pre-shock density and $L_{\rm j,iso} \equiv L_{\rm j}f_{\rm b}^{-1}$.

If a fraction $\epsilon_{e}$ and $\epsilon_{b}$ of the post-shock energy is partitioned into electron kinetic and magnetic energy, respectively, the resulting (electron) random Lorentz factor and magnetic field strength are given by
\be
\tilde{\gamma}_{e} = \frac{\epsilon_{e}m_{\rm p}(\tilde{\Gamma}_{\rm j}-1)}{\zeta_{e}m_{e}} + 1 {\mathop \simeq \limits_{\Gamma_{\rm j} \gg \Gamma_{\rm s}}} \frac{\epsilon_{e}m_{\rm p}\Gamma_{\rm j}}{2\zeta_{e}m_{e}\Gamma_{\rm s}};
\ee
\be
\tilde{B} = (8\pi\epsilon_{\rm b}\tilde{\epsilon}_{\rm sh})^{1/2} \approx \left(\frac{2\epsilon_{\rm b}L_{\rm j,iso}}{\Gamma_{\rm s}^{2}R_{\rm sh}^{2}c}\right)^{1/2} {\mathop \simeq \limits_{\Gamma_{\rm j} \gg \Gamma_{\rm s}}} \frac{\epsilon_{b}^{1/2}L_{\rm j,iso}^{1/2}}{2^{1/2}\Gamma_{\rm s}^{3}c^{3/2}t_{\rm j}}
\ee
where $\zeta_{e}$ is the fraction of electrons accelerated.

The peak synchrotron photon energy as seen by the observer is then
\begin{eqnarray}
E_{\rm peak,is} &\approx& \frac{eB\hbar\gamma_{e}^{2}\Gamma_{\rm s}}{m_{e}c} {\mathop \simeq \limits_{\Gamma_{\rm j} \gg \Gamma_{\rm s}}} 4.7{\,\rm MeV}\epsilon_{e}^{2}\epsilon_{b}^{1/2}\zeta_{e}^{-1} \times \nonumber \\
&&\left(\frac{L_{\rm j,iso}}{10^{51}\,{\rm erg\,s^{-1}}}\right)^{1/2}\left(\frac{t_{\rm j}}{10{\,\rm s}}\right)^{-1}\Gamma_{\rm j}^{2}\Gamma_{\rm s}^{-4}
\label{eq:epeakis}
\end{eqnarray}

\end{appendix}

\bibliographystyle{mn2e}
\bibliography{../biblio/bibliography}

\label{lastpage}

\end{document}